\documentclass[rmp,twocolumn,tightenlines,letterpaper,superscriptaddress,nofootinbib,showkeys,aps,floatfix]{revtex4-1}

\usepackage{amsmath}	
\usepackage{bm}		    
\usepackage{amssymb}	
\usepackage{old-arrows} 
\usepackage{xfrac}	    
\usepackage{mathrsfs}   
\usepackage{bbold}      
\usepackage{xcolor}
\usepackage{graphicx} 	
\usepackage{tensor}		
\usepackage{lipsum}     
\usepackage{hyperref} 	
\usepackage[all]{hypcap}
\hypersetup{
    colorlinks=true,	
    linkcolor=blue,		
    citecolor=blue,		
    filecolor=magenta,	
    urlcolor=black	    
}

\newcommand{\del}[1]{\nabla_{\! \! #1}}
\newcommand{\metric}{{\fontfamily{times}\selectfont\textit{g}}}
\newcommand{\metdet}{g}
\newcommand{\sqrtg}{\sqrt{-\metdet}}
\newcommand{\dete}{e}
\newcommand{\s}[1]{\mathrm{s}_{#1}}
\newcommand{\metricMink}{\eta}
\newcommand{\per}{\ . \ }
\newcommand{\com}{\ , \ }

\newcommand{\ie}{{\it i.e.}}
\newcommand{\eg}{{\it e.g.}}
\newcommand{\bbone}{\mathbb{1}}
\newcommand{\ssfrac}[2]{\sfrac{#1\mkern-2.2mu}{#2}}
\newcommand{\half}{\ssfrac{1}{2}}
\newcommand{\onehalf}{\ssfrac{1}{2}}
\newcommand{\threehalf}{\ssfrac{3}{2}}
\newcommand{\sixth}{\ssfrac{1}{6}}
\newcommand{\halfBM}{\texorpdfstring{$\ssfrac{1}{2}$}{1/2}}

\newcommand{\pprime}{{\prime\prime}}
\newcommand{\ii}{\mathrm{i}}
\newcommand{\dd}{\mathrm{d}}
\newcommand{\ee}{\mathrm{e}}

\newcommand{\Hcal}{\mathcal{H}}
\newcommand{\Rcal}{\mathcal{R}}
\newcommand{\xvec}{{\bm x}}
\newcommand{\yvec}{{\bm y}}
\newcommand{\pvec}{{\bm p}}
\newcommand{\kvec}{{\bm k}}
\newcommand{\qvec}{{\bm q}}
\newcommand{\rvec}{{\bm r}}
\newcommand{\dvec}{\bm \nabla}

\newcommand{\khat}{\hat{\kvec}}

\newcommand{\Lcal}{\mathcal{L}}
\newcommand{\normord}[1]{\xcentcolon\mathrel{#1}\xcentcolon}
\newcommand{\xcentcolon}{%
  \mathrel{\vbox{\hbox{$:$}\kern.2ex}}%
}
\newcommand{\aphi}{\Phi}
\newcommand{\bphi}{\phi}
\newcommand{\cphi}{\varphi}
\newcommand{\dphi}{\Phi}
\newcommand{\rhohat}{\hat{\rho}}
\newcommand{\drho}{\delta \hspace{-0.03cm} \rho}
\newcommand{\drhohat}{\delta \hspace{-0.03cm} \hat{\rho}}
\newcommand{\pressure}{P}
\newcommand{\barpressure}{\bar{\pressure}}
\newcommand{\dpressure}{\delta \hspace{-0.03cm} \pressure}
\newcommand{\CMB}{\text{\sc cmb}}
\newcommand{\BBN}{\text{\sc bbn}}

\newcommand{\FLRW}{\text{\sc flrw}}

\newcommand{\EH}{\text{\sc eh}}
\newcommand{\M}{\text{\sc m}}
\newcommand{\IN}{\text{\sc in}}
\newcommand{\OUT}{\text{\sc out}}
\newcommand{\BH}{\mathrm{BH}}
\newcommand{\MPl}{M_\mathrm{Pl}}
\newcommand{\Mpl}{\MPl}

\newcommand{\Hinf}{H_\mathrm{inf}}
\newcommand{\ta}{\tilde{a}}
\newcommand{\tm}{\tilde{m}}
\newcommand{\tH}{\tilde{H}}
\newcommand{\tk}{\tilde{k}}
\newcommand{\teta}{\tilde{\eta}}
\newcommand{\END}{e}
\newcommand{\etae}{\eta_\END}
\newcommand{\te}{t_\END}
\renewcommand{\ae}{a_\END}

\newcommand{\He}{H_\END}
\newcommand{\ke}{k_\END}
\newcommand{\RH}{\text{\sc rh}}
\newcommand{\etaRH}{\eta_\RH}
\newcommand{\tRH}{t_\RH}
\newcommand{\aRH}{a_\RH}
\newcommand{\taRH}{\tilde{a}_\RH}
\newcommand{\HRH}{H_\RH}
\newcommand{\kRH}{k_\RH}
\newcommand{\TRH}{T_\RH}
\newcommand{\wRH}{w_\RH}
\newcommand{\gRH}{g_{\ast\RH}}
\newcommand{\eq}{\mathrm{eq}}

\newcommand{\aeq}{a_\eq}

\newcommand{\Heq}{H_\eq}
\newcommand{\keq}{k_\eq}

\newcommand{\eref}[1]{Eq.~(\ref{#1})} 
\newcommand{\erefs}[2]{Eqs.~(\ref{#1})~and~(\ref{#2})} 
\newcommand{\Eref}[1]{Equation~(\ref{#1})} 
\newcommand{\fref}[1]{Fig.~\ref{#1}}
\newcommand{\Fref}[1]{Figure~\ref{#1}}
\newcommand{\sref}[1]{Sec.~\ref{#1}}
\newcommand{\Sref}[1]{Section~\ref{#1}}
\newcommand{\aref}[1]{Appendix~\ref{#1}} 
\newcommand{\Aref}[1]{Appendix~\ref{#1}} 

\newcommand{\rref}[1]{\textcite{#1}} 
\newcommand{\Hz}{\,\mathrm{Hz}}
\newcommand{\GeV}{\,\mathrm{GeV}}
\newcommand{\MeV}{\,\mathrm{MeV}}

\newcommand{\eV}{\,\mathrm{eV}}

\newcommand{\Mpc}{\,\mathrm{Mpc}}

\newcommand{\km}{\,\mathrm{km}}

\renewcommand{\sec}{\,\mathrm{sec}}
\newcommand{\gram}{\,\mathrm{gram}}

\newcommand{\ket}[1]{\bigl|#1\bigr>}
\newcommand{\expval}[3]{\bigl< #1 \bigr| #2 \bigl| #3 \bigr>} 

\newcommand{\ba}[1]{\begin{align} #1 \end{align}}
\newcommand{\bes}[1]{\begin{equation}\begin{split} #1 \end{split}\end{equation}}
\newcommand{\bsa}[2]{\begin{subequations}\label{#1}\begin{align} #2 \end{align}\end{subequations}}
\newcommand{\nn}{\nonumber \\}
\definecolor{brown(web)}{rgb}{0.65, 0.16, 0.16}
\newcommand{\para}[1]{\phantom{.} \\ \noindent \textsf{\color{brown(web)}{#1.}}}

\begin{document}

\title{Cosmological gravitational particle production \\ and its implications for cosmological relics}
\author{Edward W.~Kolb}
\email{rocky.kolb@uchicago.edu}
\affiliation{Kavli Institute for Cosmological Physics and Enrico Fermi Institute, The University of Chicago, 5640 South Ellis Avenue, Chicago, Illinois 60637 USA}
\author{Andrew J.~Long}
\email{andrewjlong@rice.edu}
\affiliation{Department of Physics and Astronomy, Rice University, Houston, Texas 77005 USA}
\date{\today}

\begin{abstract}
Cosmological gravitational particle production (CGPP) is the creation of particles in an expanding universe due solely to their gravitational interaction.  These particles can play an important role in the cosmic history through their connection to various cosmological relics including dark matter, gravitational-wave radiation, dark radiation, and the baryon asymmetry.  This review explains the phenomenon of CGPP as a consequence of quantum fields in a time-dependent background, catalogs known results for the spectra and cosmological abundance of gravitationally produced particles of various spins, and explores the phenomenological consequences and observational signatures of CGPP.  
\end{abstract}

\keywords{Cosmology, Quantum Fields in Curved Spacetime, Inflation, Particle Dark Matter, General Relativity Equations and Solutions}

\pdfstringdefDisableCommands{\let\HyPsd@CatcodeWarning\@gobble}
\maketitle

\tableofcontents

\section{Introduction}
\label{sec:Introduction}

The high temperatures of the early Universe offer an environment in which to study the properties of known elementary particles, and to explore the consequences of hypothetical new particle species.   An important feature of the big-bang laboratory for particle physics is that the high temperatures of the early Universe are capable of producing particles of mass far beyond the reach of terrestrial accelerators.   However, there are two caveats to the expectation that new particle species were present in the primordial plasma.
\begin{enumerate}
\item Utilizing the hot primordial plasma as a source of new particles usually assumes that the particles have interactions with standard-model (SM) particles.   But what about a new particle species without interactions with SM particles?  (When we speak of interactions with SM particles we do not include gravitational interactions.) Indeed, many ideas for beyond-the-standard-model (BSM) physics involve hidden sectors, secluded particles, the shadow world, or reclusive particle species with interactions too weak to be populated in the primordial plasma.  In the extreme, these new particles may interact only with SM particles via the gravitational force, and their cosmological origin must have been through gravity.  That is the possibility we explore in this review, although many of our results would remain unchanged if the interactions of the new species with SM particles were sufficiently weak.   
\item For a particle species to be produced from the primordial plasma, the temperature of the Universe must have been high enough to produce the particles.  However, the maximum temperature of the radiation-dominated Universe may be as low as a few mega-electron-volts without coming into conflict with observations \cite{deSalas:2015glj}.  But the mechanism of cosmological gravitational particle production (CGPP) may possibly produce particles of mass as large as $10^{14} \GeV$, even if the maximum temperature is much less than the mass of the particle or the mass scale of inflation.
\end{enumerate} 

There are several proposed scenarios for CGPP.  This review focus on gravitational particle production from the vacuum as the result of a time evolution of the background gravitational field due to the expansion of the Universe.  (When we refer to CGPP, we refer to this mechanism.) Cosmological gravitation particle production is an example of a ``semiclassical'' process in the sense that the external gravitational field is not quantized, but a ``spectator'' field is quantized, and the production of the spectator field is a quantum effect.  

For the calculation of CGPP we assume a standard inflationary evolution for the expansion of the early Universe (see \sref{sub:infcos}) with the initial energy density of the Universe determined by the potential energy of a scalar field known as the \textit{inflaton}.  The dynamics of the inflaton field determines the evolution of the expansion of the Universe. 

Since gravity is weak, CGPP is typically inefficient, and only a small fraction of the inflaton's energy is transferred to the spectator field during inflation and at the end of inflation.  This means that the backreaction from CGPP on the inflaton dynamics can usually be neglected.  Although CGPP will have a minuscule effect in the early Universe, the spectator species may come to dominate the mass density of the Universe at late times when they behave classically. 

The most familiar example of particle production in an external field is the ``Schwinger effect,'' in which a sufficiently strong external electric field leads to the emergence of electron-positron pairs from the vacuum.  The effect, which was first predicted by \rref{Heisenberg:1935qt} based upon work of \rref{Sauter:1931zz}, was completely understood in 1951 by \rref{Schwinger:1951nm}, with a complete theoretical description using QED.  The basic idea is that an external electric field interacts with virtual particles, accelerating electrons in one direction and positrons in the opposite direction.  If the external field is strong enough to accelerate particles to energies greater than its mass over a distance smaller than or equal to the Compton wavelength of the particle, $|\vec{E}| > |\vec{E}_\mathrm{crit}| = m_e^2c^3/e\hbar$, then virtual particles can be ``pulled out'' of the vacuum and propagate as real particles.   The rate for this is proportional to $\exp(-\pi |\vec{E}_\mathrm{crit}|/|\vec{E}|)$.  Note that the electromagnetic coupling $e$ appears in the denominator of the exponential, implying that the effect is nonperturbative: it cannot be captured using perturbation theory to construct a power series around $e = 0$.

We can use the Schwinger effect as an exemplar to envision pair production in the expanding Universe \cite{Audretsch:1978qu,Martin:2007bw}.  Imagine a pair of virtual particles being pulled apart due to the expansion of the Universe.  As space expands, the recessional velocity of the particles increases with distance ($\mathrm{v}=Hd$; Hubble's law).  At a distance equal to $m^{-1}$, the velocity would be $H/m$.  For this to be of the order of unity requires $H$ to exceed a critical value $H_\mathrm{crit} \sim m$.  Therefore, if $H \gtrsim m$, the particles will obtain relativistic velocities within a Compton wavelength and particle creation is possible.  In analogy with the Schwinger effect, one might expect particle production proportional to $\exp(- \pi H_\mathrm{crit}/H) \sim \exp(-\pi m/H)$.  Indeed, although we discuss exceptions and details in this review, this will serve as a good guide.

The important aspects of CGPP, as opposed to standard scenarios for cosmological particle production, is that CGPP operates even when the spectator particle does not couple directly to standard-model particles or the inflaton, it is possible to produce particles of mass in excess of the largest temperature of the radiation-dominated Universe, and CGPP does not depend on coupling of the spectator particle to the particle physics standard model.

Early-Universe production from the plasma of SM particles is the most common assumption for the origin of particle dark matter.  However, to date dark matter has stubbornly evaded detection through interactions with standard-model particles via direct detection, indirect detection, or accelerator production or detection.  Perhaps nature is telling us that dark matter may not interact with SM particles other than by gravity.  If this is so, the primordial origin of dark matter must presumably be through the gravitational force.  Dark matter is significant motivation to study early-Universe production of particles solely through  gravitational interactions.  It is an especially natural mechanism to explain the origin of dark matter, which is known only to interact gravitationally and may not have any nongravitational interactions with visible matter.  CGPP provides a compelling explanation for the origin of stable relics and may play a role in other phenomena such as the matter-antimatter asymmetry; see \sref{sub:baryogenesis} and references therein.

Inflationary cosmology is a remarkable success of modern cosmology, with extraordinary agreement between theory and observation.  Inflation provides a connection between cosmological inhomogeneities measured by cosmic microwave background (CMB) anisotropies and large-scale structure (LSS) observations, and predicts a gravitational-wave background.  The origin of the inhomogeneities and gravitational waves traces to CGPP of the inflaton field(s) and the metric field.  It is natural to consider the possibility of quantum production of other fields.  We elaborate upon the historical development of CGPP in \sref{sec:History}.

Although there has been much recent work on CGPP, which we summarize in \sref{sec:Recent}, it is a phenomenon that has yet to be fully exploited, so we present this review to provide a summary of key results, both classic and recent, and to present a top-to-bottom overview of how the CGPP calculation is performed, so as to provide interested readers with the tools that they need to exploit the results of CGPP for new applications in cosmology.

\begin{figure}[t]
\begin{center}
\includegraphics[width=0.48\textwidth]{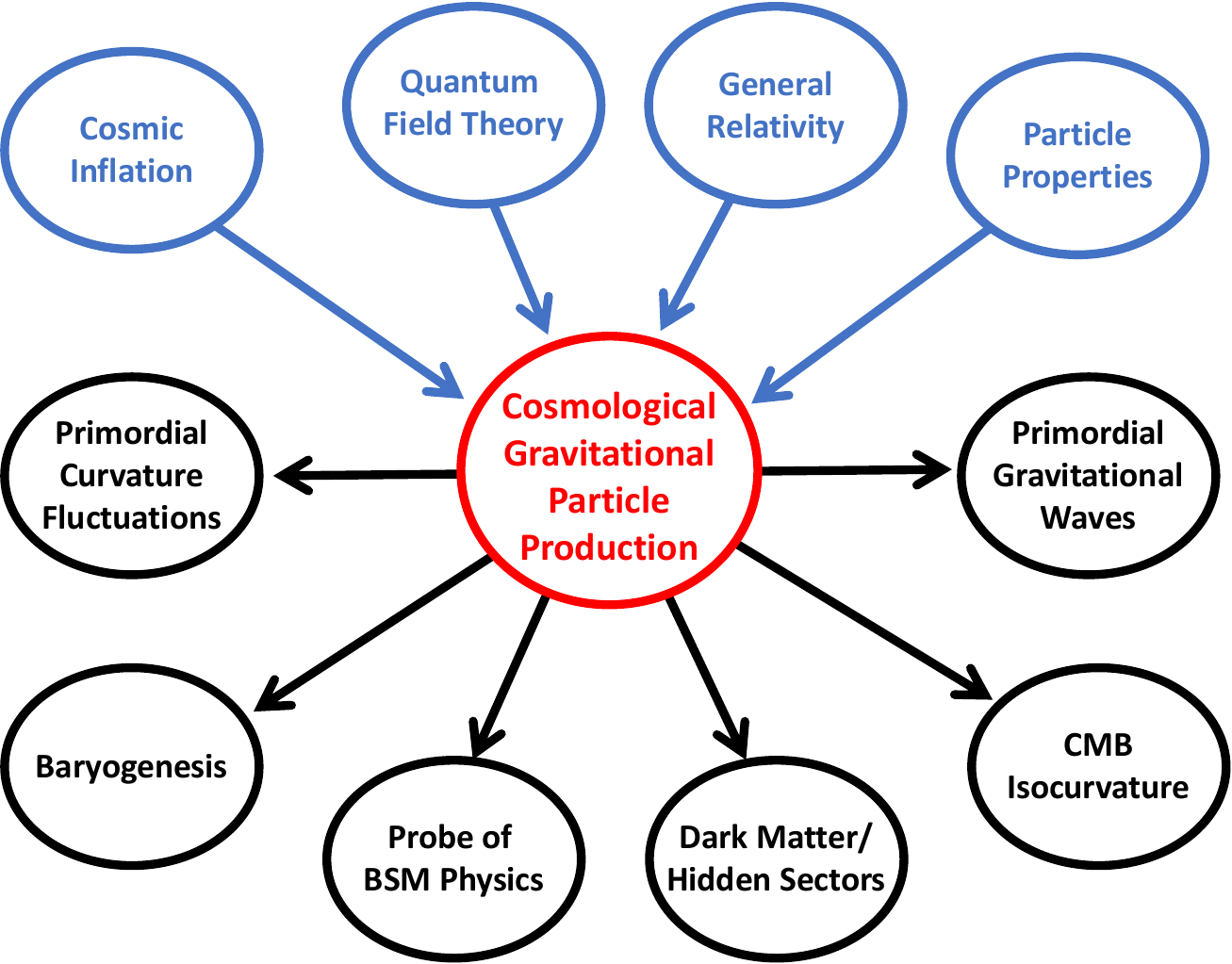} 
\caption{
\label{fig:bubbles}
CGPP takes as input cosmic inflation, quantum field theory, general relativity, and the properties of a particle, and outputs observable or potentially observable consequences. 
}
\end{center}
\end{figure}

The input to CGPP is inflation (or an alternative cosmological history), quantum field theory, general relativity, and the properties of the spectator particle.  The output of CGPP includes the generation of primordial temperature fluctuation, primordial gravitational waves, a possible arena for baryogenesis, a probe of BSM physics, the origin of dark matter and particles in a hidden sector, and CMB isocurvature fluctuations.  The CGPP input and output are illustrated in \fref{fig:bubbles}.

The remainder of the review is outlined as follows.  \Sref{sec:History} contains a short summary of CGPP's long history.  In \sref{sec:Theory} we establish the theoretical framework for CGPP using scalar field theory to develop the extension of quantum field theory (QFT) to curved spacetime.  Following that, after reviewing some salient features of inflationary cosmology, in \sref{sec:Scalar} we present numerical results and analytic approximations for CGPP of spin-$0$ particles.  \Sref{sec:Spin} contains a discussion of CGPP for particles with nonzero spin.  In \sref{sec:Recent} we review some recent work involving CGPP for the production of cosmological relics and observational signatures.  \Aref{app:general_relativity} contains a refresher on general relativity (GR), including the frame-field formalism needed for CGPP of fermions, and establishes the sign choices employed in the body of the review.  In \aref{app:FLRW_spacetime} we summarize key results for Friedmann-Lema\^{i}tre-Robertson-Walker (FLRW) spacetimes.   In \aref{app:dS_spacetime} we present the solutions to the mode equation in de Sitter space.  Finally, \aref{app:jwkb_method} reviews the Jeffreys-Wentzel-Kramers-Brillouin (JWKB) method for solving wave equations of the form encountered in CGPP calculations.

\section{Historical development}
\label{sec:History}

The phenomenon of cosmological gravitational particle production has been explored through various studies over the course of nearly a century.  CGPP provides a robustly calculable and practically testable prediction of quantum field theory in curved spacetime.  The most well-studied application of CGPP is the study of inflationary quantum fluctuations in the scalar inflaton field and metric, which lead to inhomogeneities in the spatial distribution of matter and radiation that imprint on the anisotropies of the cosmic microwave background radiation and large-scale structure of the Universe.  Other work on CGPP has sought to apply this phenomenon in explaining the origin of cosmological relics such as dark matter.  In this section we offer an overview of the key historical developments in the study of CGPP.  

The first paper discussing the creation of particle pairs in the expanding universe was Schr\"{o}dinger's 1939 paper ``The Proper Vibrations of the Expanding Universe'' \cite{Schrodinger:1939:PVE}.  In this prescient paper Schr\"{o}dinger was the first to realize that the expansion of the Universe can mix positive- and negative-frequency mode solutions of the wave equation.  Schr\"{o}dinger described this as ``mutual adulteration'' of positive- and negative-frequency modes in what he regarded as an ``alarming phenomenon,'' although he never stated why he considered it alarming.  He recognized mutual adulteration as production (or annihilation) of matter ``merely by the expansion of the Universe.''  He stated that this was a phenomenon of ``outstanding importance,'' although he did not specify why.  The paper certainly does not contain a full field-theoretical calculation (it was, after all, in 1939), and there are a few technical missteps.  Notably Schr\"{o}dinger suggested that CGPP should occur for photons, but in the Einstein-Maxwell theory (massless photons minimally coupled to gravity) the presence of conformal Weyl invariance prohibits CGPP; we provide this argument in \sref{sub:conformal_symmetry}.  Although Schr\"{o}dinger claimed that the phenomenon was of great importance, it does not seem to have been referenced until 1960 and never referenced by Schr\"{o}dinger.  

Research on particle creation in a background gravitational field did not restart in earnest until the thesis of Leonard Parker in 1965 \cite{Parkerthesis} and papers following his thesis.  Parker's treatment was the first field-theoretical analysis.  In the conclusion of a 1968 paper \cite{Parker:1968mv}, Parker wrote ``\ldots for the early stages of a Friedmann expansion it [particle creation] may well be of great cosmological significance, especially since it seems inescapable if one accepts quantum field theory and general relativity.''  Parker did not speculate on what the great cosmological significance might be.  He continued 
to develop the concept, first alone \cite{Parker:1969au,Parker:1971pt,Parker:1972kp} and then with collaborators including \rref{Parker:1974qw,Fulling:1974zr,Fulling:1974pu,Ford:1977in,Ford:1978ip,Ford:1977dj}.  

At the same time \rref{Hawking:1974rv} was developing the theory of particle creation by black holes, another example of gravitational particle production in a curved spacetime, albeit not \textit{cosmological} gravitational particle production resulting from the expanding Universe.  \rref{Birrell:1982ix} provided a good review of this early literature.  

A related phenomenon developed at around the same time is the well-known Unruh effect (sometimes referred to as the Unruh-Davies-Dewitt-Fulling effect) \cite{Fulling:1972md,Davies:1974th,Unruh:1976db,Dewitt:1979}.  An observer that moves with constant proper acceleration in Minkowski spacetime would measure a thermal bath of particles in the vacuum state of inertial observers.  For a review see \rref{Crispino:2007eb}.  The Unruh effect demonstrates how the presence or absence of particles depends on not only the state of the system but also the observer, in this case in regard to inertial versus noninertial observers.  It is relevant to the CGPP discussion in the sense that we are going to discuss how two different observers, one who uses the early-time positive- and negative-frequency modes (Schr\"{o}dinger's proper vibrations) and one who uses the late-time positive- and negative-frequency modes, can disagree on the number of particles in the system. 

Soon after Parker's 1968 paper, in the Soviet Union \rref{Zeldovich:1970} considered particle production in a cosmological setting with a background Kasner metric, a homogeneous, anisotropic metric of the form $(\dd s)^2 = (\dd t)^2 - \sum_{i=1}^3t^{2p_i}(\dd x^i)^2$.  It is a vacuum solution of Einstein's equation if the Kasner exponents $p_i$ satisfy $\sum_{i=1}^3 p_i = \sum_{i=1}^3 p_i^2 =1$.  Zel'dovich appreciated  cosmological particle creation, and he used the Kasner model to explore whether particle creation could isotropize an initially anisotropic universe, and by extension, explain how physical processes could turn an initial anisotropic and inhomogeneous universe into the observed homogeneous and isotropic Universe.  Zel'dovich continued this line of research in a series of papers, both alone and with Alexei Starobinski \cite{Zeldovich:1971mw,Zeldovich:1977vgo}.  Zel'dovich indeed identified a problem of great significance that is now generally assumed to be solved by inflation.  Others turned to calculations of particle creation in homogeneous and isotropic cosmologies \cite{Frolov:1976ya,Mamaev:1976zb,Grib:1976pw}.  A more complete set of references to work from the Soviet Union was given in the book \textit{Quantum Effects in Strong External Fields} by \rref{Grib:1980aih}. 

In the 1980s CGPP found a new home in the recently proposed at the time theory of cosmological inflation~\cite{Guth:1980zm,Starobinsky:1980te,Linde:1981mu,Albrecht:1982wi}.  This early period of accelerated expansion was initially lauded for efficiency at simultaneously solving the horizon, flatness, and monopole problems that had confounded early efforts to connect cosmology with elementary particle physics.  Before long, studies of quantum fluctuations in the scalar inflaton field and the metric led to predictions for density perturbations (as well as gravitational-wave radiation) that provided an explanation for the inhomogeneity of the large-scale structure of the Universe and the subtle anisotropies in the cosmic microwave background radiation.  The first papers explicitly to link density perturbations to particle creation were given by \rref{Mukhanov:1981xt},  \rref{Sasaki:1983kd,Mukhanov:1988jd}; see also \rref{Kodama:1985bj}.  This early literature was summarized well in the review by \rref{Mukhanov:1990me}.  Precision measurements of the large-scale structure and cosmic microwave background have furnished a wealth of information about the composition and evolution of the Universe, and their connections with the epoch of inflation are made possible through the phenomenon of CGPP.  

Besides generating primordial density perturbations, and possibly also gravitational-wave radiation, the phenomenon of CGPP during inflation may have served another purpose of great significance.  Namely, it may have provided an origin for cosmological relics such as dark matter, the matter-antimatter asymmetry, or the primordial plasma itself through gravitational reheating.  This idea was first explored by \rref{Ford:1986sy} and \rref{Turner:1987vd} who sought to explain the origin of light scalar (axion) dark matter via CGPP.  A series of studies in 1998 \cite{Kuzmin:1998uv,Kuzmin:1998kk,Chung:1998ua,Chung:1998bt,Chung:1998zb} emphasized the novel role of CGPP for creating superheavy dark matter with mass comparable to the inflationary Hubble scale $m \approx \Hinf$.  They also expanded the scope of interest to include CGPP for fermionic dark matter.  A natural candidate for CGPP is the gravitino that arises in theories of supergravity \cite{Lemoine:1999sc,Giudice:1999yt,Giudice:1999am,Kallosh:1999jj,Kolb:2021xfn} since it may couple weakly to ordinary matter and may not thermalize with the plasma.  

The past several years have seen a growing interest in the phenomenon of CGPP and its implications for cosmological relics.  Several recent developments are discussed throughout the review, particularly in \sref{sec:Recent}.  These include the production of ultralight spin-$1$ particles and connections to isocurvature-free dark photon dark matter, the enhanced production of massive spin-$\threehalf$ particles due to instabilities associated with a vanishing effective sound speed, the production of particles via thermal freeze-in with a gravitational mediator, the reinterpretation of CGPP at the end of inflation as inflaton annihilations, and efforts to detect dark matter's gravitational interactions in the laboratory. 

\section{Theoretical framework}
\label{sec:Theory}

In this section we offer a review of the theoretical framework underlying CGPP, including aspects of quantum field theory in curved spacetime.  Complementary discussions can be found in classic textbooks by \rref{Birrell:1982ix}, \rref{Mukhanov:2005sc}, \rref{Mukhanov:2007zz}, and \rref{Parker:2009uva}, as well as more recent reviews \cite{Ford:2021syk,Armendariz-Picon:2023gyl} and books \cite{Baumann:2022mni}.  Our goals here are (1) to illuminate how the framework of quantum field theory in curved spacetime is employed to derive the equations of motion for the time-dependent Fourier mode amplitudes $\chi_k(\eta)$, (2) to clarify how the cosmological expansion induces a mixing of positive- and negative-frequency mode functions, (3) to explain why this mixing can be interpreted as particle production, and (4) to address the emergence of ultraviolet divergences in the calculation of observables.  In this section we use scalar quantum field theory to develop formalism and illustrate notation.

\subsection{Quantum scalar field in an FLRW spacetime}
\label{sub:scalar_in_FLRW}

\para{Covariant action for a scalar field}  
Let $\aphi(x)$ be a real scalar field, which has spin-$0$ particles as its quantum excitations, and let the classical spacetime metric be denoted by $\metric_{\mu\nu}(x)$.  We consider a massive scalar that interacts only with gravity.  See \aref{app:general_relativity} for a refresher on general relativity and an explanation of our conventions.  We consider a theory defined by the action 
\bes{\label{eq:action}
    & S[\aphi(x),\metric_{\mu\nu}(x)] = \int \! \dd^4x \, \sqrtg \, \Bigl[ \Lcal - \tfrac{1}{2} \Mpl^2 R \Bigr] \\ 
    & \Lcal = 
    \tfrac{1}{2} \metric^{\mu\nu} \del{\mu} \aphi \del{\nu} \aphi 
    - \tfrac{1}{2} m^2 \aphi^2 
    + \tfrac{1}{2} \xi \aphi^2 R 
    \per
}
This theory has three model parameters: the reduced Planck mass $\Mpl=(8\pi G)^{-1/2}$, the scalar field's mass $m$, and the scalar field's coupling to the Ricci scalar $\xi$.  In the context of inflationary cosmology $\aphi$ is not the inflaton but rather an additional scalar spectator field.

\para{Nonminimal coupling to gravity}  
The dimensionless parameter $\xi$ is the coefficient of the dimension-4 operator coupling $\aphi$ to the Ricci scalar.  For $\xi = 0$ we say that the scalar field is minimally coupled to gravity; for $\xi = \ssfrac{1}{6}$ it is conformally coupled to gravity; and for other nonzero values of $\xi$ it is merely nonminimally coupled to gravity.  From the perspective of effective field theory, $\aphi^2 R$ is just one of many possibly operators that parametrize $\aphi$'s gravitational interactions; see \sref{sub:nonminimal} for a discussion.  While this nonminimal coupling is not needed for CGPP (one is free to set $\xi = 0$), our motivation for including the $\aphi^2 R$ operator is primarily pedagogical.  The theory of a conformally coupled scalar ($\xi = \ssfrac{1}{6}$) leads to equations of motion that are similar to the ones arising in theories of minimally coupled fermions and vectors.  We build intuition for these equations and their solutions by first investigating the simpler scalar context.  

\para{FLRW spacetime}  
We study the phenomenon of CGPP in cosmological spacetimes, which are approximately homogeneous, isotropic, and expanding.  Such spacetimes are described by the Friedmann-Lema\^itre-Robertson-Walker (FLRW) metric: 
\ba{
    \metric_{\mu\nu}^\FLRW(x) & = a^2(\eta) \, \mathrm{diag}(1,-1,-1,-1)
}
where the spacetime coordinate four-vector $x^\mu = (\eta, \xvec)$ has been written in terms of the conformal time coordinate $\eta$, related to coordinate time $t$ by $a\dd\eta=\dd t$, and the comoving spatial coordinate three-vector $\xvec = (x,y,z)$.  The time-dependent function $a(\eta)$ is called the scale factor, and the spatial expansion rate $H(\eta) = a^\prime(\eta) / a^2(\eta)$ is called the Hubble parameter.  Primes denote derivatives with respect to conformal time, and $\dvec = (\partial_x, \partial_y, \partial_z)$ denotes derivatives with respect to comoving spatial coordinates.  One can evaluate the usual tensors that arise in GR, finding for the Ricci scalar $R(\eta) = - 6 a^\pprime/a^3 = - 12 H^2 - 6 H^\prime/a$.  See \aref{app:FLRW_spacetime} for a refresher on the FLRW spacetime. 

\para{Canonical kinetic term in FLRW spacetime}  
In the FLRW spacetime the scalar field kinetic term is not canonical, since $\sqrtg \, \metric^{\mu\nu} \propto a^2$.  When a change of variables $\aphi(\eta,\xvec) = a^{-1}(\eta) \, \bphi(\eta,\xvec)$ is performed, the kinetic term is rendered canonically normalized and the action becomes 
\bes{\label{eq:SFLRW}
    S^\FLRW[\bphi(\eta,\xvec)] & = \int \! \dd \eta \int \! \dd^3\xvec \, \Bigl[ \tfrac{1}{2} (\bphi^\prime)^2 - \tfrac{1}{2} |\dvec \bphi|^2 
    \\ & \qquad 
    - \tfrac{1}{2} a^2 m_\mathrm{eff}^2 \bphi^2 - \tfrac{1}{2} \partial_\eta( a H \bphi^2) \Bigr] \, ,\\
    m_\mathrm{eff}^2(\eta) & = m^2 + (\tfrac{1}{6} - \xi) R(\eta) 
    \per
}
Note that the canonically normalized field variable $\bphi(\eta,\xvec)$ acquires a time-dependent effective squared mass $a^2(\eta) m_\mathrm{eff}^2(\eta)$ that may be either positive or negative. 

\para{Hamiltonian}  
The scalar field's dynamics are equivalently captured in the Hamiltonian formalism.  Denoting the field and its conjugate momentum density by $\bphi(\eta,\xvec)$ and $\pi(\eta,\xvec)$, the Hamiltonian can be written as
\bes{
    H(\eta) & = \int \! \dd^3\xvec \, \Bigl[ 
    \tfrac{1}{2} \pi^2 
    + \tfrac{1}{2} |\dvec \bphi|^2 
    + \tfrac{1}{2} aH \, \bigl( \pi \bphi + \bphi \pi \bigr)
    \\ & \qquad 
    + \tfrac{1}{2} \bigl( a^2 m_\mathrm{eff}^2 + 2 a^2 H^2 + a H^\prime \bigr) \bphi^2 \Bigr] 
    \per
}
We write $\pi\phi+\phi\pi$ because after quantization $[\pi,\phi]\neq0$.

\para{Canonical quantization}  
We work in the Heisenberg picture in which operators evolve subject to the Heisenberg equations of motion $\ii\, \partial_\eta \mathcal{O} = [\mathcal{O},H]$ and states are static $\ii\, \partial_\eta \ket{\psi} = 0$.  Now $\bphi$ and $\pi$ are regarded as quantum operators.  Canonical quantization is accomplished by imposing equal-time commutation relations 
\bes{
    \bigl[ \bphi(\eta,\xvec), \, \pi(\eta,\yvec) \bigr] & = \ii \, \delta(\xvec - \yvec) \\ 
    \bigl[ \bphi(\eta,\xvec), \, \bphi(\eta,\yvec) \bigr] & = \bigl[ \pi(\eta,\xvec), \, \pi(\eta,\yvec) \bigr] = 0 
    \per
}
The operator equations of motion are 
\bes{
    \pi^\prime & = \nabla^2 \bphi - aH \pi - \bigl( a^2 m_\mathrm{eff}^2 + 2 a^2 H^2 + a H^\prime \bigr) \bphi \\ 
    \bphi^\prime & = \pi + aH \bphi 
    \com
}
and they combine to give 
\ba{\label{eq:wave_eqn}
    \bphi^\pprime - \dvec^2 \bphi + a^2 m_\mathrm{eff}^2 \bphi = 0
    \com
}
which takes the form of a second-order linear wave equation with a time-dependent effective mass.  For any solution of this equation, the corresponding conjugate momentum density is simply $\pi = \bphi^\prime - aH \bphi$, so we are free to focus on solving the wave equation. 

\para{Evolution equations}  
Solutions of the linear wave equation take the form 
\bes{\label{eq:ansatz}
    & \bphi(\eta,\xvec) = \int \! \! \frac{\dd^3\kvec}{(2\pi)^3} \Bigl[ a_\kvec \, U_\kvec(\eta,\xvec) + a_\kvec^\dagger \, V_\kvec(\eta,\xvec) \Bigr] \\ 
    & \quad \text{with} \quad U_\kvec(\eta,\xvec) = \chi_k(\eta) \, \ee^{\ii \kvec \cdot \xvec} 
    \quad \text{and} \quad V_\kvec = U_\kvec^\ast
    \per
}
The complex fields $U_\kvec(\eta,\xvec)$ and $V_\kvec(\eta,\xvec)$, called the mode functions, are labeled with a comoving wavevector $\kvec$ with comoving wave number $k = |\kvec|$.  The mode functions are orthonormal basis functions that span the space of solutions of the wave equation.  The complex functions $\chi_k(\eta)$ are called the Fourier mode amplitudes (or sometimes the mode functions).  A ladder operator is assigned to each mode function: $a_\kvec$ are paired with $U_\kvec(\eta,\xvec)$ and $a_\kvec^\dagger$ are paired with $V_\kvec(\eta,\xvec)$.  The ladder operators are required to obey the usual algebra of commutation relations 
\bes{\label{eq:ladder_commutator}
    \bigl[ a_\kvec, \, a_\qvec^\dagger \bigr] & = (2\pi)^3 \, \delta(\kvec - \qvec) \\ 
    \bigl[ a_\kvec, \, a_\qvec \bigr] & = \bigl[ a_\kvec^\dagger, \, a_\qvec^\dagger \bigr] = 0 
    \per
}

Since the wave equation is second order, for each $\kvec$ it admits a two-dimensional space of solutions, and much of the analysis of CGPP is based on identifying a `good' basis of mode functions to span this space such that $U_\kvec(\eta,\xvec)$ corresponds to positive-frequency modes and $V_\kvec(\eta,\xvec)$ corresponds to negative-frequency modes.  We discuss this issue further in \sref{sub:solving}.  In lieu of specifying explicit expressions for the mode functions at this point, we assume for the moment that they form a complete and orthonormal basis that is expressed by the Wronskian conditions  
\bes{\label{eq:wronskian}
    & \int \! \dd^3\xvec \, \bigl( U_\kvec \, V_\qvec^\prime - V_\qvec \, U_\kvec^\prime \bigr) = \ii \, (2\pi)^3 \delta(\kvec-\qvec) \\ 
    & \qquad \text{implying:} \quad 
    \chi_k \, \chi_k^{\prime\ast} - \chi_k^\ast \, \chi_k^\prime = \ii
    \per
}
The wave equation leads to equations of motion for the Fourier mode amplitudes called the mode equation, 
\ba{\label{eq:mode_equation}
    \chi_k^\pprime(\eta) + \omega_k^2(\eta) \, \chi_k(\eta) = 0 
    \per
}
For each $k$ this is a harmonic oscillator equation with time-dependent comoving squared angular frequency 
\bes{\label{eq:omegak_scalar}
    \omega_k^2(\eta) 
    & = k^2 + a^2(\eta) m^2 + \bigl( \tfrac{1}{6} - \xi \bigr) a^2(\eta) R(\eta)
    \per
}
Note that $\omega_k^2(\eta)$ need not be positive. \Sref{sub:solving} discusses the solutions of \eref{eq:mode_equation}.  If $\xi = \ssfrac{1}{6}$ then the scalar is said to be conformally coupled to gravity.  

The equation of motion presented here \eqref{eq:mode_equation} resembles the Mukhanov-Sasaki equation \cite{Baumann:2022mni}, and the two equations capture similar physics: the evolution of scalar field perturbations during an inflationary phase.  However, they differ insofar as the Mukhanov-Sasaki equation accounts for the backreaction of the inflaton perturbations on the scalar metric perturbations through the Einstein equation, whereas our derivation neglects this backreaction since our scalar spectator field does not dominate the energy density of the Universe.  

\subsection{Mode functions and Bogolubov transformations}
\label{sub:solving}

The mode functions $\chi_k(\eta)$ evolve according to the mode equation \eqref{eq:mode_equation}.  We are interested in solving these equations for cosmologically relevant spacetimes, which are characterized by a quasi-de Sitter (qdS) phase of inflation followed by reheating, radiation domination, etc.  To understand how  such solutions are obtained, we begin by focusing on the conceptually and computationally simpler cases of Minkowski spacetime and general asymptotically flat spacetimes.  After discussing Bogolubov transformations and CGPP in this context, we address cosmologically relevant spacetimes. 

\para{Minkowski spacetime}  
In Minkowski spacetime the FLRW scale factor is constant $a(\eta) = \bar{a}$ and there is no cosmological expansion, $H(\eta) = \bar{H} = 0$ and $R(\eta) = \bar{R} = 0$.  The squared angular frequency is also constant, $\omega_k^2(\eta) = \bar{\omega}_k^2 = k^2 + \bar{a}^2 m^2$.  For each wavevector $\kvec$, the mode equation \eqref{eq:mode_equation} is a linear, second-order differential equation in a complex variable, and its solutions form a four-dimensional vector space.\footnote{Solutions of the wave equation \eqref{eq:wave_eqn} form only a two-dimensional vector space per Fourier mode, since $\bphi \in \mathbb{R}$, but the parametrization in \eref{eq:ansatz} introduces a twofold redundancy.}  Every solution can be written as a linear combination of a pair of basis functions with complex coefficients normalized to satisfy \eref{eq:wronskian}.  The canonical basis functions $\ee^{-\ii \bar{\omega}_k \eta} / \sqrt{2 \bar{\omega}_k}$ and $\ee^{\ii \bar{\omega}_k \eta} / \sqrt{2 \bar{\omega}_k}$ are called the positive- and negative-frequency modes, respectively.  The positive-frequency mode functions are 
\bes{\label{eq:chik_Minkowski}
    \chi_k(\eta) = \frac{\ee^{- \ii \bar{\omega}_k \eta}}{\sqrt{2 \bar{\omega}_k}} 
    \quad \text{or} \quad 
    U_\kvec(\eta,\xvec) = \frac{\ee^{- \ii (\bar{\omega}_k \eta - \kvec \cdot \xvec)}}{\sqrt{2 \bar{\omega}_k}} 
    \com
}
and they correspond to plane waves propagating in the direction of $\kvec$ with phase velocity $\bar{\omega}_k / k$.  When this set of mode functions is used to construct the field operator via \eref{eq:ansatz}, the associated ladder operators $a_\kvec$ define a vacuum state $\ket{0}$ such that $a_\kvec \ket{0} = 0$ for all $\kvec$.  Minkowski spacetime is privileged insofar as all inertial observers agree on the absence of particles in this vacuum state and the presence of particles in excited states \cite{Birrell:1982ix}.  In other words, there is no CGPP in Minkowski spacetime. 

\para{Building intuition} 
Minkowski spacetime is special, because it is static.  Therefore, it has time translations as an isometry, and there is an associated Killing vector field $\ii \, \partial/\partial\eta$ that has the positive-frequency mode functions \eqref{eq:chik_Minkowski} as its positive-eigenvalue eigenfunctions.  For cosmological spacetimes that expand or contract, it is generally not possible to specify positive-frequency mode functions that hold for all time.  This leads to the situation whereby observers at different times may decompose the field operator onto different bases of mode functions and ladder operators.  Consequently, the vacuum state at some initial time can be an excited state at another time.  These concepts are illustrated in the following discussion.  

Another way to illustrate this point is to recognize that nonconformal fields on a time-dependent spacetime background develop a time-dependent Hamiltonian.  For instance each Fourier mode of the scalar field $\phi_\kvec(\eta) = a_\kvec \, \chi_k(\eta)$ has the dynamics of a quantum simple harmonic oscillator with the time-dependent natural frequency $\omega_k^2(\eta)$.  If the frequency were constant, the ground-state wave function would be a Gaussian.  If the frequency were varied slowly, the wave function would adjust so as to track the adiabatically varying ground state, and the system would remain almost unexcited.  Alternatively, if the frequency were varied rapidly, the wave function would be unable to keep up, and the system would find itself in an excited state with respect to the new Hamiltonian.  When these concepts are carried over to quantum field theory, the excited state corresponds to particle production. 

\para{Asymptotically flat spacetimes}  
Next we consider spacetimes that are only asymptotically flat (Minkowski-like) toward early and late time.  Suppose that the scale factor $a(\eta)$ limits to $a^\IN$ as $\eta \to - \infty$ and to $a^\OUT$ as $\eta \to \infty$.  Denote the corresponding limits of the comoving angular frequency by 
\ba{
    \lim_{\eta \to -\infty} \omega_k(\eta) = \omega_k^\IN 
    \quad \text{and} \quad 
    \lim_{\eta \to \infty} \omega_k(\eta) = \omega_k^\OUT 
    \com
}
which are assumed to be real and positive.  In the asymptotic regimes where the spacetime is approximately Minkowski, the positive-frequency mode functions take the form appearing in \eref{eq:chik_Minkowski}, with $\bar{\omega}_k$ replaced by $\omega_k^\IN$ or $\omega_k^\OUT$.  This observation motivates us to identify solutions of the mode equations, denoted by $\chi_k^\IN(\eta)$ and $\chi_k^\OUT(\eta)$, that are asymptotic to the Minkowski positive-frequency mode functions at either early or late times, 
\bes{\label{eq:chiIN_chiOUT_asymptotic}
    \chi_k^\IN(\eta) & \sim \frac{1}{\sqrt{2 \omega_k^\IN}} \, \ee^{-\ii \omega_k^\IN \eta} 
    \quad \text{as} \quad \eta \to - \infty \\ 
    \chi_k^\OUT(\eta) & \sim \frac{1}{\sqrt{2 \omega_k^\OUT}} \, \ee^{-\ii \omega_k^\OUT \eta} 
    \quad \text{as} \quad \eta \to \infty 
    \per
}
Using either set of mode functions and a corresponding set of ladder operators denoted by $a_\kvec^\IN$ and $a_\kvec^\OUT$, one can construct the field operator as in \eref{eq:ansatz}, 
\ba{\label{eq:Eq16}
    \bphi(\eta,\xvec) 
    & = \int \! \! \frac{\dd^3\kvec}{(2\pi)^3} \Bigl( a_\kvec^\IN \, \chi_k^\IN(\eta) + a_{-\kvec}^{\IN \dagger} \, \chi_k^{\IN \ast}(\eta) \Bigr) \ee^{\ii \kvec \cdot \xvec} \\ 
    & = \int \! \! \frac{\dd^3\kvec}{(2\pi)^3} \Bigl( a_\kvec^\OUT \, \chi_k^\OUT(\eta) + a_{-\kvec}^{\OUT \dagger} \, \chi_k^{\OUT \ast}(\eta) \Bigr) \ee^{\ii \kvec \cdot \xvec} \nonumber 
    \per
}
The equivalence of the expressions in \eref{eq:Eq16} and the freedom to move between them is captured by the Bogolubov transformations.  

\para{Bogolubov transformations} 
In general, pairs of orthonormal basis functions are related to one another through linear transformations that leave the field operator unchanged.  Consider the matrix 
\ba{
    \begin{pmatrix} \alpha_k & \beta_k \\ \beta_k^\ast & \alpha_k^\ast \end{pmatrix} \in \mathrm{SU}(1,1) 
    \com
}
where the complex constant entries $\alpha_k$ and $\beta_k$ satisfy $|\alpha_k|^2 - |\beta_k|^2 = 1$.  Each Fourier mode of the field operator can be written as   
\bes{
    & a_\kvec \, \chi_k(\eta) + a_{-\kvec}^\dagger \, \chi_k^\ast(\eta) 
    \\ & \quad 
    = \begin{pmatrix} a_\kvec & a_{-\kvec}^\dagger \end{pmatrix} \begin{pmatrix} \chi_k(\eta) \\ \chi_k^\ast(\eta) \end{pmatrix} 
    \\ & \quad 
    = \begin{pmatrix} a_\kvec & a_{-\kvec}^\dagger \end{pmatrix} \begin{pmatrix} \alpha_k^\ast & -\beta_k \\ -\beta_k^\ast & \alpha_k \end{pmatrix} \begin{pmatrix} \alpha_k & \beta_k \\ \beta_k^\ast & \alpha_k^\ast \end{pmatrix} \begin{pmatrix} \chi_k(\eta) \\ \chi_k^\ast(\eta) \end{pmatrix} 
    \\ & \quad 
    =  \underbrace{\biggl[ \begin{pmatrix} \alpha_k^\ast & -\beta_k^\ast \\ -\beta_k & \alpha_k \end{pmatrix} \begin{pmatrix} a_\kvec \\ a_{-\kvec}^\dagger \end{pmatrix} \biggr]^T}_{\equiv \  \begin{pmatrix} \tilde{a}_\kvec & \tilde{a}_{-\kvec}^\dagger \end{pmatrix} }  
    \underbrace{\begin{pmatrix} \alpha_k & \beta_k \\ \beta_k^\ast & \alpha_k^\ast \end{pmatrix} \begin{pmatrix} \chi_k(\eta) \\ \chi_k^\ast(\eta) \end{pmatrix}}_{\equiv \ \begin{pmatrix} \tilde{\chi}_k(\eta) \\ \tilde{\chi}_k^\ast(\eta) \end{pmatrix}} 
    \\ & \quad
    = \tilde{a}_\kvec \, \tilde{\chi}_k(\eta) + \tilde{a}_{-\kvec}^\dagger \ \tilde{\chi}_k^\ast(\eta) 
    \nonumber 
}
where the last line defines a new basis of ladder operators and mode functions, denoted by a tilde.  One can verify that the ladder operators with tildes obey the commutation relations \eqref{eq:ladder_commutator} and the mode functions with tildes obey the Wronskian condition \eqref{eq:wronskian} and mode equations \eqref{eq:mode_equation}.  It follows that the field operator admits a family of equivalent representations that are related to one another through $\mathrm{SU}(1,1)$ matrices called Bogolubov transformations.  Note that for fermionic fields the commutation relations are replaced by anticommutation relations, and the Bogolubov transformations are implemented through $\mathrm{SU}(2)$ matrices with $|\alpha_k|^2 + |\beta_k|^2 = 1$. 

Since the $\IN$ and $\OUT$ mode functions and their complex conjugates both form complete bases of solutions of the same linear differential equation, it must be possible to write them as linear combinations of one another.  This is accomplished using a Bogolubov transformation with 
\bes{\label{eq:Bogo_IN_to_OUT}
    \chi_k^\IN(\eta) & = \alpha_k \, \chi_k^\OUT(\eta) + \beta_k \, \chi_k^{\OUT \ast}(\eta) \\ 
    a_\kvec^\IN & = \alpha_k^\ast \, a_\kvec^\OUT - \beta_k^\ast \, a_{-\kvec}^{\OUT \dagger} 
    \com
}
where $\alpha_k$ and $\beta_k$ now denote the set of Bogolubov coefficients specifically linking the $\IN$ and $\OUT$ mode functions.  Inverting these relations and using the Wronskian condition allow the Bogolubov coefficients to be calculated from the $\IN$ and $\OUT$ mode functions 
\bes{\label{eq:Bogo_coeffs}
    \alpha_k & = \ii \bigl( \chi_k^{\OUT \ast} \, \chi_k^{\IN \prime} - \chi_k^{\OUT \prime \ast} \, \chi_k^{\IN} \bigr) \\ 
    \beta_k & = \ii \bigl( \chi_k^{\OUT \prime} \, \chi_k^\IN - \chi_k^\OUT \, \chi_k^{\IN \prime} \bigr) 
    \per
}
Note that the individual factors and terms on the right-hand sides may carry time dependence, but this dependence must cancel when the static Bogolubov coefficients are constructed.  

\para{Bunch-Davies vacuum}  
The set of ladder operators $a_\kvec^\IN$ may be used to define a vacuum state and construct a Fock space of multiparticle states.  The vacuum state, denoted by $|0^\IN\rangle$, is defined by 
\ba{
    a_\kvec^\IN \ket{0^\IN} = 0 
    \quad \text{for all $\kvec$}
    \per
}
That is, it is the state annihilated by all of the $\IN$ lowering operators.  The vacuum state is normalized as $\left\langle0^\IN |0^\IN\right\rangle =1$.  Despite the potentially confusing notation, one should not interpret $|0^\IN\rangle$ as an initial vacuum state that evolves into another state at late times.  Remember that we are working in the Heisenberg picture in which states are static; a system prepared in the state $|0^\IN\rangle$ remains in this state at all time.  Anticipating the application of this discussion to inflationary cosmology, we refer to $|0^\IN\rangle$ as the Bunch-Davies vacuum \cite{Bunch:1978yq}.  In studies of inflationary perturbations, it is customary to assume that the Universe is prepared in the Bunch-Davies vacuum state.  

\para{Particle number}
A system prepared in the $\IN$ vacuum state may contain particles with respect to the $\OUT$ ladder operators.  The $\OUT$ number operator is defined by 
\ba{
    N^\OUT = \int \! \! \frac{\dd^3\kvec}{(2\pi)^3} \, a_\kvec^{\OUT \dagger} a_\kvec^\OUT 
    \per
}
If the $\IN$ and $\OUT$ operators are related through a Bogolubov transformation \eqref{eq:Bogo_IN_to_OUT} with parameters $\alpha_k$ and $\beta_k$, then the expected number of particles measured by the $\OUT$ number operator in the $\IN$ vacuum state is given by 
\ba{\label{eq:vacN}
    \expval{0^\IN}{N^\OUT}{0^\IN} = V \int \! \! \frac{\dd^3\kvec}{(2\pi)^3} \, |\beta_k|^2 
    \per
}
\Eref{eq:vacN} has an expected IR divergence $V = (2\pi)^3 \, \delta(\kvec - \kvec)$ associated with the homogeneous nature of particle production and the system's infinite volume.  Note that if the system prepared not in the Bunch-Davies vacuum state but instead in an excited state, then the subsequent gravitational production could be Bose enhanced or Fermi suppressed depending on the quantum statistics of the spectator particles; see \rref{Lozanov:2019jxc} for a discussion of this scenario.

\para{Number density spectrum}
In an FLRW spacetime, the comoving number density of particles that the $\OUT$ number operator measures in the $\IN$ vacuum is given by 
\ba{\label{eq:a3n}
    a^3 n 
    = \int \! \! \frac{\dd^3\kvec}{(2\pi)^3} \, |\beta_k|^2 
    \com
}
and $n$ is the physical number density.  One can interpret $|\beta_k|^2$ as the occupation number of a Fourier mode with the comoving wavevector $\kvec$.  For bosons $0 \leq |\beta_k|^2$ whereas for fermions $0 \leq |\beta_k|^2 \leq 1$.  Since $\beta_k$ is independent of the wavevector's orientation in the isotropic FLRW spacetime, the angular integration is trivial.  
One finds that 
\bes{\label{eq:dnk}
    a^3n 
    = \int \frac{\dd k}{k} \, a^3 n_k 
    \qquad \text{with} \qquad 
    a^3 n_k = \frac{k^3}{2\pi^2} \, |\beta_k^2| 
    \com
}
where $a^3 n_k$ is the comoving number density spectrum, \ie, the comoving number density of particles per logarithmic wave number interval.  

\para{Pair production}
CGPP results from the behavior of quantum fields on a time-dependent background.  Since the background breaks time-translation invariance, the field's energy need not be conserved, but since the background maintains spatial-translation invariance, the field's momentum must be conserved.  Consequently, CGPP cannot induce single particle production; at leading order it corresponds to back-to-back pair production.  

\para{CGPP discussion}  
We now summarize the essential ingredients thus far.  In a curved spacetime that is asymptotically flat, there is a natural notion of positive- and negative-frequency modes at early and late times since the spacetime in those regions is approximately Minkowski.  However, the temporary departure from Minkowski at intermediate time leads to a mixing of positive- and negative-frequency modes: the solution $\chi_k^\IN(\eta)$ that is asymptotic to the positive-frequency mode at early time can be written as a linear combination of the function $\chi_k^\OUT(\eta)$ that is asymptotic to the positive-frequency mode at late time and the function $\chi_k^{\OUT \ast}(\eta)$ that is asymptotic to the negative-frequency mode at late time.  This mixing of the mode functions is captured by a Bogolubov transformation, which implies a linear map between the associated sets of ladder operators.  As such, a vacuum state with respect to the $\IN$ ladder operators may be an excited state with respect the $\OUT$ ladder operators, which is to say that it contains particles.  That is the basis of the phenomenon of cosmological gravitational particle production.  The mathematical formalism has broader applicability to theories with a time-dependent Hamiltonian; the defining feature of CGPP is that the time dependence is induced by the cosmological expansion.  

\para{Asymptotically adiabatic spacetimes}  
Ultimately we are interested in calculating CGPP for cosmological spacetimes that describe an early epoch of inflation followed by reheating and a subsequent hot big-bang cosmology.  For such spacetimes (at least the ones that we consider) the FLRW scale factor $a(\eta)$ is always growing, the spacetime is not asymptotically flat, and how to identify the positive- and negative-frequency modes and how to define particles becomes ambiguous.  For a general spacetime this ambiguity may be insurmountable, but typical cosmological spacetimes have the appealing property that the comoving angular frequency $\omega_k(\eta)$ is slowly varying (adiabatic) at both early and late time.  In particular, one can define a dimensionless parameter to measure the departure from adiabaticity 
\ba{\label{eq:Ak_def}
    A_k(\eta) = \omega_k^\prime(\eta) / \omega_k^2(\eta) 
    \com
}
and verify that $A_k \to 0$ sufficiently quickly as $\eta \to \pm \infty$.  

For asymptotically adiabatic spacetimes a possible definition of the $\IN$ and $\OUT$ mode functions is given by 
\bes{\label{eq:chik_IN_OUT_def}
    \chi_k^\IN(\eta) & 
    \sim \frac{\ee^{-\ii \int^\eta \! \dd \eta^\prime \omega_k(\eta^\prime)}}{\sqrt{2 \omega_k(\eta)}}  
    \quad \text{as} \quad \eta \to - \infty \\ 
    \chi_k^\OUT(\eta) & 
    \sim \frac{\ee^{-\ii \int^\eta \! \dd \eta^\prime \omega_k(\eta^\prime)}}{\sqrt{2 \omega_k(\eta)}}  
    \quad \text{as} \quad \eta \to + \infty 
    \per
}
In fact, \eref{eq:chik_IN_OUT_def} defines the positive-frequency mode functions of zeroth adiabatic order, and the corresponding ladder operators $a_\kvec^\IN$ define an adiabatic $\IN$ vacuum $|0^\IN\rangle$ of zeroth adiabatic order.  However, provided that $\omega_k(\eta)$ is sufficiently slowly varying at early and late times, the corrections from keeping terms of higher adiabatic order are vanishing as $\eta \to \pm \infty$, implying that $|0^\IN\rangle$ is an $\IN$ vacuum of infinite adiabatic order.  See \aref{app:jwkb_method} for a discussion of the JWKB method and adiabatic series.  
  
\para{Reparametrization}  
To find the solution $\chi_k^\IN(\eta)$ it is useful to adopt the parametrization \cite{Kofman:1997yn} 
\bes{\label{eq:scalar_chikIN_ansatz}
    & \chi_k^\IN(\eta) = 
    \tilde{\alpha}_k(\eta) \, \frac{\ee^{-\ii \dphi_k(\eta)}}{\sqrt{2 \omega_k(\eta)}} 
    + \tilde{\beta}_k(\eta) \, \frac{\ee^{\ii \dphi_k(\eta)}}{\sqrt{2 \omega_k(\eta)}} 
    \\ & \quad \text{where} \quad 
    \dphi_k(\eta) = \int^\eta \! \dd \eta^\prime \omega_k(\eta^\prime) 
    \com
}
which introduces the new mode functions $\tilde{\alpha}_k(\eta)$ and $\tilde{\beta}_k(\eta)$.  The condition in \eref{eq:chik_IN_OUT_def} requires $\tilde{\alpha}_k(\eta) \to 1$ and $\tilde{\beta}_k(\eta) \to 0$ as $\eta \to - \infty$, and $|\tilde{\alpha}_k(\eta)|^2 - |\tilde{\beta}_k(\eta)|^2 = 1$ is maintained by the equations of motion.  The function $\chi_k^\IN(\eta)$ will solve the mode equation provided that the new mode functions obey 
\bes{\label{eq:EOMforalphabeta}
    \partial_\eta \tilde{\alpha}_k(\eta) & = \tfrac{1}{2} A_k(\eta) \, \omega_k(\eta) \, \tilde{\beta}_k(\eta) \, \ee^{2 \ii \int^\eta \! \dd \eta^\prime \omega_k(\eta^\prime)} \\ 
    \partial_\eta \tilde{\beta}_k(\eta) & = \tfrac{1}{2} A_k(\eta) \, \omega_k(\eta) \, \tilde{\alpha}_k(\eta) \, \ee^{-2 \ii \int^\eta \! \dd \eta^\prime \omega_k(\eta^\prime)} 
    \com
}
where $A_k(\eta)$ is the adiabaticity parameter from \eref{eq:Ak_def}.  

This parametrization is especially useful because the late-time behavior of $\tilde{\beta}_k(\eta)$ gives the Bogolubov coefficient $\beta_k$ without the need to evaluate $\chi_k^\OUT(\eta)$.  This can be understood as follows.  The Bogolubov coefficient $\beta_k$ can be calculated from \eref{eq:Bogo_coeffs} if the mode functions $\chi_k^\IN(\eta)$ and $\chi_k^\OUT(\eta)$ are known.  Since the left- and right-hand sides of \eref{eq:Bogo_coeffs} are independent of time, we are free to evaluate the equation as $\eta \to +\infty$.  Although $\chi_k^\OUT(\eta)$ is unknown, its asymptotic behavior at late times is known through the defining relation \eqref{eq:chik_IN_OUT_def}.  After some algebra and while making use of Eqs.~\eqref{eq:scalar_chikIN_ansatz}~and~\eqref{eq:EOMforalphabeta}, one finds that 
\ba{\label{eq:betak_from_betatildek}
    \beta_k 
    & = \lim_{\eta \to \infty} \ii \bigl( \chi_k^{\OUT \prime} \, \chi_k^\IN - \chi_k^\OUT \, \chi_k^{\IN \prime} \bigr) \\ 
    & = \lim_{\eta \to \infty} \Bigl[ \tilde{\beta}_k - \tfrac{\ii}{4} A_k \Bigl( \tilde{\alpha}_k \, \ee^{- 2 \ii \int^\eta \! \dd \eta^\prime \omega_k(\eta^\prime)} + \tilde{\beta}_k \Bigr) \Bigr] \nn 
    & = \lim_{\eta \to \infty} \tilde{\beta}_k(\eta) 
    \per
    \nonumber 
}
In the last equality of the previous equation, it is assumed that $A_k(\eta)$ vanishes sufficiently quickly at late times to neglect the second term.  This calculation reveals that the late-time limit of the mode function $\tilde{\beta}_k(\eta)$ coincides with the Bogolubov coefficient $\beta_k$ linking the $\IN$ and $\OUT$ mode functions, which is related to the amount of CGPP through \eref{eq:dnk}.  One can also invert \eref{eq:scalar_chikIN_ansatz} to find 
\bsa{eq:betachixhiprime}{
    \tilde{\alpha}_k(\eta) & = \frac{\omega_k \chi_k^\IN + \ii \partial_\eta \chi_k^\IN}{\sqrt{2 \omega_k}} \, \ee^{\ii \int^\eta \! \dd \eta^\prime \omega_k(\eta^\prime)} \\ 
    \tilde{\beta}_k(\eta) & = \frac{\omega_k \chi_k^\IN - \ii \partial_\eta \chi_k^\IN}{\sqrt{2 \omega_k}} \, \ee^{- \ii \int^\eta \! \dd \eta^\prime \omega_k(\eta^\prime)} \\ 
    |\tilde{\beta}_k(\eta)|^2 & = \frac{\omega_k}{2}|\chi_k^\IN|^2+\frac{|\partial_\eta \chi_k^\IN|^2}{2\omega_k} -\frac{1}{2} 
    \per
}
The last expression in \eref{eq:betachixhiprime} assumes that $\omega_k(\eta) \in \mathbb{R}$ and uses the Wronskian condition \eqref{eq:wronskian}.

\para{Correlation function \& power spectrum} 
The field operator's equal-time two-point correlation is given by
\bes{
    & \expval{0^\IN}{\aphi(\eta,\xvec) \aphi(\eta,\yvec)}{0^\IN} 
    \\ & \qquad 
    = \frac{1}{a^2(\eta)} \int \! \! \frac{\dd^3 \kvec}{(2\pi)^3} \, \bigl| \chi_k^\IN(\eta) \bigr|^2 \, \ee^{\ii \kvec \cdot (\xvec - \yvec)} 
}
for $\xvec \neq \yvec$.  The corresponding power spectrum is 
\ba{\label{eq:Delta_Phi_sq}
    \Delta_\aphi^2(\eta,k) & = \frac{k^3}{2\pi^2} \int \! \dd^3 \rvec \, \expval{0^\IN}{\aphi(\eta,\xvec) \aphi(\eta,\yvec)}{0^\IN} \, \ee^{-\ii \kvec \cdot \rvec} \nn 
    & = \frac{1}{a^2(\eta)} \frac{k^3}{2\pi^2} \, \bigl| \chi_k^\IN(\eta) \bigr|^2 
    \com
}
where $\yvec = \xvec + \rvec$.  Note that 
\bes{
    \bigl|\chi_k^\IN(\eta)|^2 
    & = \frac{1}{2\omega_k(\eta)} \Bigl( |\tilde{\alpha}_k(\eta)|^2 + |\tilde{\beta}_k(\eta)|^2 
    \\ & \hspace{1.5cm}
    + 2 \mathrm{Re}\Bigl[ \tilde{\alpha}_k^\ast(\eta) \, \tilde{\beta}_k(\eta) \, \ee^{2 \ii \dphi_k(\eta)} \Bigr] \Bigr) 
    \per
}
Typically the small-scale modes ($k \to \infty$) remain adiabatic, corresponding to $\omega_k(\eta) \approx k$, $\tilde{\alpha}_k(\eta) \approx 1$, and $\tilde{\beta}_k(\eta) \approx 0$, which leads to an asymptotically blue-tilted power spectrum $\Delta_\aphi^2(\eta,k) \sim k^2 / 4 \pi^2 a^2$ as $k \to \infty$.  

\para{When is a particle a particle?}
To close this section, we address several common points of confusion.  The phrase \textit{cosmological gravitational particle production} uses the word \textit{particle} and elicits the image of particle pair creation.  However, when we talk about the inflationary quantum fluctuations of the inflaton field and metric, we do not usually speak of particle creation.  However, both calculations use the Bogolubov formalism to study the evolution of quantum fields in an expanding universe.  Is there some sense in which CGPP creates particles whereas inflationary quantum fluctuations creates something else (such as field fluctuations)?  In other words, when can we say that the excitations of a quantum field correspond to a collection of particles? Several conditions are required in order to talk about the number density of particles interpreted in a Minkowskian context.
\begin{enumerate}
\item Our background cosmological model assumes  at sufficiently early times in the qdS phase, say $\eta<\eta_i$, that $\omega_k(\eta)$ is approximately constant and the evolution of the mode is approximately adiabatic, and at sufficiently late times, say $\eta>\eta_f$, the evolution is again approximately adiabatic.   But for intermediate times $\eta_i<\eta<\eta_f$ the particle interpretation is ambiguous because positive and negative-frequency solutions are mixing.  The evolution is adiabatic when $|A_k| = |\omega_k^\prime / \omega_k^2| \ll 1$:  This condition is also necessary for the identification of $\tilde{\beta}_k$ with the Bogolubov coefficient $\beta_k$ (see \eref{eq:betak_from_betatildek}) and for the decomposition into positive- and negative-frequency modes to be performed unambiguously \cite{Birrell:1982ix}.  In the aforementioned calculations, we assumed that $A_k(\eta) \to 0$ sufficiently quickly as $\eta \to \pm \infty$, which allowed for an unambiguous identification of particles associated with the $\IN$ and $\OUT$ positive-frequency mode functions  \cite{Birrell:1982ix}. 
\item The spacetime curvature must not vary across the spatial size of the wave packet of momentum $k/a$ \cite{Mukhanov:2007zz}.  In the FLRW cosmology this requires  $k/a>H$: if this condition is not satisfied, the curvature of spacetime (of order $H$) will vary across the size of the wave packet.
\item The mode must be oscillatory, which implies that $\omega_k^2>0$.  If this condition is not satisfied the mode equation does not have oscillatory solutions.
\end{enumerate}
In summary, Fourier modes of the field having comoving wave number $k$ should admit a particle description at late times when the mode evolution is adiabatic, when the modes are inside the horizon, and when the modes are oscillatory.  At intermediate times, a particle description may not be appropriate.  Similar issues arise in the study of particle production during preheating; see \rref{Kofman:1997yn} for a discussion.  

\subsection{Energy, renormalization, and observables}
\label{sub:energy_density}

\Sref{sub:solving} reviewed how to calculate the number of gravitationally produced particles.  If we are interested in physics such as baryogenesis, \ie, explaining the origin of the matter-antimatter asymmetry, then this number density is the observable quantity of interest.  If we are instead interested in relic particles that survive today as dark matter or dark radiation, which are probed through their gravitational interactions on cosmological scales, then an energy density is the quantity of interest.  In this section we define the energy carried by gravitationally produced particles and discuss how to calculate it while handling ultraviolet (UV) divergences.  We are interested in both the average energy density and the energy inhomogeneities since the latter are probed using the CMB isocurvature.  

\para{Semiclassical gravity}  
In the framework of semiclassical gravity the gravitational field is treated as classical and the radiation and matter fields are treated as quantum; it is expected to be a good approximation for systems with small fluctuations \cite{Kuo:1993if}.  Gravity's response to radiation and matter is governed by a variant of Einstein's equation, 
\ba{
    R_{\mu\nu} - \tfrac{1}{2} R \metric_{\mu\nu} = \Mpl^{-2} \, \langle T_{\mu\nu} \rangle 
    \per
}
On the left-hand side is the Einstein tensor $G_{\mu\nu}(x)$ for the classical gravitational field, and on the right-hand side is an expectation value of the stress-energy tensor $T_{\mu\nu}(x)$ for the theory's various quantum fields.  During inflation the inflaton field provides the largest contribution to $\langle T_{\mu\nu} \rangle$ and the contribution from the gravitationally produced particles is much smaller; i.e., the backreaction from CGPP can be neglected.  However, if the gravitationally produced particles survive as dark matter at late times, then they provide the largest contribution to $\langle T_{\mu\nu} \rangle$ during the matter-dominated era.  For the scalar field theory whose action appears in \eref{eq:action}, its stress-energy tensor is 
\bes{\label{eq:stressenergyscalar}
    T_{\mu\nu}(x) 
    & = (1-2\xi) \, \bigl( \partial_\mu \aphi \partial_\nu \aphi \bigr) 
    \\ & \quad 
    + (2\xi - \tfrac{1}{2}) \, \bigl( \metric_{\mu\nu} \metric^{\alpha\beta} \partial_\alpha \aphi \partial_\beta \aphi \bigr)
    \\ & \quad 
    + 2 \xi \, \bigl( \metric_{\mu\nu} \aphi \Box \aphi \bigr)
    - 2 \xi \, \bigl( \aphi \del{\mu} \partial_\nu \aphi \bigr)
    \\ & \quad 
    + \xi \, \bigl( G_{\mu\nu} \aphi^2 \bigr)
    + \tfrac{1}{2} \, \bigl( \metric_{\mu\nu} m^2 \aphi^2 \bigr)
    \per
}
Since this is a composite operator, constructed from products of field operators, one encounters UV divergences upon calculating its expectation value that need to be regulated and renormalized. 

\para{Energy density}  
The energy density measured by an observer with four-velocity $U^\mu$ is 
\ba{\label{eq:rhox}
    \rho(x) = \expval{0^\IN}{T_{\mu\nu} U^\mu U^\nu}{0^\IN}
    \per
}
In the observer's rest frame $U^\mu = (\sqrt{1/\metric_{00}},0,0,0)$ since $\metric_{\mu\nu} U^\mu U^\nu = 1$ normalizes the four-velocity.  Decomposing the field operator onto the $\IN$ basis of ladder operators and mode functions leads to 
\bes{\label{eq:SED}
    \rho(\eta) 
    & = a^{-4}\int \! \! \frac{\dd^3 \kvec}{(2\pi)^3} \Bigl[ 
    \tfrac{1}{2} |\chi_k^{\IN\prime}|^2 
    + \tfrac{1}{2} \omega_k^2 |\chi_k^\IN|^2 
    \\ & \qquad 
    + \tfrac{1}{2} (1-6\xi) \, \bigl( a^2 H^2 - \tfrac{1}{6} a^2 R \bigr) \, |\chi_k^\IN|^2 
    \\ & \qquad 
    -  (1-6\xi) \, aH \ \mathrm{Re}\bigl[ \chi_k^\IN \, \chi_k^{\IN\prime\ast} \bigr] 
    \Bigr]
    \per
}
Further using the reparametrization in \eref{eq:scalar_chikIN_ansatz} and assuming $\omega_k(\eta) \in \mathbb{R}$ leads to 
\ba{
    \rho(\eta) 
    & = a^{-4}\int \! \! \frac{\dd^3 \kvec}{(2\pi)^3} \Bigl[ 
    \frac{\omega_k}{2} \bigl( |\tilde{\alpha}_k|^2 + |\tilde{\beta}_k|^2 \bigr)
    \\ & \quad 
    + \frac{1}{4\omega_k} (1-6\xi) \, \bigl( a^2 H^2 - \tfrac{1}{6} a^2 R \bigr) \, \bigl( |\tilde{\alpha}_k|^2 + |\tilde{\beta}_k|^2 \bigr) 
    \nn & \quad 
    + \frac{1}{2\omega_k} (1-6\xi) \, \bigl( a^2 H^2 - \tfrac{1}{6} a^2 R \bigr) \, \mathrm{Re}\bigl[ \tilde{\alpha}_k^\ast \tilde{\beta}_k \, \ee^{2 \ii \dphi_k} \bigr] 
    \nn & \quad 
    + (1-6\xi) \, aH \ \mathrm{Im}\bigl[ \tilde{\alpha}_k^\ast \tilde{\beta}_k \, \ee^{2 \ii \dphi_k} \bigr] 
    \Bigr]
    \per
    \nonumber 
}
Note that several terms vanish if the scalar is conformally coupled to gravity ($\xi = \ssfrac{1}{6}$).  The value of this energy density toward late time $\eta \to \infty$ is a measure of the amount of particle production.  

\para{UV divergence}  
The integral in \eref{eq:SED} has a UV divergence associated with the large $k = |\kvec|$ part of the integration domain.  To identify the divergent behavior, it suffices to consider the $\eta \to - \infty$ limit, since the $\IN$ mode functions are given by \eref{eq:chik_IN_OUT_def}.  One finds that the various terms in \eref{eq:SED} have either quadratic or quartic divergences as $k \to \infty$.  

The appearance of this UV divergence should not be surprising.  After all, a similar divergence arises when one calculates a free field's energy density in Minkowski spacetime: each Fourier mode is a free harmonic oscillator with ground-state energy $\bar{\omega}_k/2$, and the number of modes grows like $k^3$ leading to a quartic divergence $k^3 \bar{\omega}_k \sim k^4$ as $k \to \infty$.  To handle this divergence in Minkowski QFT, the standard approach is to introduce a counterterm that renormalizes the cosmological constant (\ie, the vacuum energy).  However, the situation is more subtle in a curved spacetime where  $\omega_k(\eta)$ is not static, and the divergent part of the integral varies in time.  

\para{Adiabatic regularization}  
We now discuss two strategies for defining and calculating a renormalized energy density.  Adiabatic regularization (AR) is a prescription for removing UV divergences that arise in the study of quantum field theory in curved spacetime \cite{Parker:1974qw,Fulling:1974zr,Fulling:1974pu}.  See \rref{Parker:2009uva} for an introduction and \aref{app:jwkb_method} for supplemental details.  To employ AR it is first necessary to solve the equation of motion using the JWKB method, which allows the mode functions to be expressed as a series $\chi_k(\eta) = \mathrm{exp}[\ii \sum_{n=-1}^\infty \varepsilon^n s_n(\eta)]$ known as an adiabatic series.  Observables constructed from the mode functions such as the energy density are therefore naturally decomposed into a series.  AR calls for the removal of any order in the series that contains a UV divergence.  For instance, the previously mentioned quartic divergence arises at leading order in the adiabatic series and, once the order containing a divergence has been identified, AR requires all terms at that order to be dropped, whether or not they contain a UV divergence.  In this way one is able to define renormalized observables that are free of divergences.  Although the machinery of adiabatic regularization is general and powerful, it is also cumbersome to construct the adiabatic series, and for these reasons we adopt a different prescription for renormalization.  

\para{Renormalization via normal ordering}  
When one encounters a quartic divergence in the Minkowski spacetime energy density, a simple renormalization procedure is to define the physical energy density using the normal-ordered energy operator.  One can adopt a similar procedure in the FLRW spacetime \cite{Chung:2004nh}.  Define the renormalized energy density as 
\ba{
    \rho_\mathrm{ren}(x) = \expval{0^\IN}{\normord{T_{\mu\nu} U^\mu U^\nu}}{0^\IN} 
    \com
}
where the normal ordering is performed with respect to the $\OUT$ basis of ladder operators.  Normal-ordered products of $\IN$-basis ladder operators are given by  \bsa{eq:00_norm_ord_aa}{
	\normord{a_\kvec^\IN a_{\qvec}^\IN} & = a_\kvec^\IN a_{\qvec}^\IN + (2\pi)^3 \, \delta(\kvec + \qvec) \, \alpha_k^\ast \beta_k^\ast \\ 
	\normord{a_\kvec^\IN a_{\qvec}^{\IN \dagger}} & = a_{\qvec}^{\IN \dagger} a_\kvec^\IN - (2\pi)^3 \, \delta(\kvec - \qvec) \, |\beta_k|^2 \\ 
	\normord{a_\kvec^{\IN \dagger} a_{\qvec}^\IN} & = a_\kvec^{\IN \dagger} a_{\qvec}^\IN - (2\pi)^3 \, \delta(\kvec - \qvec) \, |\beta_k|^2 \\ 
	\normord{a_\kvec^{\IN \dagger} a_{\qvec}^{\IN \dagger}} & = a_\kvec^{\IN \dagger} a_{\qvec}^{\IN \dagger} + (2\pi)^3 \, \delta(\kvec + \qvec) \, \alpha_k \beta_k 
	\per
}
Using these relations, one can show that the renormalized energy density is 
\begin{widetext}
\bes{\label{eq:rhobar}
	\rho_\mathrm{ren}(\eta) 
	& = a^{-4} \int \! \! \frac{\dd^3 \kvec}{(2\pi)^3} \biggl\{ 
    \mathrm{Re} \Bigl[ 
    \alpha_k^\ast \beta_k^\ast \, \Bigl(
    \bigl( \partial_\eta \chi_k^\IN \bigr)^2
	+ \omega_k^2 \, \bigl( \chi_k^\IN \bigr)^2 
	- 2 \bigl( 1 - 6 \xi \bigr) \, aH \, \chi_k^\IN \, \partial_\eta \chi_k^\IN 
    \\ & \hspace{3cm}
	+ \bigl( 1 - 6\xi \bigr) \bigl( a^2 H^2 - \frac{1}{6} a^2 R \bigr) \bigl( \chi_k^\IN \bigr)^2 
	\Bigr) 
    \Bigr] 
    - |\beta_k|^2 \, \Bigl[ 
	|\partial_\eta \chi_k^\IN|^2 
	+ \omega_k^2 \, |\chi_k^\IN|^2 
    \\ & \hspace{3cm}
	- \bigl( 1 - 6 \xi \bigr) \, aH \, ( \chi_k^\IN \, \partial_\eta \chi_k^{\IN \ast} + \chi_k^{\IN \ast} \partial_\eta \chi_k^\IN )
	+ \bigl( 1 - 6\xi \bigr) \bigl( a^2 H^2 - \tfrac{1}{6} a^2 R \bigr) |\chi_k^\IN|^2 
	\Bigr]
    \biggr\} 
    \com
}
and upon using \eref{eq:scalar_chikIN_ansatz} it becomes 
\bes{
	\rho_\mathrm{ren}(\eta) 
	& = a^{-4} \int \! \! \frac{\dd^3 \kvec}{(2\pi)^3} \biggl\{ 
    \omega_k \Bigl( 2 \, \mathrm{Re}\bigl[ \alpha_k^\ast \tilde{\alpha}_k \beta_k^\ast \tilde{\beta}_k \bigr] - |\tilde{\alpha}_k|^2 |\beta_k|^2 - |\tilde{\beta}_k|^2 |\beta_k|^2 \Bigr) 
    \\ & \hspace{2cm}
    - \frac{1}{2\omega_k} \bigl( 1 - 6\xi \bigr) \bigl( a^2 H^2 - \tfrac{1}{6} a^2 R \bigr) \bigl( |\tilde{\alpha}_k|^2 + |\tilde{\beta}_k|^2 \bigr) |\beta_k|^2 
    \\ & \hspace{2cm}
    + \frac{1}{2\omega_k} \bigl( 1 - 6\xi \bigr) \bigl( a^2 H^2 - \tfrac{1}{6} a^2 R \bigr) \, \mathrm{Re}\bigl[ 2\alpha_k^\ast \tilde{\alpha}_k \beta_k^\ast \tilde{\beta}_k + \bigl( \alpha_k \tilde{\alpha}_k^{\ast 2} \beta_k - 2 \tilde{\alpha}_k^\ast \tilde{\beta}_k |\beta_k|^2 + \alpha_k^\ast \tilde{\beta}_k^2 \beta_k^\ast \bigr) \ee^{2 \ii \dphi_k} \bigr] 
    \\ & \hspace{2cm}
    + \bigl( 1 - 6 \xi \bigr) \, aH \, \mathrm{Im}\bigl[ \bigl( \alpha_k \tilde{\alpha}_k^{\ast 2} \beta_k - 2 \tilde{\alpha}_k^\ast \tilde{\beta}_k |\beta_k|^2 + \alpha_k^\ast \tilde{\beta}_k^2 \beta_k^\ast \bigr) \ee^{2 \ii \dphi_k} \bigr] 
    \biggr\} 
    \per
}
\end{widetext}
The UV divergences are regulated by the factors of $\beta_k$ that appear in each term.  Modes with large $k$ remain approximately adiabatic at all times ($A_k(\eta) \ll 1$) and do not experience appreciable CGPP so $|\beta_k| \ll 1$.  Provided that $\beta_k$ vanishes sufficiently quickly as $k \to \infty$, the UV divergence is avoided.  Moreover, since $\rho_\mathrm{ren}$ vanishes in the absence of CGPP ($\beta_k = 0$), one can interpret this renormalized energy density as a measure of the energy carried by the gravitationally produced particles.  

\para{Cosmological energy density}  
For practical application a simplified expression for the renormalized energy density is available.  At late time the mode functions asymptote to the Bogolubov coefficients, $\tilde{\alpha}_k(\eta) \to \alpha_k$ and $\tilde{\beta}_k(\eta) \to \beta_k$.   In addition it is typically the case that CGPP is inefficient in the sense that $|\beta_k| \ll |\alpha_k| \approx 1$, which allows terms suppressed by $3$ or $4$ powers of $\beta_k$ to be neglected.  Finally, in a cosmological FLRW spacetime, the Hubble parameter $H(\eta)$ and Ricci scalar $R(\eta)$ are monotonically decreasing.  At sufficiently late times terms suppressed by these factors can be neglected from the energy density, and the dispersion relation can be approximated as $\omega_k^2 \approx k^2 + a^2 m^2$.  With these simplifications the energy density is approximately 
\ba{\label{eq:rho_ren}
    \rho_\mathrm{ren}(\eta) \approx a^{-4} \int \! \! \frac{\dd^3 \kvec}{(2\pi)^3} \, \sqrt{k^2 + a^2(\eta) m^2} \, |\beta_k|^2 
    \per
}
Since the angular integration is trivial, one can write the energy density spectrum $\rho_k(\eta)$ as 
\ba{\label{eq:drhok}
    \rho_\mathrm{ren} = \int \frac{\dd k}{k} \, \rho_k 
    \quad \text{with} \quad 
    \rho_k = \sqrt{(k/a)^2 + m^2} \ n_k 
    \com
}
where $E_k(\eta) = \sqrt{(k/a)^2 + m^2}$ is the energy carried per Fourier mode and $n_k(\eta) = (k^3 / 2\pi^2 a^3) |\beta_k|^2$ is the number density spectrum from \eref{eq:dnk}.  For the nonrelativistic modes, which are expected to carry most of the energy at late times, we further have $\rho_k \approx m n_k$ and $\rho_\mathrm{ren} \approx m n$, where $n$ is the number density from \eref{eq:a3n}.  

\para{Energy density inhomogeneities}
Since CGPP is fundamentally a quantum phenomenon, one should anticipate the distribution of gravitationally produced particles to be spatially inhomogeneous even if the underlying FLRW spacetime were perfectly homogeneous and isotropic.  Moreover, these energy inhomogeneities are a potential signal of CGPP, since the spatial distribution of dark matter is probed by measurements of the cosmic microwave background radiation and large-scale structure.  In other words, CGPP leads to an isocurvature component that is constrained by current cosmological observations and can be searched for in the future. 

For this discussion it is useful to identify operators with a hat.  The energy density operator $\hat{\rho}(\eta,\xvec)$ may be decomposed as
\bes{
    \rhohat(\eta,\xvec) 
    = \, \normord{\hat{T}_{\mu\nu} U^\mu U^\nu} \, 
    = \bar{\rho}(\eta) \, \hat{\bbone} + \drhohat(\eta,\xvec) 
    \per
}
The homogeneous component $\bar{\rho}(\eta) = \rho_\mathrm{ren}(\eta)$ is simply the renormalized energy density from the preceding discussion.  The inhomogeneous component $\drhohat(\eta,\xvec)$ has vanishing expectation value in the $\IN$ vacuum but contributes to the energy two-point function.  The quantity of interest is the dimensionless power spectrum of the energy density contrast 
\bes{
    \Delta_\delta^2(\eta,k) 
    & = 
    \frac{1}{\bar{\rho}^2(\eta)} 
    \frac{k^3}{2\pi^2} 
    \int \! \dd^3 \rvec \ 
    \ee^{-\ii \kvec \cdot \rvec} 
    \\ & \quad \times 
    \expval{0^\IN}{\drhohat(\eta,\xvec) \, \drhohat(\eta,\xvec+\rvec)}{0^\IN} \, 
    \per
}
The correlation function is independent of $\xvec$ and the orientation of $\kvec$ due to the homogeneity and isotropy of the $\IN$ vacuum state.  The calculation of this quantity is simplified by focusing on nonrelativistic modes with $k \ll am$ at late times such that $H \ll m$.  Under those assumptions, the leading terms are given by 
\begin{widetext}
\bes{\label{eq:Deltadelta_chi}
    \Delta_\delta^2(\eta,k)
	& \approx \frac{a^{-8}(\eta)}{2 \bar{\rho}^2(\eta)} \frac{k^3}{2\pi^2} \! \int \! \! \frac{\dd^3 \kvec^\prime}{(2\pi)^3} \ \biggl\{ 
	|\partial_\eta \chi_{k^\prime}^{\IN}|^2 \, |\partial_\eta \chi_{|\kvec^\prime-\kvec|}^{\IN}|^2 
	+ a^4 m^4 \, |\chi_{k^\prime}^{\IN}|^2 |\chi_{|\kvec^\prime-\kvec|}^{\IN}|^2 
	\\ & \hspace{1cm} 
	+ a^2 m^2 \Bigl[ \bigl( \chi_{k^\prime}^{\IN} \, \partial_\eta \chi_{k^\prime}^{\IN \ast} \bigr) \bigl( \chi_{|\kvec^\prime-\kvec|}^{\IN} \, \partial_\eta \chi_{|\kvec^\prime-\kvec|}^{\IN \ast} \bigr) + \bigl( \chi_{k^\prime}^{\IN \ast} \, \partial_\eta \chi_{k^\prime}^{\IN} \bigr) \bigl( \chi_{|\kvec^\prime-\kvec|}^{\IN \ast} \, \partial_\eta \chi_{|\kvec^\prime-\kvec|}^{\IN} \bigr) \Bigr] 
	\biggr\} 
    \com
}
and upon using \eref{eq:scalar_chikIN_ansatz} it becomes 
\ba{\label{eq:Deltadelta}
	\Delta_\delta^2(\eta,k)
	& \approx \frac{a^{-6}(\eta) m^2}{\bar{\rho}^2(\eta)} \frac{k^3}{2\pi^2} \! \int \! \! \frac{\dd^3 \kvec^\prime}{(2\pi)^3} \ \biggl\{ 
	|\tilde{\beta}_{k^\prime}|^2 \, |\tilde{\alpha}_{|\kvec^\prime-\kvec|}|^2 
	+ \mathrm{Re}\bigl[ \tilde{\alpha}_{k^\prime}^\ast \tilde{\beta}_{k^\prime} \tilde{\alpha}_{|\kvec^\prime-\kvec|} \tilde{\beta}_{|\kvec^\prime-\kvec|}^\ast \ee^{2 \ii \dphi_{k^\prime} - 2 \ii \dphi_{|\kvec^\prime-\kvec|}} \bigr]
	\biggr\} 
	\per
}
\end{widetext}
The $\kvec^\prime$ integral is free of UV divergences provided that $\tilde{\beta}_{k^\prime}$ drops off sufficiently quickly toward large $k^\prime = |\kvec^\prime|$.  

If the integrand is sufficiently strongly peaked at some $k^\prime = k_\star$ then for $k \ll k_\star$ the integrand is dominated by $k^\prime \approx k_\star$ and the integral becomes insensitive to $k$.  Evaluating the integral, one finds that $\Delta_\delta^2(\eta,k \ll k_\star) \approx (2/a^3 n) (k^3 / 2\pi^2)$, where $\bar{\rho}(\eta) = m n(\eta)$.  Intuitively a causal generation mechanism implies a finite spatial correlation length and a characteristic $k^3$ power for the energy contrast; this point was emphasized recently by \rref{Hook:2020phx,Amin:2022nlh} in other contexts.  Since CGPP leads to a blue-tilted field amplitude power spectrum for many models (with the notable exception being the minimally coupled scalar field with $m \lesssim \Hinf$), a generic prediction of CGPP is a $k^3$ density contrast.  When the inflaton and metric perturbations are also taken into account, this implies a negligible isocurvature on cosmological scales. 

The term multiplying $m^4$ in \eref{eq:Deltadelta_chi} can also be written as \cite{Liddle:1999pr,Chung:2004nh} 
\bes{\label{eq:Deltadelta_short}
	\Delta_\delta^2(\eta,k)
& \supset \frac{m^4}{\bar{\rho}^2(\eta)} \frac{k^3}{8\pi} \! \int \! \dd^3 \kvec^\prime \ \frac{\Delta_\aphi^2(\eta,k^\prime) \Delta_\aphi^2(\eta,|\kvec^\prime-\kvec|)}{k^{\prime 3} |\kvec^\prime-\kvec|^3}  
	\com
}
where we have used $\Delta_\aphi^2(\eta,k)$ from \eref{eq:Delta_Phi_sq}.  \Eref{eq:Deltadelta_short} reveals that the density contrast's power spectrum is related to a convolution over the field amplitude's power spectrum.  However, \eref{eq:Deltadelta_short} should be used carefully since the $m^4$ term in insolation contains a UV divergence that cancels out the other terms.  If one replaces $\Delta_\aphi^2$ in this formula with $\widetilde{\Delta_\aphi^2} = \Delta_\aphi^2 - k^2 / 4\pi^2 a^2$ then the integral is convergent, and $\Delta_\delta^2$ is typically well approximated by \eref{eq:Deltadelta_short} after an additional factor of $2$ is included to account for the derivative terms in \eref{eq:Deltadelta_chi}.

\section{CGPP for spin-$0$ particles}
\label{sec:Scalar}

Since CGPP is a consequence of the cosmological expansion, it is reasonable to expect this phenomenon to be most efficient at the earliest moments in our cosmic history, when the expansion rate $H$ was the largest.  For this reason CGPP is usually discussed in the context of inflation and reheating.  In this section we provide an overview of inflationary cosmology followed by a summary of key results for scalar CGPP, including a discussion of the implications for dark matter.  We focus on spin-$0$ particles in this section, and studies of CGPP for fields with nonzero spin are discussed in \sref{sec:Spin}. 

\subsection{Inflationary cosmology}
\label{sub:infcos}

We use the phrase \textit{inflationary cosmology} to denote a cosmic history in which the Universe experienced an early qdS phase of inflation, typically followed by a matter-dominated phase after the end of inflation during the epoch known as reheating, followed by the decay of the inflaton condensate into the primordial plasma that signaled the onset of a radiation-dominated phase and the start of hot big-bang cosmology.  In the years after the seminal work of \rref{Guth:1980zm,Starobinsky:1980te,Linde:1981mu,Albrecht:1982wi}, many have written about inflationary cosmology including \rref{Martin:2013tda,Dodelson:2003,Baumann:2022mni}.

\para{Scalar inflaton field}  
In this review we focus on single-field slow-roll inflation driven by a real scalar inflaton field, which we denote by $\cphi(x)$.  However the phenomenon of CGPP is generic and is expected to occur for the various known models of inflation \cite{Martin:2013tda}.  For concreteness suppose that the inflaton field is minimally coupled to gravity with the action 
\ba{
    S = \int \! \dd^4x \, \sqrtg \Bigl[ \tfrac{1}{2} \metric^{\mu\nu} \del{\mu} \cphi \del{\nu} \cphi - V(\cphi) \Bigr] 
    \per
}
The remaining model dependence lies in the specification of the inflaton potential $V(\cphi)$.  Typically $V(\cphi)$ is assumed to have a sufficiently wide and flat region in $\cphi$ that supports a few dozen $e$-foldings of inflation and that connects smoothly into a local minimum at $\cphi = v_\cphi$ where the potential is locally quadratic $V(\cphi) \approx m_\cphi^2 (\cphi-v_\cphi)^2 / 2$.  In other models the minimum is locally quartic or there is no local minimum at all, and the inflaton instead rolls off to infinity during a kination-dominated phase.  Precision observations of the CMB probe the flatness of the inflaton potential \cite{Planck:2018jri}, but only away from the minimum, whereas CGPP probes a wider range of the potential, particularly near the minimum. 

During inflation the inflaton field is assumed to dominate the stress-energy tensor $T_{\mu\nu}$ that appears in Einstein's equation.  When one focuses on the evolution of the homogeneous and isotropic background, \ie, perturbations are neglected, the field is uniform in space $\cphi(t,\xvec) = \cphi(t)$ and the metric is an FLRW spacetime $(\dd s)^2 = (\dd t)^2 - a^2(t) |\dd \xvec|^2$, where $t$ is coordinate time.  The homogeneous background field's equation of motion is $\ddot{\cphi} + 3 H \dot{\cphi} + \dd V(\cphi)/\dd\cphi = 0$ where dots denote $\partial_t = a^{-1} \partial_\eta$.  The field's homogeneous energy density and pressure 
\begin{align}
    \rho_\cphi(t)  = \tfrac{1}{2} \dot{\cphi}^2 + V(\cphi) \ ; \quad 
    \pressure_\cphi(t)  = \tfrac{1}{2} \dot{\cphi}^2 - V(\cphi) 
    \com
\end{align}
appear in Einstein's equation and source the FLRW spacetime.  The equation of state is $w_\cphi(t) = \pressure_\cphi / \rho_\cphi$.  

\para{Quasi-de Sitter phase of inflation}  
Inflation is defined as an epoch during which the comoving Hubble scale is decreasing and the spacetime expansion is accelerating, 
\ba{\label{eq:condits_for_inflation}
    \dot{d}_H(t) < 0 
    \ , \quad 
    \ddot{a}(t) > 0 
    \ , \quad \text{or} \quad 
    \epsilon(t) < 1 
    \per
}
In these relations $d_H(t) = 2 \pi / aH$ is the comoving Hubble length scale, $\ddot{a}(t)$ is the expansion's acceleration in terms of coordinate time $t$, and $\epsilon(t) = -\dot{H}/H^2$ is the first slow-roll parameter.  The spacetime is said to be quasi-de Sitter because the Hubble parameter $H(t)$ is slowly decreasing, rather than constant, throughout inflation.

Modes of fixed comoving wave number $k$ are said to leave the horizon if there is a time at which $(aH)^{-1}$ drops below $k^{-1}$.  Modes of the inflaton field that eventually imprint themselves onto the CMB must leave the horizon several dozen $e$-foldings before the end of inflation (typically $30$ to $60$), but modes on much smaller length scales (much larger $k$) will never leave the horizon.  For studies of CGPP we are often interested in modes of a spectator field that are just about leaving the horizon at the end of inflation and have comoving wave number $k \approx \ke \equiv \ae \He$.  

\para{Inflationary observables}
Cosmological inflation leaves its imprint on the temperature and polarization anisotropies of the cosmic microwave background radiation.  Therefore, CMB measurements probe the primordial scalar metric perturbations (curvature) and tensor metric perturbations (gravitational waves); definitions and additional details are given in \aref{app:cosmo_perturb}.  Precision observations of the CMB by the \textit{Planck} satellite telescope furnish measurements of the scalar power spectrum amplitude $\ln(10^{10} A_s) = 3.044 \pm 0.014$ and spectral tilt $n_s = 0.9649 \pm 0.0042$ \cite{Planck:2018jri}.  Combining with data from BICEP2, \textit{Keck Array}, and BICEP3 leads to a limit on the tensor-to-scalar ratio $r < 0.036$ at $95\%$ confidence level (CL) \cite{BICEP:2021xfz}.  The values of $A_s$ and $n_s$ imply a slightly red-tilted dimensionless power spectrum for the primordial (scalar) curvature perturbations that has an amplitude of $\Delta_\Rcal^2(k) \approx 2.099 \times 10^{-9}$ for $k \approx k_\CMB = 0.05 \, a_0 \Mpc^{-1}$.  Single-field slow-roll inflation predicts the tensor-to-scalar ratio to be 
\bes{
    r 
    & = \frac{A_t}{A_s} 
    = \frac{2 \Hinf^2}{\pi^2 \MPl^2 A_s} 
    \approx  0.16  \biggl( \frac{\Hinf}{10^{14} \GeV} \biggr)^2 
    \com
}
where $\Hinf$ is the value of the Hubble parameter while the CMB-scale modes are leaving the horizon.  Thus the observational constraint on $r$ implies that 
\ba{\label{eq:Hinf_limit}
    \Hinf < 0.5 \times 10^{14} \GeV 
    \per
}
The abundance of gravitationally produced particles typically goes as $\Hinf^2$, and \eref{eq:Hinf_limit} restricts the amount of CGPP.  If inflationary gravitational waves are detected with next-generation telescopes, the value of $\Hinf$ may lie close to the current limit \cite{Kamionkowski:2015yta}, which is particularly favorable for CGPP.  

\para{The end of inflation}
The qdS phase of inflation must end in order for the hot big-bang cosmology to commence.  In most models of single-field slow-roll inflation, the end of inflation occurs when the inflaton field reaches a sufficiently steep region of the inflaton potential.  We define the end of inflation to be the earliest time at which the conditions in \eref{eq:condits_for_inflation} fail.  For instance, at the end of inflation $\epsilon(\te) = 1$.  We denote the end of inflation with an $e$ subscript:  $\te$ for the coordinate time, $\etae$ for the conformal time, $\ae$ for the scale factor, $\He$ for the Hubble parameter, and $\cphi_e$ for the inflaton field amplitude.  It is convenient to use the end of inflation as a reference point for measuring time; for example, $N(t) = \ln a(t) / \ae$ gives the signed number of $e$-foldings that elapse between the time $t$ and the time $\te$ at the end of inflation. 

\para{Hubble-scale modes}  
Modes that are on the Hubble scale at the end of inflation have a comoving wave number of $k \approx \ke \equiv \ae \He$.  Currently this corresponds to a wavelength of $\lambda_\mathrm{phys} = 2 \pi a_0 / \ke$ and a momentum of $p_\mathrm{phys} = \ke / a_0$ where $a_0$ is the scale factor today.  Assuming the cosmological expansion described later (with $\wRH = 0$), one finds that $\lambda_\mathrm{phys} \approx (4 \times 10^{-20} \Mpc) H_{12}^{-1/3} T_9^{-1/3}$ and $p_\mathrm{phys} \approx (1 \times 10^{-18} \GeV) H_{12}^{1/3} T_9^{1/3}$ where $H_{12} = \He / 10^{12} \GeV$ and $T_9 = \TRH / 10^9 \GeV$.  Note that $10^{-20} \Mpc \approx 0.3 \, \mathrm{km}$.  For reference, cosmological observables (CMB and large-scale structure) probe much larger length scales with $k \approx (1 - 10^4) k_0$, where modes that are on the Hubble scale today have $k \approx k_0 \equiv a_0 H_0 \approx 3.33 \times 10^{-4} \, a_0 h \Mpc^{-1}$.  The expansion rate of the present-day Universe, which is known as the Hubble constant, is parametrized as $H_0 = H_{100} h$ where $H_{100} \equiv 100 \km / \mathrm{sec} / \mathrm{Mpc}$ and $h = 0.674 \pm 0.005$ \cite{Planck:2018vyg}. 

\para{Coherent oscillations}  
After inflation has ended, we assume that the inflaton field finds a local quadratic minimum of its potential and begins a phase of spatially coherent oscillations.  At this time the equation of motion for the homogeneous inflaton field is approximately $\ddot{\cphi} + 3 H \dot{\cphi} + m_\cphi^2 \cphi = 0$ and the first Friedmann equation is approximately $3 \Mpl^2 H^2 = \dot{\cphi}^2 / 2 + m_\cphi^2 \cphi^2 / 2$.  These equations are solved by $\cphi(t) \approx \cphi_e (a/\ae)^{-3/2} \cos(m_\cphi t)$ up to corrections that are suppressed by additional factors of $H^2/m_\cphi^2$ or $R/m_\cphi^2$.  The field oscillates quasiharmonically, while the amplitude decreases monotonically due to the cosmological expansion.  The time-averaged pressure is close to zero since the systems spends roughly equal time being potential and kinetic dominated, implying a vanishing equation of state $w_\cphi \approx 0$.  During the epoch of oscillations, the Universe is effectively matter dominated by a population of nonrelativistic inflaton particles.  

\para{Nonlinear effects}  
Since we assume that the inflaton field is free (apart from its gravitational interaction), it follows that the equation of motion is linear, and each of the inflaton's Fourier modes evolves independently.  However, if the inflaton has an appreciable nongravitational self-interaction, or if it interacts sufficiently strongly with other fields, then its field equation develops significant nonlinearity and the Fourier modes couple.  This nonlinearity may lead to various effects, including tachyonic instability, driven resonance, parametric resonance, and inflaton fragmentation~\cite{Traschen:1990sw,Shtanov:1994ce,Kofman:1994rk,Kofman:1997yn}.  The period during which these effects play a significant role is called \textit{preheating}, and it typically occurs soon after the end of inflation.  Preheating is an ever-developing field of study~\cite{Amin:2014eta,Lozanov:2019jxc}, thanks in part to growing computational resources that facilitate numerical simulation~\cite{Figueroa:2021yhd}.  We neglect these nonlinear effects in what follows, since CGPP is mainly sensitive to the background FLRW expansion, and it can remain effectively matter dominated despite nonlinear effects. 

\para{Numerical examples}  
The end of inflation and the inflaton's coherent oscillations play an important role in many models of CGPP.  At the end of inflation, the dominant energy component in the Universe changes from the inflaton's potential energy (with equation of state $w_\cphi \approx -1$) to a balance between its potential and kinetic energies (with a time-average equation of state $w_\cphi \approx 0$).  As a result of this shift in dominant energy component, there is a corresponding change in the cosmological expansion rate that is parameterized by $a$, $H = a^\prime/a^2$, and $R = -6a^\pprime/a^3$.  In addition, as the inflaton field oscillates around the minimum of its potential, the scale factor and its derivatives develop oscillatory components.  Besides the inflaton field and gravity, the various matter-sector fields of the theory ``feel'' this change in the cosmological expansion rate, leading to nonadiabatic evolution and CGPP.  

\begin{figure}[t]
\begin{center}
\includegraphics[width=0.48\textwidth]{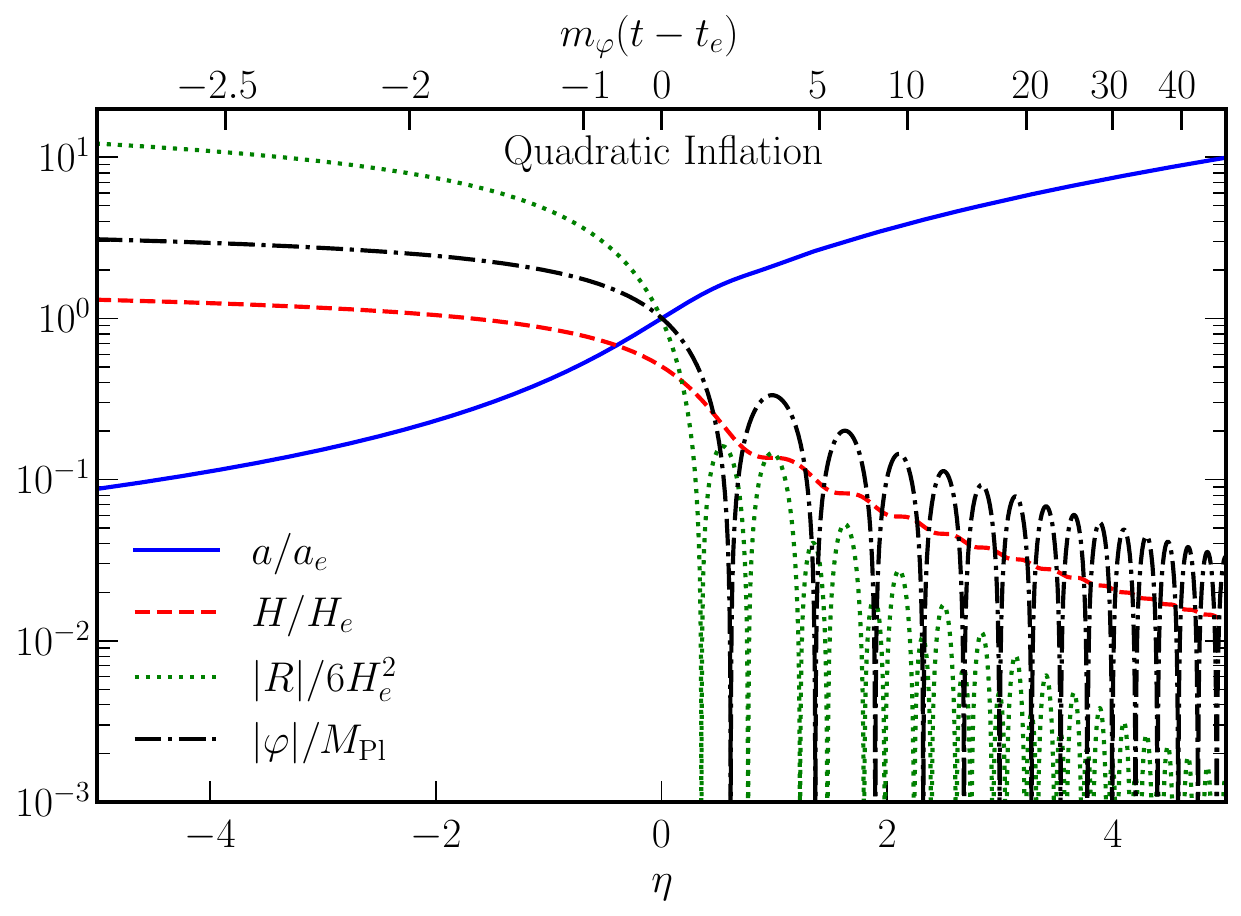} \\
\includegraphics[width=0.48\textwidth]{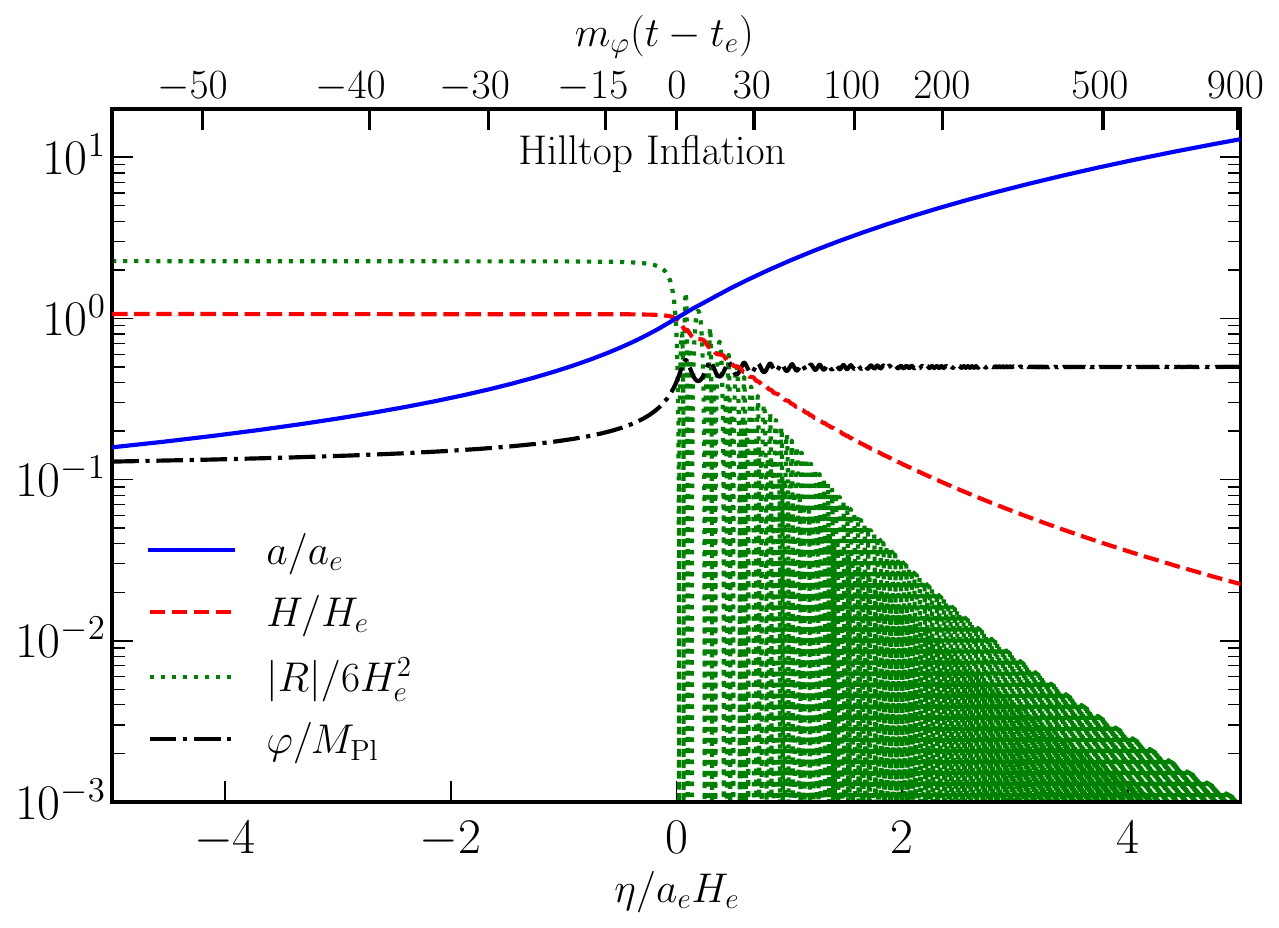} \\
\caption{
\label{fig:end_of_inflation}
Evolution of the FLRW scale factor $a(\eta)$, the Hubble parameter $H(\eta)$, the Ricci $R(\eta)$, and the inflaton field amplitude $\cphi(\eta)$ near the end of inflation.  Conformal time is normalized to vanish at the end of inflation $\etae = 0$ when the scale factor and Hubble parameter take the values $\ae$ and $\He$.  The curves were calculated by numerically solving the homogeneous inflaton equation of motion while assuming a quadratic inflaton potential (upper panel) and a hilltop inflaton potential with $v=\Mpl/2$ (lower panel).   
}
\end{center}
\end{figure}

This behavior is illustrated in \fref{fig:end_of_inflation} for two models of inflation: quadratic (chaotic) inflation with $V(\cphi) = m_\cphi^2 \cphi^2 / 2$ \cite{Linde:1983gd} and a particular implementation of hilltop inflation with $V(\cphi)= m_\cphi^2 v^2(1-\cphi^6/v^6)^2/72$ and $v=\Mpl/2$.  The evolution of the scale factor $a$, the Hubble parameter $H$, the Ricci scalar $R$, and the inflaton field $\cphi$ are shown as functions of the conformal time $\eta$.  Note that the relationships between $\eta/\ae \He$ and $m_\cphi (t-\te)$ are different in the two models because quadratic inflation has $m_\cphi \approx 2 \He$ and hilltop inflation has $m_\cphi \approx 30 \He$.  

During inflation $H$, $R$, and the inflaton field $\cphi$ are approximately constant while the growth of the scale factor $a$ accelerates.  If the spacetime were exactly de Sitter, one would find that $a(\eta) = \ae / (1 - \ae \He \eta)$, $H(\eta) = \He$, and $R(\eta) = -12 \He^2$.  The end of inflation is normalized to $\etae = 0$.  After the qdS phase the Hubble parameter begins to decrease quickly and the growth of the scale factor decelerates.  At this time the inflaton oscillates around the minimum of its potential with an effective equation of state $w_\cphi \approx 0$, corresponding to a matter-dominated phase.  If the equation of state were identically zero, then Friedmann's equations would give $a(\eta) = \ae (1 + \ae \He \eta / 2)^2$, $H(\eta) = \He / (1 + \ae \He \eta / 2)^3$, and $R(\eta) = -3 \He^2 / (1 + \ae \He \eta / 2)^6$, thereby implying that $H(\eta) = \He [a(\eta)/\ae]^{-3/2}$.  Note how the expansion changes abruptly around the end of inflation.  This is one source of nonadiabaticity that particularly impacts the evolution of modes with comoving wave number $k \approx \ae \He$ and contributes to CGPP. 

\para{Epoch of reheating} 
During the epoch of coherent oscillations, the inflaton field gradually transfers its energy into relativistic particles, so this period is also known as the epoch of reheating.  The inflaton's energy transfer is typically accomplished through a combination of perturbative decay channels, for example, $\cphi \to \mathrm{SM} + \mathrm{SM}$, and nonperturbative channels such as tachyonic instability and parametric resonance; see \rref{Amin:2014eta} for a review of nonlinear inflaton dynamics and preheating.  The inflaton's decay products scatter and thermalize to form the primordial plasma, and we denote the plasma's temperature by $T$.  Although $T$ tends to grow due to energy injection from the inflaton decay, it also tends to decrease due to the cosmological expansion with the net effect typically going as $T \propto a^{-3/8}$ \cite{Giudice:2000ex}.  When the energy stored in the growing plasma first exceeds the energy stored in the decaying inflaton condensate, the point of reheating is said to be reached.  This typically occurs when the age of the Universe $H^{-1}$ rises to meet the inflaton's lifetime $\Gamma_\cphi^{-1}$.  We denote the point of reheating with an $\RH$ subscript:  $\tRH$ for coordinate time, $\etaRH$ for conformal time, $\aRH$ for scale factor, $\HRH$ for Hubble parameter, and $\TRH$ for plasma temperature.  However, note that ``reheat'' is somewhat of a misnomer since $\TRH$ is not necessarily the maximum temperature reached after inflation \cite{Giudice:2000ex}; rather, it is the maximum temperature of the radiation-dominated Universe.  Moreover, the Universe may have never been ``heated'' before it was \textit{reheated}.

\para{Reheating temperature}  
The value of $\TRH$ is empirically undetermined.  The smallest viable reheating temperature $\TRH^\mathrm{(min)} \approx 5 \MeV$ is required for successful nucleosynthesis and neutrino thermalization \cite{deSalas:2015glj}.  The largest possible reheating temperature follows from energy conservation: if the inflaton's energy were immediately converted into a thermal bath of radiation at the end of inflation (\ie, instantaneous reheating and thermalization) then the plasma temperature would be 
\ba{\label{eq:instant_RH}
    & \TRH^\mathrm{(max)} 
    = \biggl( \frac{\rho_\cphi(\te)}{\pi^2 \gRH / 30} \biggr)^{1/4} 
    = \biggl( \frac{90 \He^2 \Mpl^2}{\pi^2 \gRH} \biggr)^{1/4} \\ 
    & \ 
    \approx 
    \bigl( 8 \times 10^{14} \GeV \bigr) 
    \biggl( \frac{\He}{10^{12} \GeV} \biggr)^{1/2} 
    \biggl( \frac{\gRH}{106.75} \biggr)^{-1/4}
    \per
    \nonumber
}
In \eref{eq:instant_RH} $\gRH$ counts the effective number of degrees of freedom in the plasma at a temperature of $\TRH$.  Since inflationary observables impose an upper limit on $H_e < H_\mathrm{inf}$ [see \eref{eq:Hinf_limit}], the reheating temperature is expected to fall below $\TRH \approx 6 \times 10^{15} \GeV$.  For a given $\He$ lowering $\TRH$ corresponds to extending the duration of the reheating epoch.  Assuming that the Universe is matter dominated during reheating ($\wRH = 0$) such that $H \propto a^{-3/2}$ for $\ae < a < \aRH$, one finds that 
\bes{\label{eq:aRH_ov_ae}
    \frac{\aRH}{\ae} 
    & \approx 
    \bigl( 8 \times 10^7 \bigr) 
    \biggl( \frac{\He}{10^{12} \GeV} \biggr)^{2/3} 
    \\ & \qquad \times 
    \biggl( \frac{\TRH}{10^9\GeV} \biggr)^{-4/3} 
    \biggl( \frac{\gRH}{106.75} \biggr)^{-1/3} 
    \per
}
A different relation is obtained for other choices of $\wRH$.  Since the Universe is radiation dominated at the point of reheating, the first Friedmann equation implies that $\pi^2 \gRH \TRH^4  / 30 = 3 \HRH^2 \Mpl^2$, which gives 
\ba{\label{eq:HRH}
    \HRH \approx 
    \bigl( 1.4 \GeV \bigr) 
    \biggl( \frac{\TRH}{10^9 \GeV} \biggr)^2 
    \biggl( \frac{\gRH}{106.75} \biggr)^{1/2} 
    \per
}
\Eref{eq:HRH} will prove useful since one can compare $m$ to $\HRH$ to determine whether a field of mass $m$ is released from Hubble drag during the reheating epoch ($\HRH < m$; ``early reheating'') or during the radiation epoch ($m < \HRH$; ``late reheating'').  

\para{Comoving entropy conservation} 
After reheating is completed ($a > \aRH$) it is customary to assume that the comoving entropy density of the primordial plasma remains constant.  The entropy density at time $t$ is written as $s(t) = (2\pi^2/45) g_{\ast S}(t) T^3(t)$ where $g_{\ast S}(t)$ is the effective number of relativistic species.  Entropy conservation after reheating implies $\aRH^3 s_\RH = a^3(t) s(t) = a_0^3 s_0$, and thus 
\bes{\label{eq:a0_ov_aRH}
	\frac{a_0}{\aRH} 
	& \approx 
	\bigl( 1 \times 10^{22} \bigr) 
	\biggl( \frac{\TRH}{10^9 \GeV} \biggr)^{} 
	\biggl( \frac{\gRH}{106.75} \biggr)^{1/3} 
    \com
}
where $T_0 \approx 0.234 \ \mathrm{meV}$ and $g_{\ast S,0} \approx 3.91$. 

\begin{figure}[t]
\begin{center}
\includegraphics[width=0.48\textwidth]{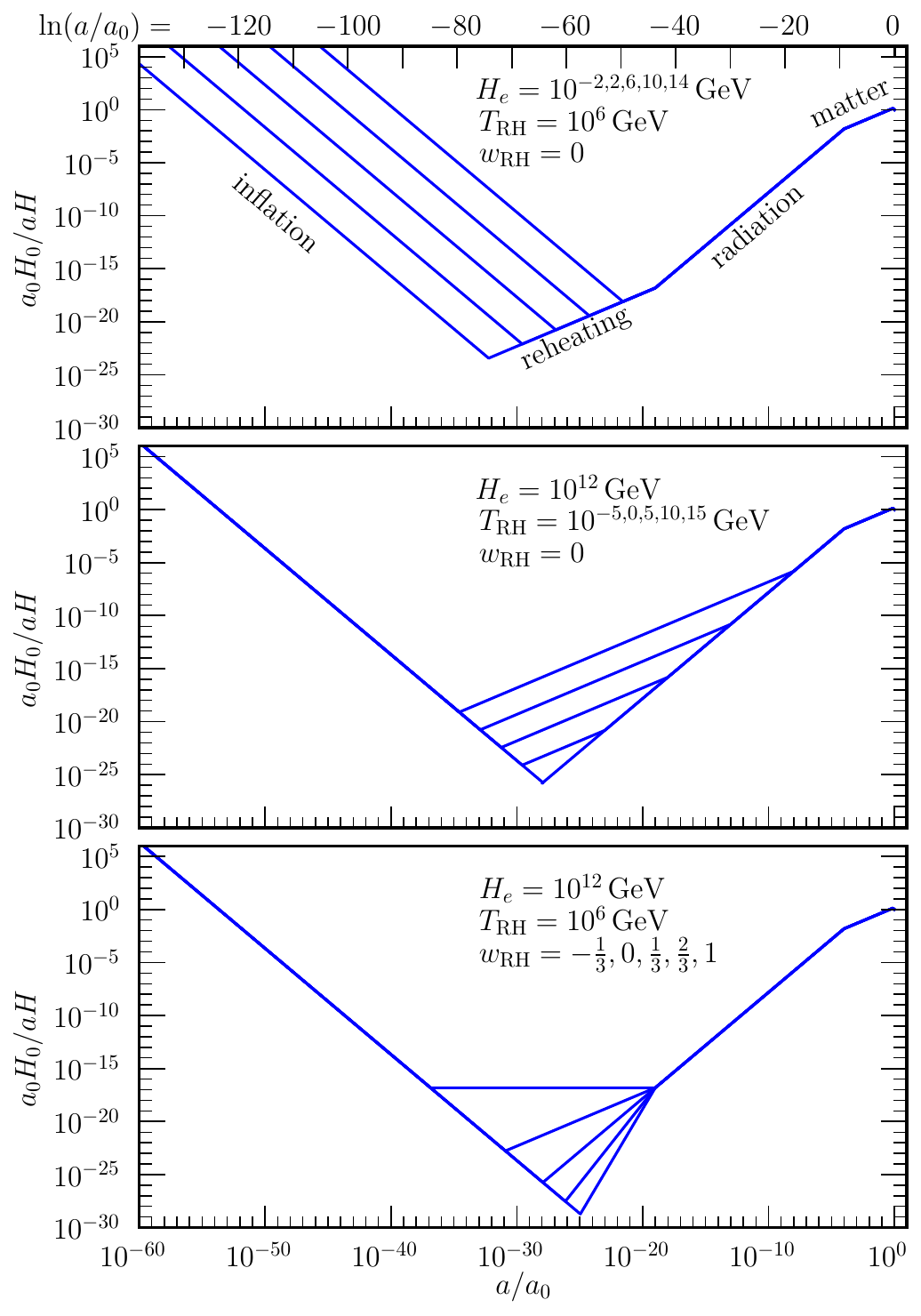} 
\caption{
\label{fig:cosmic_history}
Cosmological expansion history.  The comoving Hubble length $d_H(a) = 1 / a H(a)$ is shown as a function of the FLRW scale factor $a$.  The three panels show the impact of varying the Hubble parameter at the end of inflation $\He$ (top panel), the reheating temperature $\TRH$ (middle panel), and the equation of state during reheating $\wRH$ (bottom panel).  From top to bottom, the curves show $\log_{10} \He/\mathrm{GeV} = -2, 2, 6, 10, 14$; $\log_{10} \TRH/\mathrm{GeV} = -5, 0, 5, 10, 15$; and $\wRH = - \tfrac{1}{3}, 0, \tfrac{1}{3}, \tfrac{2}{3}, 1$.  
}
\end{center}
\end{figure}

\para{$\Lambda$CDM cosmology} 
After reheating has occurred, the hot primordial plasma is formed and the subsequent cosmological evolution is described by the $\Lambda$CDM cosmological concordance model.  See \rref{KolbTurner:1990} for an introduction to big-bang nucleosynthesis (BBN), radiation-matter equality, recombination, the cosmic microwave background (CMB), the large scale structure (LSS), and dark energy.  The full cosmological expansion history is illustrated in \fref{fig:cosmic_history}, which shows the evolution of the comoving Hubble length scale $d_H(a) = 1 / a H(a)$ as a function of the FLRW scale factor $a$ in terms of its present value.  Reading from the left in \fref{fig:cosmic_history}, the first segment corresponds to the epoch of inflation during which $H(a) = \He$ is approximately constant and $d_H(a)$ decreases, the second segment corresponds to the epoch of reheating during which $H(a) \propto a^{-3(1+\wRH)/2}$, the third segment corresponds to the radiation-dominated era during which $H(a) \propto a^{-2}$, the fourth segment corresponds to the matter-dominated era during which $H(a) \propto a^{-3/2}$, and the final segment (which is difficult to see at $a/a_0 \approx 1$) corresponds to the $\Lambda$-dominated era during which $H(a) = H_0$ is approximately constant.  Note that \fref{fig:cosmic_history} does not capture the slow evolution of $H(a)$ during inflation and the smooth transition at the end of inflation, which are shown in \fref{fig:end_of_inflation}. 

\subsection{Mode evolution and spectrum}
\label{sub:number_spectrum}

This section discusses the evolution of the mode functions $\chi_k(\eta)$ and the resultant comoving number density spectrum $a^3 n_k$.  We consider CGPP for massive spin-$0$ particles that correspond to the quantum excitations of either conformally coupled or minimally coupled scalar fields.  Our approach is to present an example of numerical results, discuss the salient features, and provide an analytical understanding.  A derivation of the analytical results will be provided by \rref{JKLT}.  We discuss conformally coupled scalars and then minimally coupled scalars.  In this section we employ the dimensionless variables 
\bes{
    \ta & = a / \ae
    \, , \quad 
    \tm = m / \He
    \, , \quad 
    \tH = H / \He 
    \, , \quad 
    \\ 
    \tk & = k / \ae \He
    \, , \quad \text{and} \quad 
    \teta = \ae \He (\eta - \etae) 
    \com
}
where $\ae$, $\He$, and $\etae$ are the values of the FLRW scale factor, the Hubble parameter, and the conformal time at the end of inflation.   Recall that Hubble-scale modes at the end of inflation have $k \approx \ke \equiv \ae \He$ corresponding to $\tk \approx 1$.  

\begin{figure}[t]
\begin{center}
\includegraphics[width=0.48\textwidth]{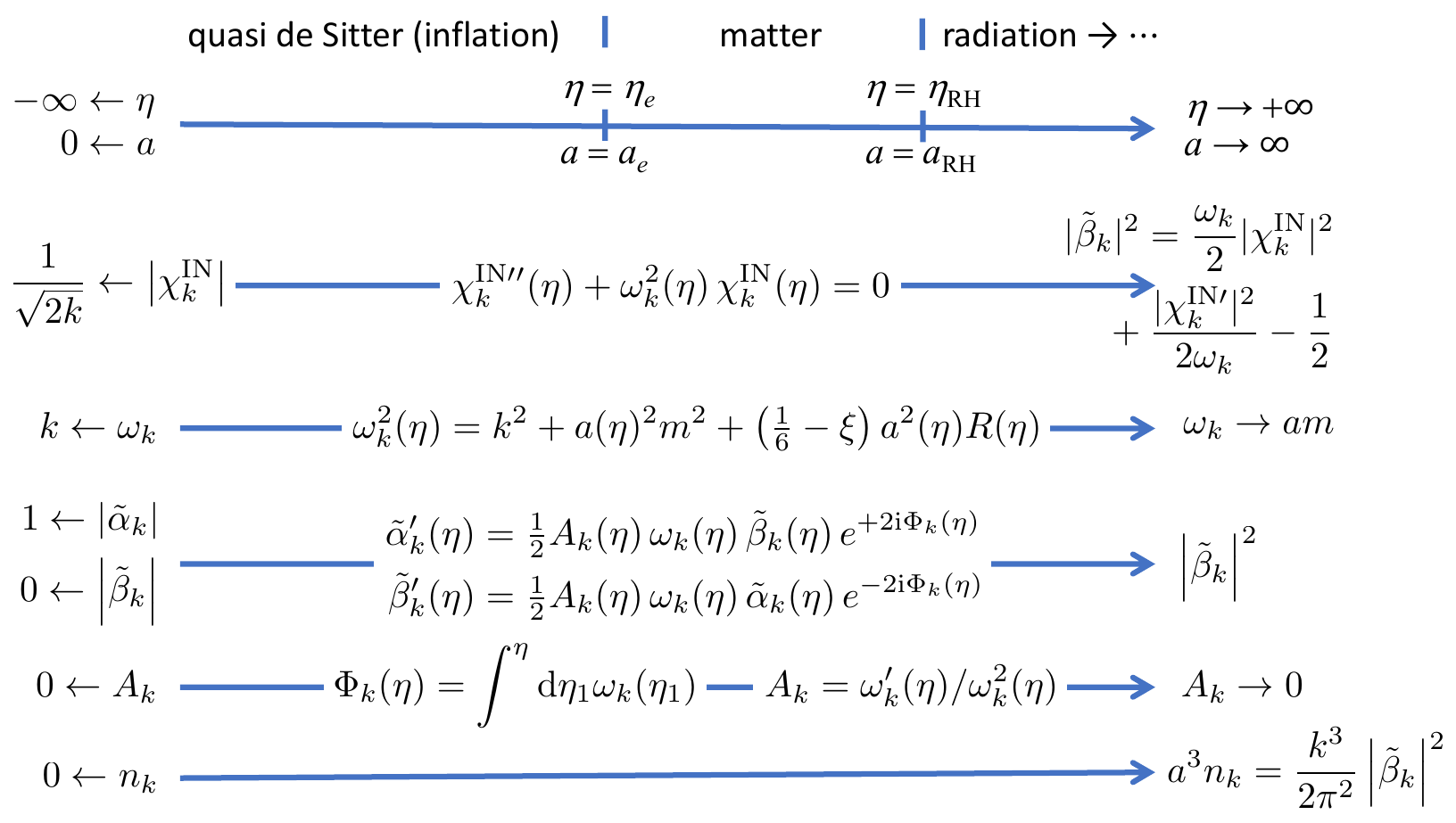} 
\caption{
\label{fig:cartoon}
Graphical summary of the CGPP calculation for inflationary cosmology.  One seeks to calculate the comoving number density spectrum $a^3 n_k$, which requires the Bogolubov coefficient $\tilde{\beta}_k$, which is related to the mode function $\chi_k^\IN$, which solves the oscillator equation with a time-dependent frequency $\omega_k(\eta)$ subject to a Bunch-Davies initial condition as $\eta \to -\infty$. 
}
\end{center}
\end{figure}

\para{Procedure}
\Fref{fig:cartoon} illustrates the key formulas in the CGPP calculation.  To begin, a model of inflation is selected.  The Friedmann equation is then solved along with the inflaton's homogeneous equation of motion to find the FLRW scale factor $a(\eta)$.  If one is solving numerically, it is necessary to select a wide enough range of $\eta$ that the modes of interest are initially inside the horizon ($a < k/\Hinf$) and eventually they either reenter the horizon or the field is released from Hubble drag ($H < m$).  Next the equation of motion for the spectator field \eqref{eq:omegak_scalar} is solved along with the initial condition \eqref{eq:chik_IN_OUT_def} that $\chi_k^\IN(\eta) \to \ee^{-\ii k \eta} / \sqrt{2k}$ as $\eta \to -\infty$.  If solving numerically, one can impose $\chi_k^\IN(\eta_i) = 1 / \sqrt{2 \omega_k(\eta_i)}$ and $\partial_\eta \chi_k^\IN(\eta_i) = - \ii \omega_k(\eta_i) \chi_k^\IN(\eta_i)$ at an early time $\eta_i$ when the modes with comoving wave number $k$ are inside the horizon.  The mode functions $\tilde{\beta}_k(\eta)$ are calculated using \eref{eq:betachixhiprime} and CGPP is completed when $|\tilde{\beta}_k(\eta)|^2$ reaches a constant, which coincides with the Boglubov coefficient $|\beta_k|^2$ linking the $\IN$ and $\OUT$ ladder operators.  Alternatively, one can solve \eref{eq:EOMforalphabeta} to obtain $\tilde{\alpha}_k(\eta)$ and $\tilde{\beta}_k(\eta)$ directly, which is most useful when $\omega_k^2$ remains positive.  Either way, \eref{eq:dnk} gives the comoving number density spectrum $a^3 n_k = (k^3 / 2\pi^2) |\beta_k|^2$ and \eref{eq:drhok} gives the energy density spectrum $\rho_k = E_k \, n_k$.  

\begin{figure}[t]
\begin{center}
\includegraphics[width=0.48\textwidth]{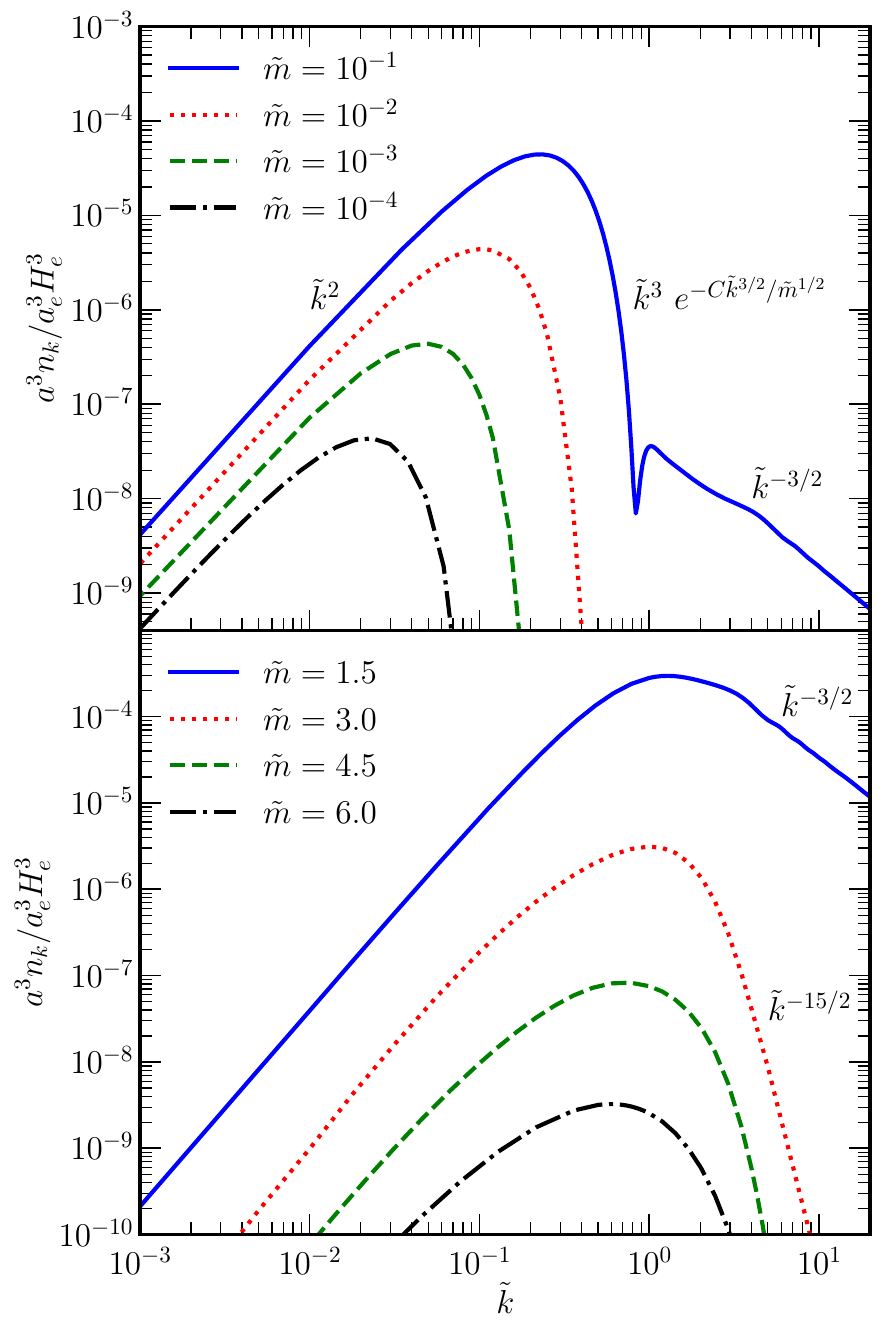} 
\caption{
\label{fig:es-conformal}
The comoving number density spectrum $a^3 n_k$ resulting from CGPP for a conformally coupled scalar field in quadratic inflation assuming late reheating ($\HRH < m$).  On the horizontal axis the comoving wave number is expressed as $\tk = k / \ae \He$ such that $\tk \approx 1$ corresponds to Hubble-scale modes at the end of inflation.  The various curves show different values of the scalar field's mass with $\tm = m / \He$. 
}
\end{center}
\end{figure}

\para{Numerical results: comoving number density spectrum} 
\Fref{fig:es-conformal} shows a selection of numerical results for CGPP of massive scalars that are conformally coupled to gravity ($\xi=\sixth$).  Some notable features can be observed: 
(1)  For small wave numbers, the spectrum scales as $n_k \propto k^2$ for low mass and scales more steeply for high mass.  
(2)  For intermediate wave numbers, the spectrum reaches a peak at a value of $k$ that depends on $m$.  
(3)  Beyond the peak the spectrum decreases exponentially with increasing $k$.  
(4)  For still larger $k$ the spectrum develops a power-law tail that progresses as $\tk^{-3/2}$ or $\tk^{-15/2}$.  
(5)  The amplitude of the spectrum grows as $n_k \propto m$ as $\tm$ approaches $\approx 1$ from below, and it decreases as $n_k \propto \ee^{-c \tm}$, where $c$ is a constant, as $\tm$ increases beyond $\approx 1$. 
In what follows we discuss each of these features.  

\para{Low-$k$ power law}
The behavior of the spectrum toward small $\tk$ is a simple power law.  Modes with small wave number ($\tk < 1$) leave the horizon during inflation and reenter later during the reheating or radiation era.  Consequently, the evolution of the mode functions $\chi_k(\eta)$ is insensitive to the dynamics of the inflaton field at the end of inflation but instead can be studied by treating the spacetime as a de Sitter one with a constant Hubble parameter during inflation.  It is well known that the mode equations for a scalar spectator field in de Sitter spacetime are Hankel functions; see \aref{app:dS_spacetime}.  Using the de Sitter solution leads to 
\ba{
    \frac{a^3n_k}{\ae^3\He^3} \simeq \frac{1}{8\pi^2} \tk^{3 - 2 \mathrm{Re}[\nu]} \tm^{1/3} 
    \qquad \text{for $\tk < \tm^{1/3} < 1$} 
    \com
}
where the index $\nu = \sqrt{1/4 - \tm^2}$ is given by \eref{eq:nu_def} with $\xi = \sixth$.  For $\tm < 1/2$ the spectrum scales as $\tk^2$, which agrees with the scaling seen in \fref{fig:es-conformal}.  For a larger mass $\tm > 1/2$, the index $\nu$ is imaginary, and the spectrum is expected to scale as $\tk^3$.  \Fref{fig:es-conformal} displays a redder spectrum closer to $\tk^2$ since the Hubble parameter decreases by a factor of almost $10$ during quadratic inflation, rather than remaining constant as in de Sitter; see \fref{fig:end_of_inflation}.

\para{Dominant modes: $k_\star$}
For models with $\tm < 1$, the spectrum is largest for values of the comoving wave number around $\tk \approx \tk_\star^\mathrm{(late)} \equiv \tm^{1/3}$.  These correspond to modes that leave the horizon during inflation and reenter the horizon at the same time when the Hubble parameter $H(\eta)$ drops below the mass $m$ and the field is released from Hubble drag.  In brief, 
\ba{\label{eq:kstar_def}
    H(\eta_\star) = m
    \quad \text{and} \quad 
    k_\star = a(\eta_\star) H(\eta_\star) = a(\eta_\star) m 
    \per
}
The meaning of the superscript ``late'' is discussed later.  

\para{A blue-tilted spectrum}
It is straightforward to calculate the spectrum analytically for small $\tk$ where the de Sitter approximation is applicable.  However, we draw attention to the fact that the spectrum rises with increasing $\tk$, corresponding to a blue spectral tilt.  The modes that carry most of the energy have $\tk \approx \tm^{1/3}$, which is possibly $\approx 1$, corresponding to modes that are on the Hubble scale at the end of inflation.  Therefore, to calculate the total number of produced particles, it is necessary to more accurately model the dynamics of the inflaton field at the end of inflation as the Universe transitions from a quasi-de Sitter phase of inflation into an effectively matter-dominated phase of reheating.  \rref{Chung:1998zb} noted that assuming an instantaneous transition from the de Sitter phase into matter domination can lead to a spurious overestimate of CGPP; see also \rref{Li:2019ves}. 

\para{Exponential drop}
For models with $\tm < 1$ the spectrum begins to drop exponentially for the modes with $\tk > \tm^{1/3}$ that are past the peak.  This behavior can be derived analytically via several related approximations, including the steepest descent method \cite{Chung:1998bt} and the Stokes phenomenon \cite{Li:2019ves,Hashiba:2021npn}; see also \rref{Birrell:1982ix,Chung:1998zb,Chung:2018ayg} for supplementary discussions.  The result is found to be 
\ba{
    \frac{a^3 n_k}{\ae^3 \He^3} \approx \frac{\tk^3}{2\pi^2} \, \ee^{-C \tk^{3/2}/\tm^{1/2}} 
    \quad \mathrm{for}\ \ \tm^{1/3} < \tk < \frac{m_\cphi \kappa}{\He}
    \com
}
where $C\sim 2\pi$ agrees well with the numerical results and $\kappa = (1 - m^2 / m_\cphi^2)^{1/2}$.  The upper boundary $m_\cphi \kappa / \He$ is discussed next. 

To develop intuition consider the mode equations in terms of $\tilde{\alpha}_k(\eta)$ and $\tilde{\beta}_k(\eta)$ from \eref{eq:EOMforalphabeta}.  If one anticipates that CGPP will be inefficient for $\tm > 1$, the solution will have $|\tilde{\beta}_k| \ll 1$ and $|\tilde{\alpha}_k| \approx 1$.  Setting $\tilde{\alpha}_k(\eta) \approx 1$ yields 
\ba{
    \tilde{\beta}_k(\eta) \approx \int^\eta \! \dd\eta^\prime \, \tfrac{1}{2} A_k(\eta^\prime) \, \omega_k(\eta^\prime) \, \ee^{-2 \ii \int^{\eta^\prime} \! \dd \eta^\pprime \omega_k(\eta^\pprime)} 
    \per
}
Moreover, for models with $\tm > 1$ the nonrelativistic modes have $\omega_k \approx a m$.  At late times $\tilde{\beta}_k \to \beta_k$, leading to 
\ba{\label{eq:Eq64}
    \beta_k \approx \int_{-\infty}^\infty \! \dd t^\prime \, \frac{1}{2} \frac{\dot{\omega}_k(t^\prime)}{\omega_k(t^\prime)} \, \ee^{-2 \ii m t^\prime} 
    \com
}
where $a \dd \eta = \dd t$.  \Eref{eq:Eq64} reveals that $\beta_k$ is the Fourier transform of $\dot{\omega}_k / 2 \omega_k$ evaluated for an angular frequency of $2m$.  Since the time dependence is induced by the cosmological expansion, which varies on the timescale $1/H$, one does not expect $\dot{\omega}_k / \omega_k$ to include a mode that varies on the much shorter timescale of $1/m$, which is the origin of the exponential suppression.  

\para{High-$k$ power law tail}
For the highest wave number modes shown in \fref{fig:es-conformal}, the spectrum develops a power-law tail that falls off like $\tk^{-3/2}$ or $\tk^{-15/2}$, depending on $\tm$.  This behavior is summarized here and further discussed in \sref{sub:inflaton_annihilations}.  At the end of inflation, the inflaton field oscillates around the minimum of its potential.  If $m_\cphi > \He$ then these oscillations are rapid, which evades the preceding argument by providing support for the Fourier transform of $\dot{\omega}_k / \omega_k$ at frequency of $2m \approx 2 m_\cphi$.  Moreover, the system can be understood to consist of a collection of cold inflaton particles.  These particles interact gravitationally with one another and with the scalar spectator.  In particular, they can annihilate as $\cphi \cphi \to \chi \chi$ through an $s$-channel graviton exchange provided that $m \equiv m_\chi < m_\cphi$.  The $\chi$ particles produced in this way carry a momentum of $p_\chi \approx m_\cphi$, which begins to decrease due to cosmological redshift.  The earliest particles to be produced have a comoving wave number (\ie, comoving momentum) of $k \approx \ae m_\cphi$, and the last particles to be produced just before reheating is completed have a comoving wavenumber of $k \approx \aRH m_\cphi$.  By considering the scattering rate, one can show that the spectrum decreases as a power law across this range of wave numbers.  If $m < m_\cphi$ then the two-to-two channel is kinematically accessible and it dominates, giving 
\ba{
    \frac{a^3 n_k}{\ae^3 \He^3} \approx C \tm^4 \tk^{-3/2}
    \quad \text{for $\frac{m_\cphi \kappa}{\He} < \tk < \taRH \frac{m_\cphi \kappa}{\He}$}
    \com
}
where $C = (9 / 1024 \pi) (m_\cphi / \He)^{-5/2} \kappa^{5/2}$ \rref{Chung:2018ayg} is applicable for quadratic inflation.  However, if $m_\cphi < m < 2 m_\cphi$ then the two-to-two channel is kinematically forbidden and the 4-to-2 channel dominates giving $n_k \propto \tk^{-15/2}$ instead.  To understand the change in this power-law behavior in \fref{fig:es-conformal}, recall that quadratic inflation has $m_\cphi \approx 2 \He$, so the break from $\tk^{-3/2}$ to $\tk^{-15/2}$ happens at around $\tm = 2$.  

Although not pronounced in \fref{fig:es-conformal}, there are oscillations in $a^3 n_k$ in the region of $k^{-3/2}$ scaling.  These oscillations are the result of quantum interference between the  $2\cphi\to\chi\chi$ and $4\cphi\to\chi\chi$ channels \cite{Basso:2022tpd}.  The interference is more pronounced in models with a $\cphi^3$ term at the bottom of the inflaton potential such as hilltop inflation, where there is interference between $2\cphi\to\chi\chi$ and $3\cphi\to\chi\chi$.

\para{Conformally coupled scalar: summary}
To summarize, the numerical results in \fref{fig:es-conformal} for CGPP of a massive scalar field that is conformally coupled to gravity are well described by the following analytical approximations, 
\ba{\label{eq:a3nk_conformal}
    \frac{a^3n_k}{\ae^3\He^3} 
    \approx \left\{ 
    \begin{array}{ll}
    \!\! \dfrac{1}{8 \pi^2} \tk^2 \tm^{1/3} & \ 0 < \tk < \tm^{1/3} \\[1ex]
    \!\! \dfrac{1}{2\pi^2} \tk^{3} \ee^{-C\tk^{3/2}/\tm^{1/2}} & \ \tm^{1/3} < \tk < \dfrac{m_\cphi \kappa}{\He} \\[1ex]
    \!\! C \tm^4 \tk^{-3/2} & \ \dfrac{m_\cphi \kappa}{\He} < \tk < \taRH \dfrac{m_\cphi \kappa}{\He} \com
    \end{array} \right. 
    \com
}
provided that $m < m_\cphi$.  

\begin{figure}[t]
\begin{center}
\includegraphics[width=0.48\textwidth]{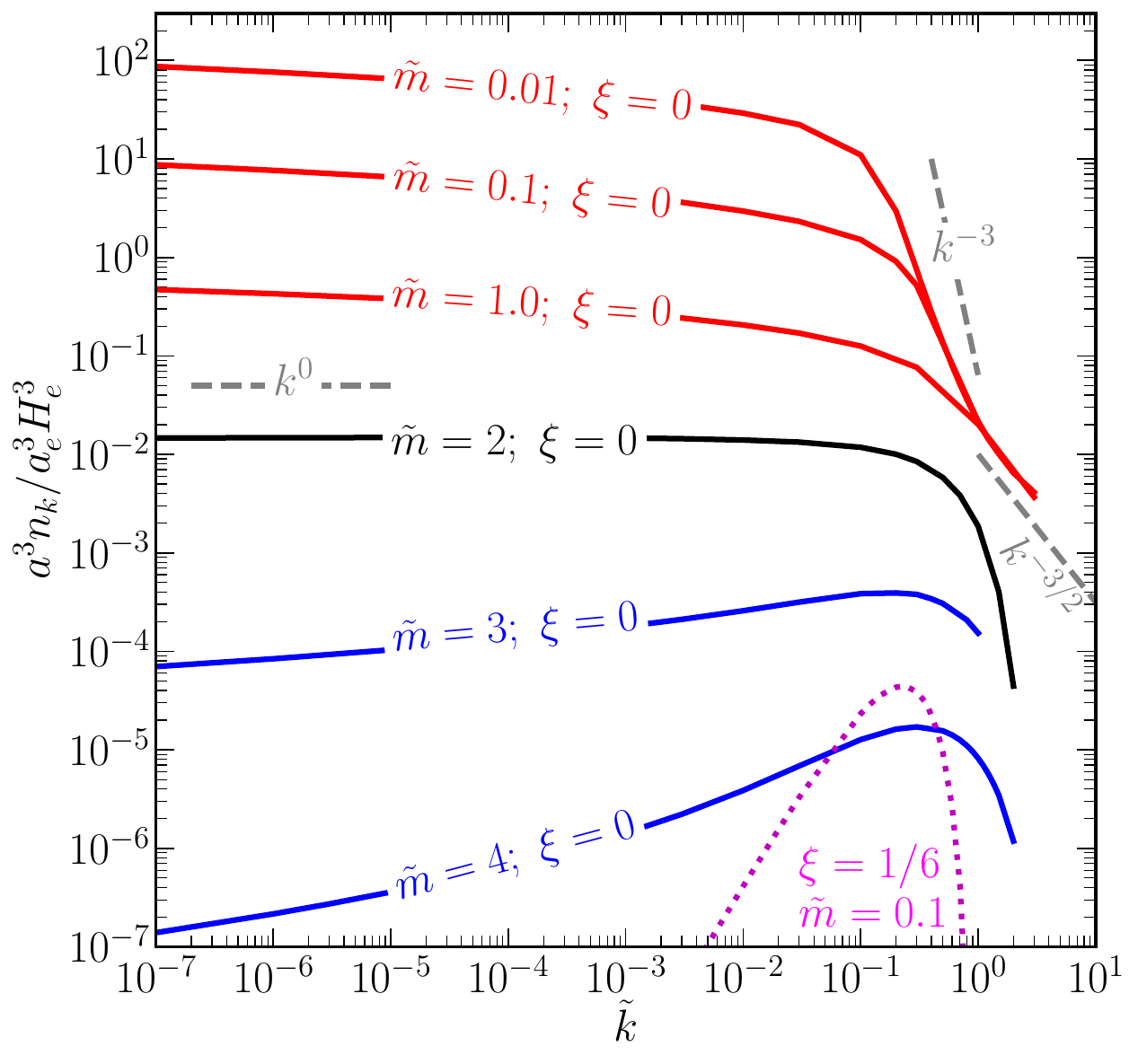} 
\caption{
\label{fig:es-minimal}
Same as \fref{fig:es-conformal}, but here we show spectra for a minimally coupled scalar (solid curves).  For comparison, the dotted curve shows a conformally coupled scalar.  
}
\end{center}
\end{figure}

\para{Numerical results:  Comoving number density spectrum}
\Fref{fig:es-minimal} shows a selection of numerical results for CGPP of massive scalars that are minimally coupled to gravity ($\xi=0$).  Several notable features can be observed: 
(1)  For the same mass (see $m/\He = 0.1$) the spectrum for a minimally coupled scalar is much larger than that for a conformally coupled scalar.  
(2)  For small wave numbers, the spectrum is nearly flat but slightly red tilted for low masses $m/\He \lesssim 1$, but it becomes increasingly blue tilted for large masses $m/\He \gtrsim 1$.  
(3)  For larger wave numbers, the spectrum decreases as either a power law or an exponential.  
In what follows we discuss each of these features.  

\para{Contrasting conformal and minimal}
The distinct behavior between the gravitational production of conformally and minimally coupled scalars (which are light during inflation) can be understood by inspecting their respective equations of motion.  Both mode equations take the form $\chi_k^\pprime(\eta) + \omega_k^2(\eta) \chi_k(\eta) = 0$, where the angular frequency is given by \eref{eq:omegak_scalar}: $\omega_k^2 = k^2 + a^2 m^2 + (1/6 - \xi) a^2 R$.  For the conformally coupled scalar, $\xi = \sixth$ and the $a^2 R$ term is absent; however, for the minimally coupled scalar, $\xi = 0$ and the $a^2 R$ term appears with a positive coefficient.  During inflation, the Hubble parameter is approximately constant ($H^\prime \approx 0$) and the Ricci scalar is negative ($R \approx - 12 H^2$).  Therefore, whereas $\omega_k^2(\eta)$ always remains positive for the conformally coupled scalar, it may become negative for the minimally coupled scalar, specifically, for a light scalar ($m < \Hinf$) and for modes outside the horizon ($k < aH$).  For the tachyonic modes with $\omega_k^2 < 0$, the mode functions grow as $\chi_k(\eta) \propto a(\eta)$.  This correspond to a frozen amplitude ($\aphi \propto a^0$) for the original field variable before its kinetic term is canonically normalized; see \eref{eq:SFLRW}.  This growth of $\chi_k(\eta)$ for modes outside the horizon is primarily responsible for the enhanced CGPP for minimally coupled scalars relative to conformally coupled scalars.  

\para{Low-$k$ behavior}
Per the previous discussion regarding the conformally coupled scalar, the evolution of modes with $\tk < 1$ is captured by the de Sitter approximation, and a simple analytic form for the spectrum can be derived.  Using the expressions from \aref{app:dS_spacetime} gives 
\ba{
    \frac{a^3n_k}{\ae^3\He^3} \simeq \frac{1}{8\pi^2} \tm^{-1} \tk^{3 - 2 \mathrm{Re}[\nu]} 
    \qquad \text{for $\tk < \tm^{1/3} < 1$}
}
where the index $\nu = \sqrt{9/4 - \tm^2}$ is given by \eref{eq:nu_def} with $\xi = 0$.  For $\tm \ll 9/4$ the exponent approaches $0$ and the scaling approaches $\tk^0$, corresponding to a flat spectrum.  The predicted flatness is linked with the assumption of a constant Hubble parameter throughout inflation.  For a model in which the Hubble parameter decreases significantly, such as quadratic inflation used to derive the numerical results in \fref{fig:es-minimal}, the spectrum is red tilted instead.  

\para{Intermediate-$k$ behavior}
For models in which the minimally coupled scalar is light ($\tm < 1$), the spectrum has a break at $\tk \approx \tk_\star^\mathrm{(late)} \equiv \tm^{1/3}$.  These modes are entering the horizon at the same time when the field is released from Hubble drag; see \eref{eq:kstar_def}.  Modes with larger values of $\tk$ enter the horizon earlier during the epoch of reheating, which is assumed to be effectively matter-dominated, while they are still relativistic.  By tracking the evolution of these modes, one finds that 
\bes{
    \frac{a^3 n_k}{\ae^3 \He^3} 
    \approx \frac{1}{8\pi^2} \tk^{-3} 
    \qquad \text{for $\tm^{1/3} < \tk < \frac{m_\cphi \kappa}{\He}$} 
    \per
}
However, in models with a heavy scalar ($\tm > 1$), the modes with $\tm^{1/3} < \tk$ are already nonrelativistic.  Consequently the power law becomes an exponential. 

\para{High-$k$ behavior}
The range of modes with $\ae m_\cphi \kappa < k < \aRH m_\cphi \kappa$ are populated by the inflaton's coherent oscillations between the end of inflation and the end of reheating.  The calculation is similar to the previously discussed conformally coupled scalar.  One notable difference is that the cross section for gravitational production via inflaton annihilations $\cphi \cphi \to \chi \chi$ is enhanced relative to the conformally coupled calculation by a factor of $m_\cphi^4 / m^4$.  Consequently the spectrum is insensitive to the mass, 
\bes{
    \frac{a^3 n_k}{\ae^3 \He^3} 
    \approx C \tk^{-3/2} 
    \quad \text{for $\dfrac{m_\cphi \kappa}{\He} < \tk < \taRH \dfrac{m_\cphi \kappa}{\He}$} 
    \com
}
where $C \approx (9/256 \pi) (m_\cphi / \He)^{3/2} \kappa^{5/2}$ for quadratic inflation.   This mass insensitivity is shown in \fref{fig:es-minimal}, as the three red curves appear to converge at large $\tk$.  

\para{Minimally coupled scalar: summary}
To summarize, the numerical results in \fref{fig:es-minimal} for CGPP of a massive scalar field that is minimally coupled to gravity are well described by the following analytical approximations, 
\ba{\label{eq:a3nk_minimal}
    \frac{a^3n_k}{\ae^3\He^3} 
    \approx \left\{ 
    \begin{array}{ll}
    \dfrac{1}{8 \pi^2} \tm^{-1} \tk^0 & \ 0 < \tk < \tm^{1/3} \\[1ex]
    \dfrac{1}{8 \pi^2} \tk^{-3} & \ \tm^{1/3} < \tk < \dfrac{m_\cphi \kappa}{\He} \\[1ex]
    C \tk^{-3/2} & \ \dfrac{m_\cphi \kappa}{\He} < \tk < \taRH \dfrac{m_\cphi \kappa}{\He} \com
    \end{array} \right. 
    \com
}
provided that $m \ll \He < m_\cphi$.  As $m$ rises to approach $3\He/2$, the low-$k$ power law begins tilting toward blue as $\tk^{3 - 2 \mathrm{Re}[\nu]}$, and the peak at $\tk_\star^\mathrm{(late)} = \tm^{1/3}$ moves toward $\tk = 1$, which pinches off the intermediate $\tk^{-3}$ behavior revealing an exponential suppression.  As $m$ grows to exceed $m_\cphi$, the $\tk^{-3/2}$ power-law tail gives way to a $\tk^{-15/2}$ power-law tail associated with $4\cphi \to \chi\chi$ scattering.  

\begin{figure}[t]
\begin{center}
\includegraphics[width=0.48\textwidth]{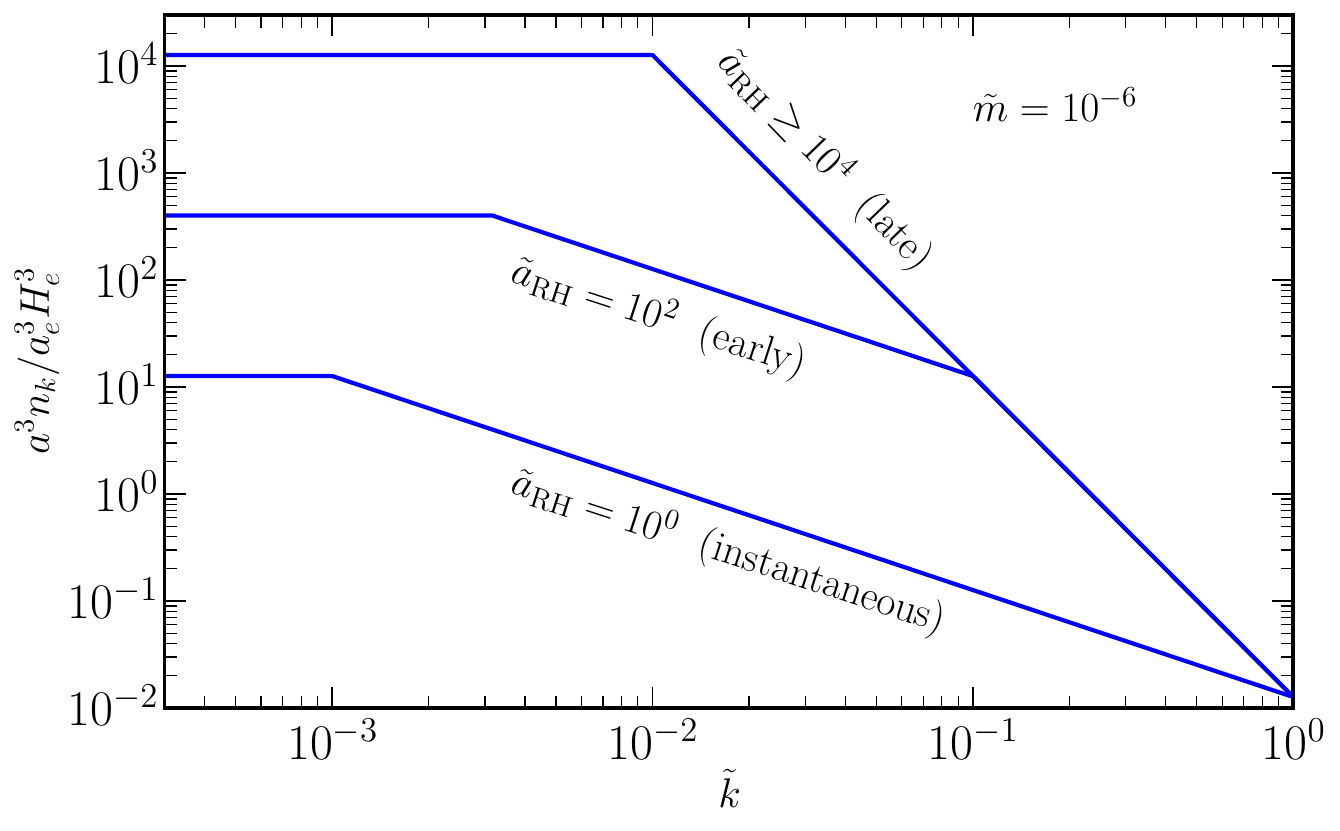} \\
\caption{
\label{fig:minimal_reheat}
Illustration of the reheating temperature's effect on the comoving number density spectrum $a^3 n_k$ for massive scalars that are light ($\tm < 1$) and minimally coupled to gravity ($\xi = 0$).  In the late reheating regime, which corresponds to $\taRH > \tm^{-2/3} = 10^4$, the spectrum is not sensitive to the reheating temperature.  In the early reheating regime ($\taRH < \tm^{-2/3} = 10^4$), the spectrum is suppressed by raising the reheating temperature (thereby reducing $\aRH$).
}
\end{center}
\end{figure}

\para{Dependence on reheating}
Finally we address how the comoving number density spectrum $a^3 n_k$ depends on the reheating temperature $\TRH$.  For the numerical results in Figs.~\ref{fig:es-conformal} and \ref{fig:es-minimal}, the spacetime background corresponds to a quasi-de Sitter phase of inflation followed by an effectively matter-dominated phase of reheating.  We have not implemented an inflaton decay rate, and the system remains matter dominated for an arbitrarily long time.  This is a reasonable approximation when $\TRH$ is low and reheating completes at later times, but the calculation should be modified when $\TRH$ is high and reheating completes at earlier times.  

To understand the dependence on reheating, compare the particle mass $m$ with the Hubble parameter at reheating $\HRH$; see \eref{eq:HRH}.  If the reheating temperature is low such that $\HRH < m$ then the field is released from Hubble drag ($H$ drops below $m$) during the epoch of reheating.  Modes entering the horizon at this time correspond to the special wave number $k_\star$ defined by \eref{eq:kstar_def}, and we have already seen that $\tk_\star^\mathrm{(late)} = \tm^{1/3}$.  However, if the reheating temperature is high such that $\HRH > m$ then the field is not released from Hubble drag until the radiation era.  \Eref{eq:kstar_def} then gives $\tk_\star^\mathrm{(early)} = \taRH^{1/4} \tm^{1/2}$, which has a different scaling with the mass $\tm = m / \He$ and which depends on the reheating temperature through the ratio $\taRH = \aRH/\ae$; see \eref{eq:aRH_ov_ae}.  In summary, the two regimes are 
\bes{\label{eq:late_early_reheating}
	\text{late RH:} & \quad \HRH < m \ , \quad \tk_\star^\mathrm{(late)} = \tm^{1/3} \\ 
	\text{early RH:} & \quad \HRH > m \ , \quad \tk_\star^\mathrm{(early)} = \taRH^{1/4} \tm^{1/2}
	\com
}
where $\HRH \lessgtr m$ is equivalent to $\taRH^{-3/2} \lessgtr \tm$

We have already provided spectra for the late reheating regime.  We now address how the spectra are modified for the early reheating regime.  For illustration we focus on a massive scalar that is light ($\tm < 1$) and minimally coupled to gravity ($\xi = 0$).  For such a model, the spectrum in the late reheating regime is given by \eref{eq:a3nk_minimal}; in particular, it is flat for $\tk < \tk_\star^\mathrm{(late)} = \tm^{1/3}$ and falls like $\tk^{-3}$ for $\tk > \tm^{1/3}$.  Here we focus only on the low-$k$ modes ($\tk < 1$) that leave the horizon during inflation.  For larger $\TRH$ the spectrum develops a second break point corresponding to the modes that reenter the horizon at the end of reheating; these modes have wave numbers around $k \approx k_\RH \equiv \aRH \HRH$ corresponding to $\tk_\RH = \aRH^{-1/2}$.  The spectrum is modified to 
\ba{\label{eq:n_k_RH}  
    \frac{a^3 n_k}{\ae^3\He^3} = \left\{ 
    \begin{array}{ll}
    \!\! \dfrac{1}{8\pi^2} \taRH^{3/4}\tm^{-1/2} & \ \tk < \taRH^{1/4} \tm^{1/2} \\[2ex]
    \!\! \dfrac{1}{8\pi^2} \taRH \tk^{-1} & \ \taRH^{1/4} \tm^{1/2} < \tk < \taRH^{-1/2} \\[2ex]
    \!\! \dfrac{1}{8\pi^2} \tk^{-3} & \ \taRH^{-1/2} < \tk < 1 \\[2ex]
    \end{array} \right. 
    \com
}
which is also illustrated in \fref{fig:minimal_reheat}.  The highest-$k$ modes enter the horizon during the reheating era; the intermediate-$k$ modes enter the horizon during the radiation era but before the field is released from Hubble drag ($H > m$); and the lowest-$k$ modes enter the horizon during the radiation era and after the field is released from Hubble drag ($H < m$).  For instantaneous reheating \eqref{eq:instant_RH}, $\taRH = 1$ and the $\tk^{-3}$ branch is absent.  

\subsection{Resultant particle abundance}
\label{sub:relic_abundance}

Depending on the application, the total amount of particles arising from CGPP is typically quantified using either the comoving number density $a^3 n$ or the cosmological energy fraction today $\Omega$.  See \sref{sec:Recent} for a discussion of applications including baryogenesis and dark matter.  

\para{Comoving number density} 
The comoving number density $a^3 n$ is calculated by integrating the spectrum $n_k$ (or $n_p$) over the comoving wave number (or momentum), 
\ba{
    a^3 n = \int_0^\infty \! \frac{\dd k}{k} \, a^3 n_k 
    \per
}
By performing this calculation for two models discussed in \sref{sub:number_spectrum}, the conformally coupled scalar and the minimally coupled scalar, one finds that 
\ba{\label{eq:a3n_results}
    a^3 n \approx \begin{cases} 
    \dfrac{\ae^3 \He^2 m}{16 \pi^2} 
    & , \ \xi = \sixth, \ \text{late RH} \\[8pt]
    \dfrac{\ae^3 \He^4}{8 \pi^2 m} \, \log \dfrac{k_\star^\mathrm{(late)}}{k_\mathrm{min}} 
    & , \ \xi=0, \ \text{late RH} \\[8pt]
    \dfrac{\ae^{9/4} \aRH^{3/4} \He^{7/2}}{8 \pi^2 m^{1/2}} \, \log \dfrac{k_\star^\mathrm{(early)}}{k_\mathrm{min}} 
    & , \ \xi=0, \ \text{early RH} 
    \end{cases}
    \per
}
For these estimates we have assumed that $m < \He$.  For the conformally coupled scalar, lowering the mass $m$ reduces the total number of particles, but for the minimally coupled scalar it raises the number.  For the minimally coupled model, the spectrum $n_k$ is independent of the comoving wave number for $k < k_\star$ where $k_\star^\mathrm{(late)} = \ae \He^{2/3} m^{1/3}$ and $k_\star^\mathrm{(early)} = \ae^{3/4} \aRH^{1/4} \He^{1/2} m^{1/2}$.  We regulate the integral by imposing an IR cutoff $k \geq k_\mathrm{min}$.  A physically motivated choice for the cutoff is $k_\mathrm{min} = a_0 H_0$ since all the modes that are outside the horizon today ($k < a_0 H_0$) appear to be homogeneous on our Hubble patch.  Since they all oscillate with the same frequency $\omega_k \approx a m$, they are indistinguishable from the $k = 0$ mode.  

\para{Comoving number conservation} 
After CGPP has completed, if particle-number-changing reactions are inactive, then the comoving number density is conserved, 
\ba{\label{eq:a3n_conservation}
    a^3 n = a^3(t) \, n(t) = a_0^3 n_0 
    \per
}
Specifically, $a^3 n$ gives the value today when the FLRW scale factor is $a_0$ and the cosmological number density is $n_0$.  For example, \eref{eq:a3n_conservation} would not hold if the particles were unstable (or at least not cosmologically long-lived).  Similarly, it would be violated if the particles had large nongravitational interactions that allowed them to thermalize with the primordial plasma or be produced via thermal freeze-in; see \sref{sub:thermal_freeze_in}.  

\para{Cosmological energy fraction} 
The cosmological energy fraction $\Omega$ is calculated by evaluating the ratio
\ba{
    \Omega = \frac{\rho_0}{\rho_c} 
    \qquad \text{or} \qquad 
    \Omega h^2 = \frac{\rho_0}{3 \Mpl^2 H_{100}^2} 
    \per
}
The denominator $\rho_c$ is the cosmological critical density today, which is given by $\rho_c \equiv 3 \MPl^2 H_0^2$ where $H_0 = H_{100} h$ is the Hubble constant, $H_{100} \equiv 100 \km / \mathrm{sec} / \mathrm{Mpc}$, and $h = 0.674 \pm 0.005$ \cite{Planck:2018vyg}.  The numerator $\rho_0$ is the spatially averaged energy density today carried by a particular species of radiation or matter.  For a spatially flat FLRW cosmology, the values of $\Omega$ for all of the cosmological components must sum to $1$, and therefore $0 < \Omega \leq 1$ gives the fractional energy carried by a particular species.  For dark matter in the Universe today, observations of the CMB yield a measurement of $\Omega_\mathrm{dm} h^2 = 0.120 \pm 0.001$ \cite{Planck:2018vyg}; see also \rref{KolbTurner:1990}.  

\para{Relating $\Omega h^2$ with $a_0^3 n_0$} 
For particles that are nonrelativistic today, the energy density $\rho_0$ and number density $n_0$ are related by a factor of mass $\rho_0 \approx m n_0$, neglecting kinetic energy.  Writing $n = (a_0^3 n_0 / \ae^3 \He^3) (\ae^3 \He^3 / a_0^3)$, the redshift factors can then be evaluated with the help of \erefs{eq:aRH_ov_ae}{eq:a0_ov_aRH} to obtain 
\bes{\label{eq:Omegah2}
    \frac{\Omega h^2}{0.12} & \approx 
    \biggl( \frac{m}{\He} \biggr) 
    \biggl( \frac{\He}{10^{12} \ \mathrm{GeV}} \biggr)^2 
    \\ & \qquad \times 
    \biggl( \frac{\TRH}{10^9 \ \mathrm{GeV}} \biggr) 
    \biggl( \frac{1}{10^{-5}} \frac{a_0^3 n_0}{\ae^3 \He^3} \biggr) 
    \per
}
Note that $n_0$ may bring an additional mass dependence.  We emphaize that \eref{eq:Omegah2} holds for any cosmological relic with mass $m$ and number density $n_0$, not just for gravitationally produced particles; however, it does assume that $\wRH = 0$ and $\aRH^3 s_\RH = a_0^3 s_0$.  

\para{Cosmological energy fraction from CGPP}
When we use the analytical results derived for the two models discussed in \sref{sub:number_spectrum}, the conformally coupled scalar and the minimally coupled scalar, the cosmological energy fraction is found to be 
\begin{widetext}
\bes{\label{eq:Omegah2_results}
   \frac{\Omega h^2}{0.1} 
    \approx 
    \begin{cases}
    7
    \Bigl( \dfrac{m}{10^{11} \ \mathrm{GeV}} \Bigr)^{2} 
    \Bigl( \dfrac{\TRH}{10^9 \ \mathrm{GeV}} \Bigr)^{} 
    & , \quad \xi = \sixth\ \\[8pt] 
    \Bigl( \dfrac{\He}{10^{14} \ \mathrm{GeV}} \Bigr)^{2}  
    \Bigl( \dfrac{\TRH}{10^2 \ \mathrm{GeV}} \Bigr)^{}
    \log \Bigl( \dfrac{k_\star^\mathrm{(late)}}{k_\mathrm{min}} \Bigr)^{}
    & , \quad \xi = 0,\ \text{late RH} \\[8pt]
    \Bigl( \dfrac{\He}{10^{14} \GeV} \Bigr)^{2}  
    \Bigl( \dfrac{m}{10^{-5} \eV} \Bigr)^{1/2} 
    \Bigl( \dfrac{g_{\ast\RH}}{106.75} \Bigr)^{-1/4}  
    \log \Bigl( \dfrac{k_\star^\mathrm{(early)}}{k_\mathrm{min}} \Bigr)^{}
    & , \quad \xi=0,\ \text{early RH} 
    \end{cases}
}
\end{widetext}
for $m < \He$.  The logarithmic factors depend on $\He$, $m$, and $\TRH$.  For $k_\mathrm{min} = a_0 H_0$ these factors are on the order of a $\mathrm{few} \times 10$.  For the conformally coupled scalar, note that the cosmological energy fraction decreases for a small mass like $\Omega h^2 \propto m^2$.  However, for the minimally coupled scalar, the energy fraction is insensitive to the mass $m$ in the late reheating regime, and it varies as $\Omega h^2 \propto m^{1/2}$ in the early reheating regime.  One can see from \eref{eq:Omegah2_results} that there is generally a two-parameter family of models for which the predicted $\Omega h^2$ matches the observed dark-matter energy fraction $\Omega_\mathrm{dm} h^2 \approx 0.12$.  

\subsection{Inhomogeneity}
\label{sub:inhomogeneity}

Here we address the spatial inhomogeneity of the gravitationally produced particles, and in \sref{sub:isocurvature} we discuss the observational probes of dark-matter spatial inhomogeneities (dark-matter and radiation isocurvature perturbations), which are probed in the CMB.  

\para{Density contrast power spectrum}
To quantify inhomogeneity in the gravitationally produced particles, it is useful to employ the dimensionless energy density contrast power spectrum $\Delta_\delta^2(k)$, which can be calculated using \eref{eq:Deltadelta}.  If these particles are intended to provide a dark-matter candidate, then the behavior of $\Delta_\delta^2(k)$ toward small values of $k$ is especially important, since cosmological observables probe the dark-matter inhomogeneity on length scales corresponding to $k \approx 10^{-4}$ to $1 \, a_0 \Mpc^{-1}$, which is generally much smaller than $\ke = \ae \He$. 

\para{Conformally coupled scalar field}
Consider a scalar spectator field that is light during inflation ($m < \He$) and conformally coupled to gravity ($\xi = \sixth$).  As remarked following \eref{eq:Deltadelta}, if the spectrum $n_k$ is sufficiently strongly peaked at some nonzero $k \approx k_\star$ then the dimensionless power spectrum has a universal scaling $\Delta_\delta^2(\eta,k \ll k_\star) \propto k^3$.  This is the case for a conformally coupled scalar, which has $n_k \propto k^2$ for small $k$ and a peak at $k \approx k_\star^\mathrm{(late)} = \ae \He (m/\He)^{1/3}$.  The resulting power spectrum is evaluated to be 
\ba{
    \Delta_\delta^2(k \ll k_\star) \approx 
    24 \frac{k^3}{\ae^3 \He^2 m} 
    \com
}
which is also $24 (k/k_\star^\mathrm{(late)})^3$.  For the small-scale modes with $k$ approaching $k_\star^\mathrm{(late)}$ from below, the power spectrum approaches $O(1)$, whereas, for the large-scale modes with smaller $k$, the power decreases as $k^3$.  

The scaling $\Delta_\delta^2(k) \propto k^3$ is especially significant if the conformally coupled scalar is intended to provide a candidate for dark matter.  Note that $\Delta_\delta^2 \approx (8 \times 10^{-71}) (k / a_0 H_0)^{3} (m / 10^{12} \GeV)^{-1} (\TRH / 10^9 \GeV)^{-1}$ is minuscule for cosmological-scale modes today (this estimate assumes that $\wRH = 0$).  As $\Delta_\delta^2 \to 0$ the dark-matter inhomogeneities correspond to adiabatic perturbations, and CMB constraints on dark-matter isocurvature are easily evaded.  For additional details see \sref{sub:isocurvature}.  

\para{Minimally coupled scalar field}
Next consider a scalar spectator field that is light during inflation ($m < \He$) and minimally coupled to gravity ($\xi = 0$).  The spectrum is approximately scale invariant on large scales, corresponding to $n_k \propto k^0$ or $\beta_k \propto k^{-3}$ for $k < k_\star$.  As a result the power spectrum integral in \eref{eq:Deltadelta} is dominated by the small $|\kvec^\prime|$ region of the integration domain.  To estimate the power spectrum, one can neglect the second term in \eref{eq:Deltadelta} and divide the integration domain into two regions $k_\mathrm{min} < |\kvec^\prime| < k$ and $k < |\kvec^\prime| < k_\star$ while approximating $\tilde{\alpha}_{|\kvec^\prime-\kvec|}$ as $\tilde{\alpha}_{k}$ in the first region and $\tilde{\alpha}_{k^\prime}$ in the second.  The resulting power spectrum is evaluated to be 
\ba{
    \Delta_\delta^2(k \ll k_\star) \sim 
    \frac{\log (k / k_\mathrm{min})}{\log^2 (k_\star / k_\mathrm{min})}
    \com
}
up to $O(1)$ factors and additional terms that are subleading for small $k$.  For the minimally coupled scalar spectator the dimensionless power spectrum is nearly scale invariant, varying only logarithmically with wave number.  

\para{Dark matter isocurvature}
If the gravitationally produced particles are to provide a viable candidate for dark matter, their spatial inhomogeneities are constrained by observations of the CMB anisotropies.  In particular, it is necessary to avoid an unacceptably large value of dark-matter isocurvature on CMB scales.  The connections with isocurvature are further discussed in \sref{sec:Recent}.  In short, the conformally coupled scalar evades isocurvature constraints since $\Delta_\delta^2 \propto k^3$ is minuscule on cosmological scales, whereas the minimally coupled scalar that is light during inflation is strongly constrained by isocurvature since $\Delta_\delta^2(k) \propto \log k$ is nearly scale invariant.  More generally whenever the spectrum is sufficiently blue tilted, \ie, $n_k \propto k^n$, with $n > 0$, the power spectrum is strongly blue tilted $\Delta_\delta^2 \propto k^3$ and evades isocurvature constraints.  

\section{Extension to fields with spin}
\label{sec:Spin}

The preceding discussion of CGPP in scalar field theories may be extended to fields with spin.  One appealing aspect of invoking CGPP to explain the origin of cosmological relics, such as dark matter, is that the exercise of model building is limited to specifying only the particle's mass and spin if a minimal coupling to gravity is assumed.  Therefore it is worthwhile to survey this limited parameter space by considering a range of spins.  

In this section we first provide a general argument that CGPP requires conformal symmetry breaking.  We then discuss several representations of the Lorentz group with nonzero spin.  For each representation we illustrate how the cosmological expansion leads to time-dependent equations of motion.  We highlight how the phenomenon of CGPP differs for each of these theories from the previously discussed scalar theories.  

\subsection{CGPP requires conformal symmetry breaking}
\label{sub:conformal_symmetry}

\para{General argument}  
The phenomenon of CGPP derives from the mixing of positive- and negative-frequency modes due to a time-dependent dispersion relation in an expanding universe.  No particle production would occur in Minkowski spacetime.  Since the homogeneous, isotropic, and expanding FLRW spacetime is related to the homogeneous, isotropic, and static Minkowski spacetime by a time-dependent conformal transformation, any fields that exhibit a conformal symmetry in their coupling to gravity do not experience CGPP.  For such fields the trace of the stress-energy tensor is the order parameter for conformal symmetry breaking, and it can be used as a diagnostic tool to assess whether CGPP will occur in a given theory.  

\para{Trace of the stress-energy tensor}  
One can show that $T_\mu^\mu$ is the order parameter for conformal symmetry breaking as follows.  Consider a conformal transformation 
\ba{
    \metric_{\mu\nu}(x) \mapsto \ee^{2 \Omega(x)} \, \metric_{\mu\nu}(x)
    \com
}
where $\Omega(x)$ is a real scalar field.  These are also known as Weyl or scale transformations since they locally stretch or shrink spacetime.  For example, a conformal transformation with $e^{2 \Omega} = a^{-2}(\eta)$ maps the FLRW metric $\metric_{\mu\nu}^\FLRW = a^2(\eta) \metricMink_{\mu\nu}$ to the Minkowski metric $\metric_{\mu\nu}^\mathrm{Mink} = \metricMink_{\mu\nu}$.  Under an arbitrary infinitesimal variation of the metric $\delta \metric_{\mu\nu}(\eta,\xvec)$, the variation in the radiation-matter action is 
\ba{
    \delta S_\M 
    = - \frac{1}{2} \int \! \dd^4 x \, \sqrtg \, T^{\mu\nu} \, \delta \metric_{\mu\nu} 
    \com
}
which defines the stress-energy tensor $T^{\mu\nu}$.  When one specializes to a conformal transformation, $\delta \metric_{\mu\nu} = 2 \, \delta\Omega \, \metric_{\mu\nu}$ gives 
\ba{
    \delta S_\M 
    = - \int \! \dd^4 x \, \sqrtg \, \tensor{T}{^\mu_\mu} \delta\Omega 
    \per
}
If the trace of the stress-energy tensor vanishes, then the matter action is invariant under conformal transformations, and the previous argument implies that no CGPP will take place.  When quantum effects are taken into account, even conformally flat spacetimes may admit CGPP via the trace anomaly $\langle \tensor{T}{^\mu_\nu} \rangle \neq 0$ \cite{Dolgov:1981nw}.  In general, conformal invariance must be broken for particle creation to occur. 

\para{Example: massless conformal scalar}  
For the previously discussed scalar field theory, the trace of the stress-energy tensor \eqref{eq:stressenergyscalar} is 
\ba{
    \tensor{T}{^\mu_\mu} = \bigl( 6 \xi - 1 \bigr) \bigl( \metric^{\mu\nu} \partial_\mu \aphi \partial_\nu \aphi + \aphi \Box \aphi \bigr) + m^2 \aphi^2 
    \per
}
Note that the trace vanishes for the model with $m^2 = 0$ and $\xi = \ssfrac{1}{6}$, and by the preceding argument there should be no gravitational particle production for a massless scalar that is conformally coupled to gravity.  This conclusion can also be drawn from the scalar's mode equation \eqref{eq:mode_equation}, which becomes $\chi_k^\pprime + k^2 \chi_k = 0$ through the static squared angular frequency $\omega_k^2 = k^2$.  In the absence of a time-dependent frequency, positive- and negative-frequency modes do not mix, and no CGPP occurs.  In the broader parameter space, observables must connect continuously to the absence of CGPP at $m^2 = 0$ and $\xi = \ssfrac{1}{6}$.  This behavior is reflected in the analytical and numerical results that appear in \sref{sub:number_spectrum}, which show a suppression of the cosmological energy fraction $\Omega h^2$ as $m^2$ and $\xi$ approach the enhanced conformal symmetry point; see also \sref{sub:nonminimal}. 

\para{Example: massless vector (photon)}  
The covariant action for a massless real vector field $A_\mu(x)$ with a minimal coupling to gravity is written as 
\ba{
    S = \int \! \dd^4x \, \sqrtg \Bigl[ - \tfrac{1}{4} \metric^{\mu\rho} \metric^{\nu\sigma} F_{\mu\nu} F_{\rho\sigma} \Bigr] 
    \com
}
where $F_{\mu\nu} = \del{\mu} A_\nu - \del{\nu} A_\mu$.  For example, $A_\mu(x)$ could be the electromagnetic field with the photon as its quantum excitation.  The corresponding stress-energy tensor is 
\ba{
    T_{\mu\nu} = \bigl( \tfrac{1}{4} \metric_{\mu\nu} \metric^{\rho\sigma} - \delta_\mu^{\rho} \delta_\nu^{\sigma} \bigr) \metric^{\alpha\beta} F_{\rho\alpha} F_{\sigma\beta} 
    \per
}
Note that the parenthetical factor vanishes when it contracts with $\metric^{\mu\nu}$.  It follows that the stress-energy tensor is traceless $T_\mu^\mu = 0$, and no CGPP occurs.  In other words, a massless vector boson that is minimally coupled to gravity is also conformally coupled to gravity.  If the conformal symmetry were broken, for example by introducing a mass $m^2 A_\mu A^\mu$ or a nonminimal coupling to gravity like $R A_\mu A^\mu$ or $R^{\mu\nu} A_\mu A_\nu$, then CGPP may occur (see later discussion).  

\subsection{Massive spin-$\onehalf$ field}
\label{sub:spin_one_half}

The analysis of CGPP for spin-$\half$ particles runs parallel to the procedure discussed previously for spin-$0$ particles.  Introductions to spinor fields in curved spacetime were given by \rref{Parker:2009uva} and \rref{Freedman:2012zz}, and their applications to CGPP were discussed by \rref{Kuzmin:1998kk,Chung:2011ck,Boyle:2018rgh}.  In this section, we present the key formulas and summarize the main results. 

\para{Action in FLRW spacetime} 
Massive spin-$\onehalf$ particles interacting with gravity are described by a spinor field $\Psi(x))$ with the action 
\ba{\label{eq:Diracactioncurved}
    S = \int \! \dd^4x \, \dete \, \Bigl[ 
    \tfrac{\ii}{4} \bar{\Psi} \, \gamma^\mu  \del{\mu} \Psi 
    - \tfrac{\ii}{4} \bar{\Psi} \overleftarrow{\del{\mu}} \gamma^\mu \Psi 
    - \tfrac{1}{2} m \bar{\Psi} \Psi 
    \Bigr] 
    \com
}
which generalizes the Dirac action \cite{Dirac:1928hu,Weyl:1929fm} to curved spacetime.  See \aref{app:general_relativity} for a refresher on spinors in curved spacetime and the frame-field formalism.  We require $\Psi = \Psi^C \equiv - \ii \gamma^2 \Psi^\ast$ such that $\Psi(x)$ is a self-conjugate Majorana spinor.  When one's attention is restricted to an FLRW spacetime, the field equation is  
\ba{
    \bigl( \ii \gamma^a \delta_a^\mu \partial_\mu - a(\eta) m \bigr) \bigl[ a^{3/2}(\eta) \Psi(\eta,\xvec) \bigr] = 0 
    \per
}
This equation of motion for $\psi = a^{3/2} \Psi$ takes the same form as the Dirac equation in Minkowski spacetime, with the exception that the mass $m$ is replaced by a time-dependent effective mass $a(\eta) m$.  For a massless spin-$\half$ particle ($m=0$) there is no explicit time dependence in the equation of motion, no mixing of positive- and negative-frequency modes, and no particle production.  One also arrives at this conclusion by inspecting the trace of this theory's stress-energy tensor, $\tensor{T}{^\mu_\mu} = a^3 m \bar{\Psi} \Psi$, which vanishes at $m = 0$, so the spin-$\half$ particle is conformally coupled to gravity; see \sref{sub:conformal_symmetry}.  More generally, one expects to find a suppression of CGPP as $m \to 0$.  

\para{Mode equations}  
Solutions of the field's equation of motion take the form 
\ba{
    & \Psi(\eta,\xvec) 
    = [a(\eta)]^{-3/2} \int \! \! \frac{\dd^3\kvec}{(2\pi)^3} \sum_{\lambda=\pm \half} 
    \\ & \qquad \qquad \times 
    \Bigl[ 
    a_{\kvec,\lambda} \, U_{\kvec,\lambda}(\eta,\xvec)  
    + a_{\kvec,\lambda}^\dagger \, V_{\kvec,\lambda}(\eta,\xvec) \Bigr] 
    \nn & 
    U_{\kvec,\pm\onehalf}(\eta,\xvec) = \begin{pmatrix} u_{A,k}(\eta) \\ \pm u_{B,k}(\eta) \end{pmatrix} \otimes h_{\kvec,\pm} \ \ee^{\ii \kvec \cdot \xvec} 
    \com 
    \nonumber 
}
where $V_{\kvec,\lambda} = - \ii \gamma^2 U_{\kvec,\lambda}^\ast$.  The ladder operators $a_{\kvec,\lambda}$ and mode functions $U_{\kvec,\lambda}(\eta,\xvec)$ are labeled with a comoving wavevector $\kvec$ and helicity $\lambda$.  The two-component spinors $h_{\kvec,\pm}$ are eigenspinors of the helicity operator $\tfrac{1}{2} \khat \cdot {\bm \sigma} h_{\kvec,\pm} = \pm \tfrac{1}{2} h_{\kvec,\pm}$ and are normalized such that $h_{\kvec,2\lambda}^\dagger h_{\kvec,2\lambda^\prime} = \delta_{\lambda \lambda^\prime}$.  We write $\kvec = (k \, \sin \theta \, \cos \phi, \ k \, \sin \theta \, \sin \phi, \ k \, \cos \theta)$ and 
\ba{\label{eq:h_spinor_def}
    h_{\kvec,+} = \begin{pmatrix} \ee^{-\ii \phi} \cos \tfrac{\theta}{2} \\ \sin \tfrac{\theta}{2} \end{pmatrix} 
    , \ \ 
    h_{\kvec,-} = \begin{pmatrix} \ee^{-\ii \phi} \sin \tfrac{\theta}{2} \\ - \cos \tfrac{\theta}{2} \end{pmatrix} 
    \com
}
up to irrelevant phases.  The time-dependent components of the spinor wave function $u_{A,k}(\eta)$ and $u_{B,k}(\eta)$ are required to obey 
\begin{equation}\label{eq:DiraculaFLRW}
    \ii \partial_\eta \begin{pmatrix} u_{A,k} \\ u_{B,k} \end{pmatrix} 
    = \begin{pmatrix} am & k \\ k & - am \end{pmatrix} \begin{pmatrix} u_{A,k} \\ u_{B,k} \end{pmatrix} 
    \com
\end{equation}
in the Dirac representation of the gamma matrices.  This linear mode equation takes the same form as the time-dependent Schr\"odinger equation for a two-level system with the matrix on the right-hand side playing the role of the Hamiltonian; its time-dependent eigenvalues are $\omega_k(\eta) = \pm [k^2+a^2(\eta)m^2]^{1/2}$.

\para{Ansatz}  
In analogy with the previously discussed scalar model, one is looking for the mode functions $u_{A/B,k}^\IN(\eta)$ and $u_{A/B,k}^\OUT(\eta)$ that solve the mode equation, respect the orthonormality conditions, and have the desired asymptotic behavior toward early and late times.  It is convenient to take the ansatz 
\ba{
    \begin{pmatrix} u_{A,k}^\IN \\ u_{B,k}^\IN \end{pmatrix} 
    = & \begin{pmatrix} 
    \sqrt{\dfrac{\omega_k + am}{2\omega_k}} \, \ee^{-\ii \dphi_k} 
    & - \sqrt{\dfrac{\omega_k - am}{2\omega_k}} \, \ee^{\ii \dphi_k} 
    \\ \sqrt{\dfrac{\omega_k - am}{2\omega_k}} \, \ee^{-\ii \dphi_k} 
    & \sqrt{\dfrac{\omega_k + am}{2\omega_k}} \, \ee^{\ii \dphi_k} \end{pmatrix} 
   \begin{pmatrix} \tilde{\alpha}_k \\ \tilde{\beta}_k \end{pmatrix} 
   \com
}
where $\dphi_k(\eta) = \int^\eta \! \dd \eta^\prime \, \omega_k(\eta^\prime)$.  The analogous relation for the scalar model appears in \eref{eq:scalar_chikIN_ansatz}.  This introduces the new mode functions $\tilde{\alpha}_k(\eta)$ and $\tilde{\beta}_k(\eta)$, which are normalized to obey $|\tilde{\alpha}_k|^2 + |\tilde{\beta}_k|^2 = 1$ at all time.  The $\IN$ mode functions are asymptotic to the leading-order Wentzel-Kramers-Brillouin approximation at early times, which corresponds to $\tilde{\alpha}_k(\eta) \to 1$ and $\tilde{\beta}_k(\eta) \to 0$ as $\eta \to -\infty$.  In terms of the new mode functions, the mode equations are 
\bes{
    \partial_\eta \tilde{\alpha}_k(\eta) 
    & = - \tfrac{1}{2} A_k(\eta) \, \omega_k(\eta) \tilde{\beta}_k(\eta) \, \ee^{2 \ii \dphi_k} \\ 
    \partial_\eta \tilde{\beta}_k(\eta) 
    & = + \tfrac{1}{2} A_k(\eta) \, \omega_k(\eta) \, \tilde{\alpha}_k(\eta) \, \ee^{-2 \ii \dphi_k} 
    \com
}
which has the same form as \eref{eq:EOMforalphabeta} for scalars.  The dimensionless variable $A_k(\eta)$, which is defined by 
\ba{\label{eq:Ak_fermion}
    A_k(\eta) = \frac{a^2 H m k}{(k^2 + a^2 m^2)^{3/2}} 
    \com
}
serves as the adiabaticity measure.  Note that this is different from $\omega_k^\prime / \omega_k^2 = a^3 H m^2 / (k^2 + a^2 m^2)^{3/2}$.  

\para{CGPP intuition} 
By inspecting the adiabaticity measure \eqref{eq:Ak_fermion}, one can develop intuition for the CGPP of spin-\halfBM\ particles.  
Since the equations of motion and adiabaticity parameter are the same for both $\lambda = \pm \onehalf$, the two polarization modes will be produced equally.  Since $A_k(\eta)$ vanishes as $m \to 0$, we can anticipate a suppression of CGPP for low-mass spin-\halfBM\ particles.  Typically CGPP for spin-$\onehalf$ particles is most efficient for masses around the inflationary Hubble scale, perhaps $m \sim \Hinf \sim 10^{14} \GeV$, which corresponds to the superheavy regime.  Similarly, since $A_k(\eta)$ vanishes as $k \to 0$, one expects CGPP to be suppressed for long-wavelength modes, implying negligible power in the energy contrast on cosmological scales and if the particles survive as dark matter, predominantly adiabatic perturbations.  Both of these behaviors mirror the results for conformally coupled scalars.  Finally, since the mode functions are normalized to obey $|\tilde{\alpha}_k|^2 + |\tilde{\beta}_k|^2 = 1$, it implies that the Bogolubov coefficient $|\beta_k| = \lim_{\eta \to +\infty} |\tilde{\beta}_k(\eta)|$ cannot exceed unity.  This is a manifestation of the Pauli exclusion principle: for fermionic CGPP Pauli blocking implies that there can be at most one particle per mode.

\para{Numerical example}
\Fref{fig:abundances} shows the results of a numerical evaluation of CGPP for a massive spinor field in quadratic inflation; see the curves labeled $s_\lambda = 1/2_{\pm 1/2}$.  As expected, the relic abundance of spin-$\half$ fermions closely tracks the abundance of conformally coupled scalars with the same mass.  Since the fermion mass is decreased below $m/\He \approx 1$ the comoving number density decreases as $a^3 n \propto m^1$ and the cosmological energy fraction decreases as $\Omega h^2 \propto m^2$.  CGPP leads to the largest relic abundance for $m/\He \approx 1$, and reasonable choices of $\He$ and $\TRH$ lead to $\Omega h^2 \approx 0.12$.  Therefore if the fermions are stable, they provide a candidate for superheavy dark matter.  Here we have considered CGPP for a Majorana spinor field, and the relic abundance would be doubled for a Dirac spinor field, which has both spin-$\half$ particles and antiparticles as its quantum excitation.  The number density spectrum $a^3 n_k$ (not shown) also tracks the conformally coupled scalar; it is peaked for modes with $k \approx \ke = \ae \He$ and suppressed for modes with $k \ll \ke$.  Consequently, $\Delta_\delta^2(k) \propto k^3$ and isocurvature constraints are easily avoided.  

\subsection{Massive spin-$1$ field}
\label{sub:spin_one}

Recently there has been a renewed interest in the cosmological production of stable massive spin-$1$ particles \cite{Dimopoulos:2006ms,Graham:2015rva,Ema:2019yrd,Ahmed:2020fhc,Kolb:2020fwh,Gross:2020zam,Arvanitaki:2021qlj,Ozsoy:2023gnl,Cembranos:2023qph,Garcia:2023obw}.  This attention is partly motivated by studies of ultralight dark photon dark matter \cite{Caputo:2021eaa}, which is a candidate for wavelike dark matter with a mass $m \ll 10 \eV$.  When applied to massive spin-$1$ particles, the phenomenon of CGPP can explain the origin of dark photon dark matter as well as its heavier cousins with $m \sim \Hinf$.  

\para{Action in FLRW spacetime}  
Massive spin-$1$ particles interacting with gravity are described by a vector field $A_\mu(x)$ with the action 
\bes{
    S
    & = \int \! \dd^4 x \, \sqrtg \, \Bigl[ 
    - \tfrac{1}{4} \metric^{\mu\rho} \metric^{\nu\sigma} F_{\mu\nu} F_{\rho\sigma} 
    + \tfrac{1}{2} m^2 \metric^{\mu\nu} A_\mu A_\nu \Bigr] 
    \com
}
which generalizes the de Broglie-Proca action to curved spacetime \cite{deBroglie:1922,deBroglie:1934,Proca:1936}.  The field strength tensor is $F_{\mu\nu} = \del{\mu} A_\nu - \del{\nu} A_\mu$.  We require $A_\mu = A_\mu^C \equiv A_\mu^\ast$ such that $A_\mu(x)$ is a self-conjugate real vector.  Here a minimal coupling to gravity is assumed; additional terms such as $R^{\mu\nu} A_\mu A_\nu$ and $R \metric^{\mu\nu} A_\mu A_\nu$ may be introduced to allow for a nonminimal coupling \cite{Golovnev:2008cf,Himmetoglu:2008zp}, which impacts CGPP \cite{Kolb:2020fwh,Ozsoy:2023gnl,Cembranos:2023qph}.  We assume $m \neq 0$ such that $A_\mu$ propagates three polarization modes (two transverse and one longitudinal) and CGPP may occur.  The model with $m = 0$ has an enhanced gauge symmetry, that reduces the number of propagating degrees of freedom to simply two transverse polarization modes, and it has an enhanced conformal symmetry that prohibits CGPP; see \sref{sub:conformal_symmetry}.  

When $\metric_{\mu\nu}(x) = \metric_{\mu\nu}^\FLRW(\eta,\xvec)$ is fixed to be an FLRW metric with scale factor $a(\eta)$, the action becomes 
\bes{
    S
    & = \int \! \dd\eta \, \dd^3 \xvec 
    \Bigl[ 
    \tfrac{1}{2} A_i^\prime A_i^\prime 
    - \tfrac{1}{2} a^2 m^2 A_i A_i
    \\ & \qquad 
    - \tfrac{1}{2} \epsilon^{ijk} \epsilon^{lmk} \partial_i A_j \partial_l A_m 
    + \tfrac{1}{2} m^2 a^2 A_0^2 
    \\ & \qquad 
    + \tfrac{1}{2} \partial_i A_0 \partial_i A_0
    - A_i^\prime \partial_i A_0 
    \Bigr] 
}
where repeated Latin indices are summed from $1$ to $3$.  Among the four components of $A_\mu$, only three of them are propagating, while the fourth component is restricted by an algebraic constraint that links it to the others.  

\para{Mode equations}  
Solutions of the field's equation of motion take the form 
\ba{
    & A_\mu(\eta,\xvec) = \int \! \! \frac{\dd^3\kvec}{(2\pi)^3} \sum_{\lambda=\pm1,0} 
    \\ & \qquad \qquad \times 
    \Bigl[ 
    a_{\kvec,\lambda} \, U_{\kvec,\lambda,\mu}(\eta,\xvec)  
    + a_{\kvec,\lambda}^\dagger \, V_{\kvec,\lambda,\mu}(\eta,\xvec) \Bigr] 
    \nn 
    & U_{\kvec,\lambda,0}(\eta,\xvec) = -\ii \frac{\kvec_i U_{\kvec,\lambda,i}^\prime}{k^2 + a^2 m^2} 
    \nn 
    & U_{\kvec,\lambda,i}(\eta,\xvec) = \chi_{k,\lambda}(\eta) \, \varepsilon_{\kvec,\lambda,i} \ \ee^{\ii \kvec \cdot \xvec} 
    \nonumber 
}
where $V_{\kvec,\lambda,\mu} = U_{\kvec,\lambda,\mu}^\ast$.  The ladder operators $a_{\kvec,\lambda}$ and mode functions $U_{\kvec,\lambda,i}(\eta,\xvec)$ are labeled with a comoving wavevector $\kvec$ and helicity $\lambda$.  The three-vectors $\varepsilon_{\kvec,\lambda,i}$ are eigenvectors of the helicity operator $\ii \khat \times {\bm \varepsilon}_{\kvec,\lambda} = \lambda {\bm \varepsilon}_{\kvec,\lambda}$ and are normalized such that $\varepsilon_{\kvec,\lambda,i}^\ast \varepsilon_{\kvec,\lambda^\prime,i} = \delta_{\lambda \lambda^\prime}$.  If $\kvec = k (\sin\theta \cos\phi, \sin\theta \sin\phi, \cos\theta)$ then 
\bes{\label{eq:eps_vector_def}
    \varepsilon_{\kvec,\pm,i} & = \frac{\ii}{\sqrt{2}} \ee^{\mp\ii \gamma} \begin{pmatrix} \mp\cos\theta \cos\phi + \ii \sin\phi \\ \mp\cos\theta \sin\phi - \ii \cos\phi \\ \pm\sin\theta \end{pmatrix} \\ 
    \varepsilon_{\kvec,0,i} & = \begin{pmatrix} \sin\theta \cos\phi \\ \sin\theta \sin\phi \\ \cos\theta \end{pmatrix} 
    \com
}
up to an irrelevant phase $\gamma$.  Note that $\lambda = 0$ can be identified as the longitudinal polarization mode and $\lambda = \pm 1$ as the two transverse polarization modes.  The mode functions are required to solve the equations of motion 
\bes{\label{eq:vector_mode_equations}
    & \bigl( \partial_\eta^2 + \omega_T^2 \bigr) \, \chi_{k,\pm} = 0 \\ 
    & \bigl( \partial_\eta^2 + \omega_L^2 \bigr) \biggl( \sqrt{\frac{a^2 m^2}{k^2 + a^2 m^2}} \, \chi_{k,0} \biggr) = 0 
    \per
}
where the time-dependent squared frequencies are 
\ba{
    \omega_T^2 & = k^2 + a^2 m^2 \\ 
    \omega_L^2 & = k^2 + a^2 m^2 + \frac{1}{6} \frac{k^2 \, R}{k^2 + a^2 m^2} + 3 \frac{k^2 \, a^4 H^2 m^2}{(k^2 + a^2 m^2)^2} \nonumber 
    \per
}

\para{CGPP intuition} 
Since the vector field's equations of motion \eqref{eq:vector_mode_equations} take the same form as a scalar field's equations of motion \eqref{eq:mode_equation}, the Bogolubov coefficients are calculated in the same way.  Note that the transverse modes obey the same equations of motion as the modes of a conformally coupled scalar field with $\xi=\ssfrac{1}{6}$; see \eref{eq:omegak_scalar}.  Thus, one should anticipate the same behavior in regard to CGPP, for example, a suppression of particle production as $m \to 0$.  However, the longitudinal polarization mode has a time-dependent equation of motion, even for $m = 0$, and one expects some amount of particle production.  Remember that sending $m \to 0$ in the de Broglie-Proca action yields the theory of a massless vector and a massless scalar with $\omega_k^2 = k^2 + a^2 R/6$. 

\begin{figure}[t]
\begin{center}
\includegraphics[width=0.48\textwidth]{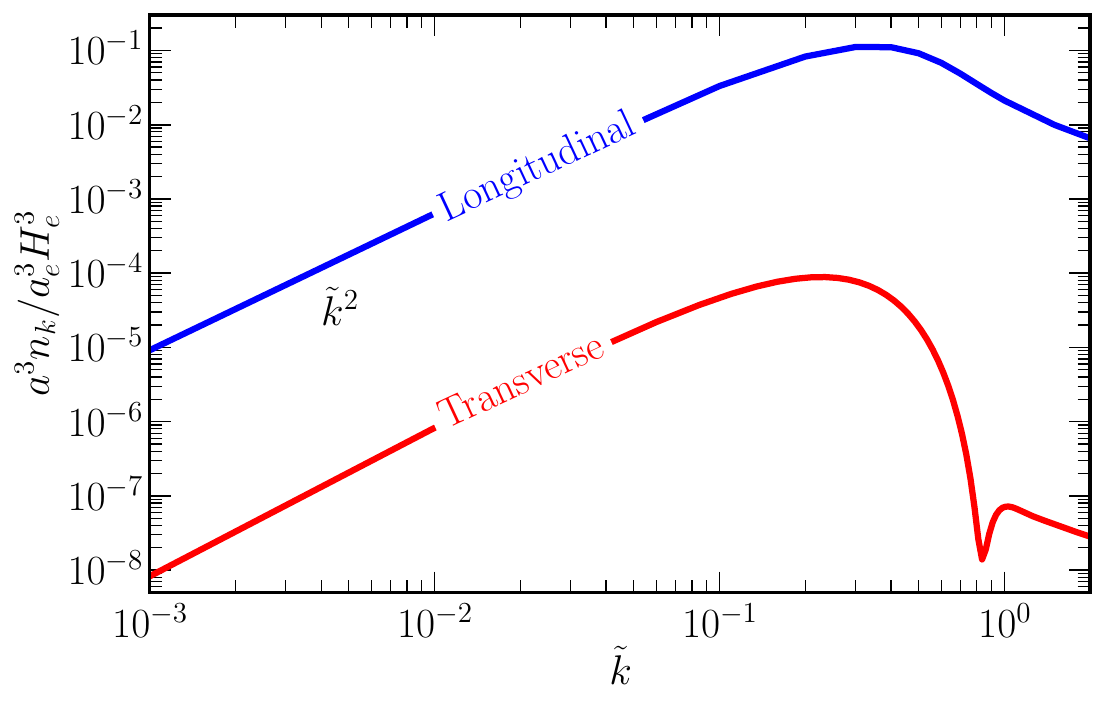} 
\caption{
\label{fig:spectrum}
The comoving number density spectrum $a^3 n_k$ as a function of $\tk = k / \ae \He$ that results from CGPP for a massive vector field with $m = 0.1 \He$ in quadratic inflation.  The curves show the spectra for the single longitudinal polarization mode and the sum of the two transverse polarization modes.  
}
\end{center}
\end{figure}

\para{Numerical example}
Figures~\ref{fig:spectrum}~and~\ref{fig:abundances} show the numerical results when CGPP is evaluated for a massive vector field in quadratic inflation.  The comoving number density spectrum is shown in \fref{fig:spectrum} for a vector field with mass $m = 0.1 \He$.  Both the transverse and longitudinal polarization modes display a blue-tilted spectrum that rises like $\tk^2$ for small values of $\tk = k / \ae \He$.  The spectrum reaches a peak at scales $\tk \approx \tk_\star^\mathrm{(late)} \equiv \tm^{1/3}$, corresponding to modes that are entering the horizon at the same time when the field is released from Hubble drag; see \eref{eq:kstar_def}.  The suppression of $n_k$ toward small $k$ translates into a similar suppression in the dimensionless power spectrum $\Delta_\delta^2(k)$ and allows the model to evade constraints on dark-matter isocurvature \cite{Graham:2015rva}; see \sref{sub:inhomogeneity}. 

The relic abundance of gravitationally produced vectors is shown in \fref{fig:abundances}.  The longitudinal polarization modes are populated more efficiently than the transverse polarization modes, which corresponds to a larger comoving number density $a^3 n$ and cosmological energy fraction $\Omega h^2$.  For models with $m \ll \He$, the production of transverse polarization modes progresses as $a^3 n \propto m$, similar to a conformally coupled scalar ($s_\xi = 0_{1/6}$.  However, the production of longitudinal polarization modes progresses as $a^3 n \propto m^{-1}$, similar to a minimally coupled scalar ($s_\xi = 0_0$); see \eref{eq:a3n_results}.  These calculations assume late reheating \eqref{eq:late_early_reheating}, and for early reheating the scaling changes to $a^3 n \propto m^0$ for the longitudinal polarization modes \cite{Graham:2015rva,Ahmed:2020fhc,Kolb:2020fwh}.  Both cases are captured by the relations in \eref{eq:Omegah2_results}, and setting the logarithm factors equal to $1$ gives a good estimate of the particle abundance. 

\subsection{Massive spin-$\threehalf$ field}
\label{sub:spin_three_half}

A long-standing interest in massive spin-$\threehalf$ particles and their role in cosmology has largely derived from the spin-$\threehalf$ gravitino, which is the superpartner to the graviton in theories of supergravity; for an introduction see \rref{Freedman:2012zz}.  The gravitino's weak coupling to ordinary matter allows these particles to be cosmologically long-lived, and thus their overproduction in the early Universe through thermal or nonthermal channels presents a serious constraint on theories of supergravity in the context of cosmology \cite{Krauss:1983ik,Ellis:1984eq,Kawasaki:1994af}.  The implications of CGPP for gravitinos and spin-$\threehalf$ particles more generally were explored in several landmark papers \cite{Lemoine:1999sc,Giudice:1999yt,Giudice:1999am,Kallosh:1999jj,Bastero-Gil:2000lgf} and developed further in later studies \cite{Hasegawa:2017hgd,Garcia:2020hyo,Kolb:2021xfn,Dudas:2021njv,Antoniadis:2021jtg,Kaneta:2023uwi,Casagrande:2023fjk}. 

\para{Action in FLRW spacetime}  
Massive spin-$\threehalf$ particles interacting with gravity are described by a vector-spinor field $\Psi_\mu(x)$ with the action 
\bes{
    & S
    = \int \! \dd^4 x \, \dete \, \Bigl[ 
    \tfrac{\ii}{4} \bar{\Psi}_\mu \gamma^{\mu\nu\rho} \del{\nu} \Psi_\rho 
    - \tfrac{\ii}{4} \bar{\Psi}_\mu \overleftarrow{\del{\nu}} \gamma^{\mu\nu\rho} \Psi_\rho 
    \\ & \qquad \qquad \qquad 
    + \tfrac{1}{2} m \bar{\Psi}_\mu \gamma^{\mu\nu} \Psi_\nu 
    \Bigr] 
    \com
}
which generalizes the Rarita-Schwinger action to curved spacetime \cite{Rarita:1941mf}.  See \aref{app:general_relativity} for a review of the frame-field formalism.  We require $\Psi_\mu = \Psi_\mu^C \equiv - \ii \gamma^2 \Psi_\mu^\ast$ such that $\Psi_\mu(x)$ is a self-conjugate vector-spinor.  When attention is restricted to an FLRW spacetime, the field's equation of motion becomes 
\bes{
    & \Bigl[ 
    \ii \gamma^{abc} \delta_b^\mu \partial_\mu 
    - \ii a H ( \gamma^a \eta^{c0} - \gamma^c \eta^{a0} ) 
    \\ & \qquad 
    + am \gamma^{ac} 
    \Bigr] \bigl( a^{1/2} \Psi_c \bigr) = 0 
}
where $\Psi_c = \delta_c^\mu \Psi_\mu$.  Latin indices are raised and lowered with the Minkowski metric $\eta_{ab}$ and inverse $\eta^{ab}$.  

\para{Mode equations}  
Solutions of the field's equation of motion take the form [see also \rref{Kaneta:2023uwi}]
\ba{\label{eq:Psia_Ansatz}
    & \Psi_a(\eta,\xvec) = [a(\eta)]^{-1/2} \int \! \! \frac{\dd^3\kvec}{(2\pi)^3} \sum_{\lambda=\pm \threehalf, \pm \onehalf} 
    \\ & \qquad \qquad \times 
    \Bigl[ 
    a_{\kvec,\lambda} \, U_{\kvec,\lambda,a}(\eta,\xvec)  
    + a_{\kvec,\lambda}^\dagger \, V_{\kvec,\lambda,a}(\eta,\xvec) \Bigr] 
    \nn &
    U_{\kvec,\lambda,0}(\eta,\xvec) = \frac{R/3 + H^2 - 3 m^2}{3 (H^2 + m^2)} \gamma^0 \gamma^i U_{\kvec,\lambda,i} 
    \nn & 
    U_{\kvec,\pm\threehalf,i}(\eta,\xvec) = \begin{pmatrix} u_{A,\threehalf,k}(\eta) \\ \pm u_{B,\threehalf,k}(\eta) \end{pmatrix} \otimes h_{\kvec,\pm} \, \varepsilon_{\kvec,\pm,i} \, \ee^{\ii \kvec \cdot \xvec}  
    \nn & 
    U_{\kvec,\pm\half,i}(\eta,\xvec) = \biggl\{ 
    \sqrt{\frac{1}{3}} \begin{pmatrix} u_{A,\onehalf,k}(\eta) \\ \mp u_{B,\onehalf,k}(\eta) \end{pmatrix} \otimes h_{\kvec,\mp} \, \varepsilon_{\kvec,\pm,i} 
    \nn & \qquad 
    - \ii \ee^{\mp\ii \gamma} \sqrt{\frac{2}{3}} \begin{pmatrix} u_{A,\onehalf,k}(\eta) + \tfrac{k}{am - \ii aH} u_{B,\onehalf,k}(\eta) \\[5pt] \mp u_{B,\onehalf,k}(\eta) \pm \frac{k}{am + \ii aH} u_{A,\onehalf,k}(\eta) \end{pmatrix} 
    \nn & \qquad 
    \otimes h_{\kvec,\pm} \, \varepsilon_{\kvec,0,i}
    \biggr\} \, \ee^{\ii \kvec \cdot \xvec} 
    \nonumber
}
where $V_{\kvec,\lambda,a} = - \ii \gamma^2 U_{\kvec,\lambda,a}^\ast$.  The ladder operators $a_{\kvec,\lambda}$ and mode functions $U_{\kvec,\lambda,a}(\eta,\xvec)$ are labeled with the comoving wave number $\kvec$ and helicity $\lambda$.  The mode functions carry a spacetime index $a$ and a suppressed spinor index.  The temporal components are restricted by a constraint equation, and the spatial components are decomposed onto an orthonormal basis of helicity eigenspinors $h_{\kvec,\pm}$ and three-vectors $\varepsilon_{\lambda,\kvec,i}$.  In the Dirac representation of the gamma matrices, the equations of motion are 
\bes{
    \ii \partial_\eta \begin{pmatrix} u_{A,\threehalf,k} \\ u_{B,\threehalf,k} \end{pmatrix} 
    & = \begin{pmatrix} am & k \\ k & - am \end{pmatrix} \begin{pmatrix} u_{A,\threehalf,k} \\ u_{B,\threehalf,k} \end{pmatrix} \\ 
    \ii \partial_\eta \begin{pmatrix} u_{A,\half,k} \\ u_{B,\half,k} \end{pmatrix} 
    & = \begin{pmatrix} am & c_s k \\ c_s^\ast k & - am \end{pmatrix} \begin{pmatrix} u_{A,\half,k} \\ u_{B,\half,k} \end{pmatrix} 
    \com
}
where the time-dependent complex sound speed is 
\ba{
    c_s(\eta) & = \frac{m^2 - \tfrac{1}{3} H^2 - \tfrac{1}{9} R}{(m - \ii H)^2} 
    \per
}
Note that the helicity $\pm \threehalf$ mode functions solve the same equation of motion as a spin-$\half$ Dirac spinor field, and the analysis of CGPP is equivalent as well.  However, the helicity $\pm \half$ mode functions have a modified gradient term that arose as a consequence of the constraint that eliminated $\Psi_0$ for $\Psi_i$.  

\para{CGPP intuition} 
A novel feature of the spin-$\threehalf$ field in the FLRW spacetime is that the helicity $\pm \half$ polarization modes develop an effective sound speed $c_s(\eta)$ that varies in time in response to the cosmological expansion \cite{Giudice:1999yt,Hasegawa:2017hgd}.  If the FLRW expansion is driven by a perfect fluid with an energy density $\rho(\eta)$ and pressure $P(\eta)$, then Friedmann's equations may be used to write the squared sound speed as 
\ba{\label{eq:spin32_sound_speed}
    |c_s(\eta)|^2 = \frac{(3 m^2 \Mpl^2 - \pressure)^2}{(3 m^2 \Mpl^2 + \rho)^2} 
    \per
}
Although this quantity cannot be negative, it may drop to zero if there exists a time $\eta_\ast$, or several such times, when $\pressure(\eta_\ast) = 3 m^2 \Mpl^2$.  During inflation $\pressure < 0$ since $w = \pressure / \rho \approx -1$, and $|c_s|^2 > 0$ remains positive, but after inflation $\pressure(\eta)$ oscillates around zero, and if these oscillations have a sufficiently large amplitude the sound speed may momentarily vanish.  Typically small-scale modes with a large comoving wave number $k$ have a nearly adiabatic evolution, since $\omega_k^2 \approx k^2$ is approximately static, and their CGPP is suppressed.  However, a vanishing sound speed momentarily lifts the suppression for CGPP in high-wave-number modes, allowing for ``catastrophic'' particle production up to the cutoff of the theory \cite{Hasegawa:2017hgd,Kolb:2021xfn}.  These dynamics are similar to gradient instabilities in other systems where the squared sound speed is negative; for a few examples see \rref{Lue:1998mq,Alexander:2004us,Gumrukcuoglu:2015nua,Comelli:2015pua}.  If the gradient instability is to be avoided, one requires the mass of the spin-$\threehalf$ particle to be larger than the Hubble parameter at the end of inflation: $m > \sqrt{\pressure_\mathrm{max} / 3 \MPl^2} \approx \He$.  Unless the mass varies in time, one is typically considering superheavy CGPP in the context of spin-$\threehalf$ particles.  

\para{Numerical example}
\Fref{fig:abundances} shows the results of a numerical evaluation of CGPP for a massive vector-spinor field in quadratic inflation; see the curves labeled $s_\lambda = 3/2_{\pm 3/2}$ and $s_\lambda = 3/2_{\pm 1/2}$.  The spectrum of the helicity $\pm \threehalf$ polarization modes is identical to the spectrum of spin-$\onehalf$ particles since their mode equations are identical.  CGPP is more efficient for the helicity $\pm \onehalf$ polarizations.  Lowering the mass toward the threshold of the gradient instability $m \approx \He$ enhances the abundance. 

\para{Connections with gravitinos}
This discussion has focused on a spin-$\threehalf$ vector-spinor field with a constant mass parameter $m$.  In theories of supergravity with a spin-$\threehalf$ gravitino, the mass parameter may also vary in response to the evolving scale of supersymmetry breaking \cite{Nilles:2001ry,Nilles:2001fg}.  If $\dd m/\dd \eta \neq 0$, the squared sound speed \eqref{eq:spin32_sound_speed} develops an additional contribution preventing $|c_s|^2 = 0$ \cite{Kolb:2021xfn}.  Similarly the mixing of the Goldstino and inflatino can modify the equations of motion so as to avoid the gradient instability for models with $m < \He$ \cite{Antoniadis:2021jtg,Dudas:2021njv}.  For such models the CGPP spectrum is expected to differ significantly from the fixed-mass vector-spinor field.  Possible connections between gravitino CGPP and supergravity's UV embedding were discussed by \rref{Kallosh:2004yh,Kolb:2021nob,Antoniadis:2021jtg,Castellano:2021yye,Cribiori:2021gbf,Terada:2021rtp,Dudas:2021njv}.  

\subsection{Massive spin-$2$ field}
\label{sub:spin_two}

Finally we discuss the phenomenon of CGPP for a massive spin-$2$ tensor field, which is a spectator during inflation and not the graviton.  Several recent studies have explored the idea that dark matter might be a massive spin-$2$ particle \cite{Babichev:2016bxi,Babichev:2016hir,Aoki:2016zgp,Manita:2022tkl,Gorji:2023cmz}.  The properties and interactions of such particles can be studied in the framework of ghost-free nonlinear bigravity \cite{Hassan:2011zd}.  Here we first discuss the theory of a massive spin-$2$ field in an inflationary cosmology and the resultant CGPP, and afterward we discuss the connection with bigravity.  

\para{Action in FLRW spacetime}
To study the gravitational production of massive spin-$2$ particles during and after inflation, one desires an effective field theory that describes a massive tensor field and a scalar inflaton field on a fixed curved spacetime background.  A relativistically covariant action can be constructed using a symmetric two-index tensor field $v_{\mu\nu}(x)$.  Of the ten independent variables in $v_{\mu\nu}$, only five of them correspond to the polarization modes of the massive spin-$2$ particle, and the remaining variables are eliminated by constraints or gauge symmetries.  Care must be taken to construct an action that does not promote the extraneous variables into propagating degrees of freedom, which may be ghostly having a wrong-sign kinetic term \cite{Hinterbichler:2011tt,deRham:2014zqa}.  A free theory with the usual minimally coupled matter sector can be obtained by expanding the Einstein-Hilbert action plus the matter sector to quadratic order in perturbations around any background that satisfies the GR equations of motion and adding the Fierz-Pauli mass term \cite{Fierz:1939ix}.  The resulting theory of a free massive spin-$2$ field on a fixed background is ghost-free, while nonlinearities generically introduce a ghost.

The aforementioned procedure leads to an action of the form \cite{Kolb:2023dzp} 
\ba{
    S = \int \! \dd^4x \, \sqrt{-\bar{g}} \, \Bigl( \Lcal_{vv}^{(2)} + \Lcal_{\cphi_v \cphi_v}^{(2)} + \Lcal_{v \, \cphi_v}^{(2)} + \Lcal_\mathrm{int} \Bigr) 
    \com
}
where $\Lcal_{vv}^{(2)}$ are terms quadratic in the massive metric perturbation $v_{\mu\nu}(x)$ including a Fierz-Pauli mass term, $\Lcal_{\cphi_v \cphi_v}^{(2)}$ are terms quadratic in the inflaton perturbations $\cphi_u(x)$, $\Lcal_{v \, \cphi_v}^{(2)}$ are terms that mix the scalar metric perturbations with the inflaton perturbations, and $\Lcal_\mathrm{int}$ are cubic and higher-order terms corresponding to interactions.  It is important to retain the time-dependent mixing, since dropping it would lead to an expression that violates gauge symmetry and introduces a ghost.  

\para{Mode equations}  
The massive spin-$2$ field $v_{\mu\nu}(x)$ may be decomposed onto components that transform as three-scalars, three-vectors, and three-tensors under the $\mathrm{SO}(3)$ symmetry group of spatial rotations; this is the usual scalar-vector-tensor decomposition \cite{Baumann:2022mni}.  Upon eliminated components that are restricted by constraints, one obtains the equations of motion for the five polarization modes with helicities $\pm 2$, $\pm 1$, and $0$ \cite{Kolb:2023dzp}.  The Fourier components of the helicity $\pm 2$ modes obey 
\bes{\label{eq:spin2_tensor}
    & \chi_{k,s}^{\prime\prime} + \omega_k^2 \, \chi_{k,s} = 0 \qquad \text{for $\lambda = \pm 2$} \\
     & \omega_k^2(\eta) = k^2 + a^2 m^2 + \tfrac{1}{6} a^2 R 
    \per
}
Equations~\ref{eq:spin2_tensor} are equivalent to the equations of motion for a minimally coupled scalar field, which appear in \eref{eq:mode_equation}, and the analysis of CGPP is identical.  Moreover, setting $m=0$ yields the familiar equations of motion that arise in the study of primordial gravitational-wave generation during inflation.  The helicity $\pm 1$ modes obey 
\ba{\label{eq:spin2_vector}
    & \chi_{k,s}^{\prime\prime} + \omega_k^2 \, \chi_{k,s} = 0 \qquad \text{for $\lambda = \pm 1$} \\
     & \omega_k^2(\eta) = k^2 + a^2 m^2 - f^{\prime\prime}/f,\quad
    f = \frac{a^2}{\sqrt{k^2 + a^2 m^2}} \nonumber
    \per
}
For nonrelativistic modes ($k < am$) both the tensor and vector polarizations have $\omega_k^2 \approx a^2 m^2 + a^2 R / 6$, and one can anticipate similar solutions and implications for CGPP.  The helicity-$0$ modes have a much more complicated equation of motion since the scalar metric perturbations mix with the inflaton perturbation, and the system of $2$ degrees of freedom must be solved simultaneously.  Expressions for the quadratic action were given by \rref{Kolb:2023dzp}; see also \rref{Gorji:2023cmz}.  

\para{CGPP intuition} 
We emphasize that the helicity-$0$ mode of a massive spin-$2$ field in curved spacetime threatens to acquire a wrong-sign kinetic term and develop a ghost instability \cite{Fasiello:2012rw,Fasiello:2013woa}.  In an FLRW spacetime this pathology is avoided if the field's mass respects the lower limit \cite{Kolb:2023dzp}
\ba{\label{eq:Eq110}
    m^2 > 2 H^2(\eta) \bigl[ 1 - \epsilon(\eta) \bigr]
    \com
}
where $\epsilon = -H^\prime / aH^2$ is the first slow-roll parameter; \eref{eq:Eq110} is not an expansion in small $\epsilon$.  
During the qdS phase of inflation, the right-hand side of \eref{eq:Eq110} is approximately static $2 \Hinf^2$, and the Higuchi bound is regained \cite{Higuchi:1986py}.  After inflation $1 - \epsilon < 0$ is negative, and the ghost instability is avoided for any nonnegative choice of the squared mass parameter.  Models with $m > \sqrt{2} \Hinf$ that evade the ghost during inflation, typically have superheavy spin-$2$ particles, particularly if the energy scale of inflation is high, say $\Hinf \approx 10^{14} \GeV$.  Conversely, models of ultralight spin-$2$ dark matter require additional assumptions (such as a time-dependent mass or low cutoff) to evade a ghost instability during inflation. 

\para{Numerical example}
\Fref{fig:abundances} shows the results of a numerical evaluation of CGPP for a massive spin-$2$ tensor field in quadratic inflation; see the curves labeled $s_\lambda = 2_{\pm 2}$.  The helicity-$\pm 2$ mode functions obey the same equation of motion as a conformally coupled scalar field with the same mass, and the predictions for CGPP are identical.  We do not show results for the other polarization modes $\lambda = \pm 1$ and $\lambda = 0$.  

\para{Connection with ghost-free nonlinear bigravity}  
One may ask whether the quadratic action presented here can be extended into a fully nonlinear theory of massive spin-$2$ particles interacting with gravity without introducing ghosts.  This can be accomplished in the framework of ghost-free nonlinear bigravity \cite{Hassan:2011zd}; see also the review article by \rref{deRham:2014zqa}.  Theories of bigravity are formulated in terms of two dynamical metrics, typically denoted by $\metric_{\mu\nu}$ and $f_{\mu\nu}$, that interact with one another and may interact with matter.  When both metrics are expressed as background plus perturbation, the quadratic action for the perturbations describes a massless spin-$2$ field that corresponds to the usual gravitational field, as well as an additional massive spin-$2$ field that corresponds to our $v_{\mu\nu}$.  Additional model dependence enters in how the metrics couple to matter.  Different couplings to matter \cite{deRham:2014naa} may lead to strikingly different results for CGPP \cite{Kolb:2023dzp}.  

\subsection{Numerical comparison}
\label{sub:comparison}

\begin{figure}[t]
\begin{center}
\includegraphics[width=0.48\textwidth]{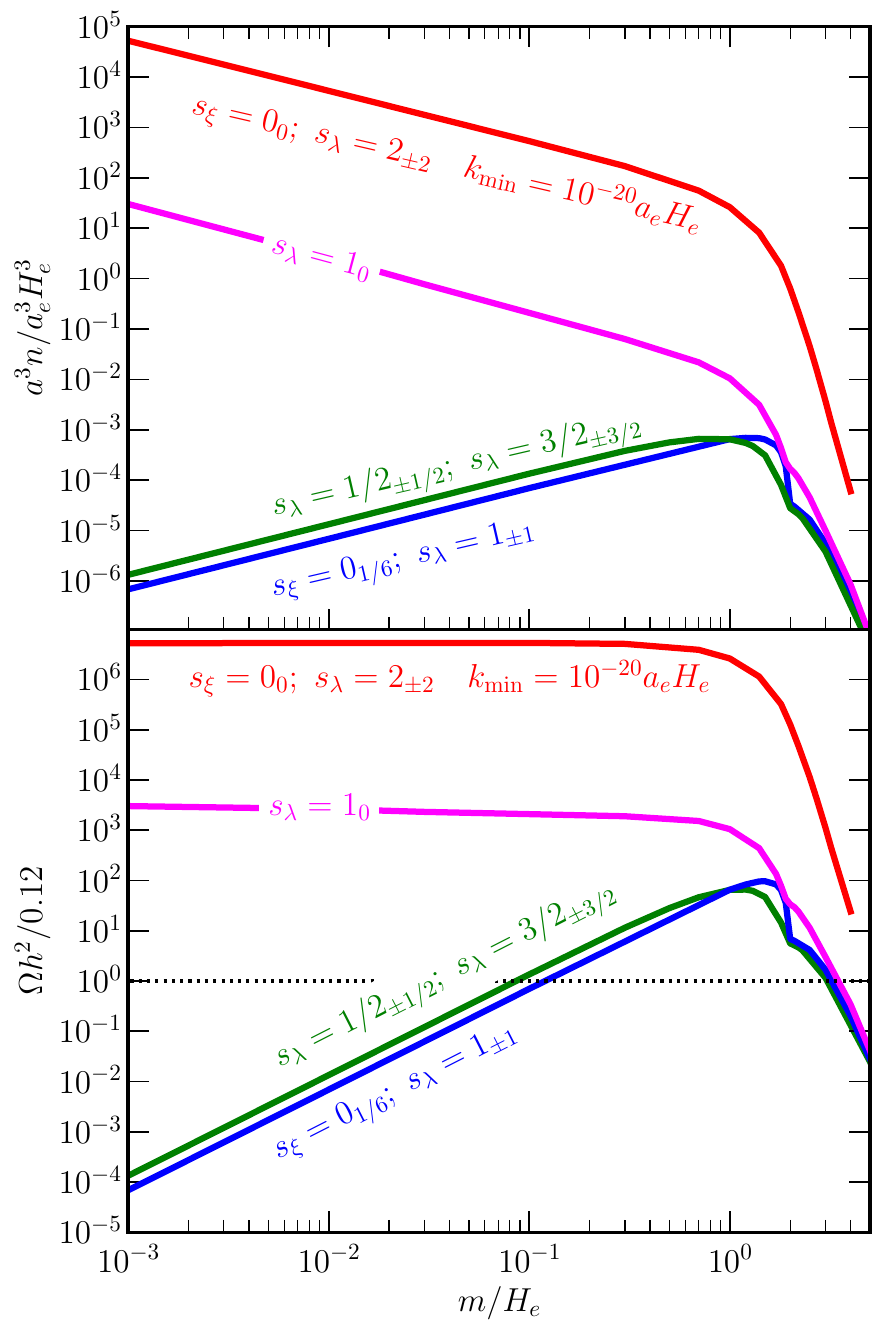} 
\caption{
\label{fig:abundances}
The predicted abundance of gravitationally produced particles per polarization state when quadratic inflation is assumed.  Each curve corresponds to different model labeled with the particle's spin $s$, helicity $\lambda$, and (for scalar fields) coupling to gravity $\xi$.  For instance, the curve labeled as $s_\lambda=1_0$ corresponds to the longitudinal polarization mode of a massive vector field.  Assuming that the particles are stable, the present-day abundances are given.  Top panel:  the comoving number density $a^3 n$ of produced particles in units of $\ae^3 \He^3$ as a function of the particle's mass $m$.  Bottom panel: the cosmological energy fraction $\Omega h^2$ assuming a reheating temperature of $\TRH = 10^9 \GeV$ and $\He=10^{12}\GeV$ (see \eref{eq:Omegah2} for the scaling with $\TRH$ and $H_e$).  The calculation assumes late reheating.  
}
\end{center}
\end{figure}

\para{Assumptions}
In this section we present the results of several numerical calculations of CGPP for particles of different spin.  For the examples presented here, we model the cosmological expansion using quadratic inflation (see \fref{fig:end_of_inflation}).  The inflaton drives a phase of quasi-de Sitter inflation and afterward oscillates around the minimum of its quadratic potential during an effectively matter-dominated epoch of reheating.  We do not explicitly model the inflaton decay, which corresponds to an assumption of late reheating \eqref{eq:late_early_reheating}.  For the particle species that experiences gravitational production, we vary its mass $m$, its spin $s$, and (in the case of scalars) its coupling to gravity $\xi$.  For particles with spin ($s \neq 0$), we further distinguish the various polarization modes (labeled as $\lambda$), which experience different levels of gravitational production.  We neglect nongravitational interactions, and assume that the particles are long-lived; see \sref{sec:Recent} for remarks on how relaxing these assumptions could modify the results.  

\para{Relic abundance}
\Fref{fig:abundances} shows the total abundance of gravitationally produced particles: the top panel shows the comoving number density $a^3 n$, and the bottom panel shows the cosmological energy fraction today $\Omega h^2$.  These quantities are related by \eref{eq:Omegah2} and we take $\TRH = 10^9 \GeV$, but more generally $\Omega h^2 \propto \TRH$ in late reheating.  

\para{Superheavy and ultralight dark matter}
For $\Omega h^2 \approx \Omega_\mathrm{dm} h^2 \approx 0.12$ gravitationally produced particles can make up all of dark matter.  For the spin-$0$ conformally coupled scalar field ($s_\xi = 0_{1/6}$), the spin-$\half$ spinor field ($s_\lambda = 1/2_{\pm 1/2}$), the spin-$\threehalf$ vector-spinor field ($s_\lambda = 3/2_{\pm 3/2}$), and the spin-$2$ tensor field ($s_\lambda = 2_{\pm 2}$), the cosmological energy fraction decreases toward small mass like $\Omega h^2 \propto m^2$.  Consequently, reaching $\Omega h^2 \approx 0.12$ is easier for a larger mass $m \approx \He$, corresponding to ``superheavy'' dark matter.  However, for the spin-$0$ minimally coupled scalar field ($s_\xi = 0_0$) and the longitudinal polarization of the spin-$1$ vector field ($s_\lambda = 1_0$), the energy fraction is insensitive to the mass $\Omega h^2 \propto m^0$.  Keep in mind that this scaling holds only while $m > \HRH$ and the system is in the late reheating regime \eqref{eq:late_early_reheating}.  For small $m$ the system is in the early reheating regime and $\Omega h^2$ decreases with decreasing $m$ [\eref{eq:Omegah2_results}].  For these models all of the dark matter can be produced gravitationally for masses as low as $m \sim 10^{-5} \eV$.  However, the minimally coupled scalar is constrained by dark-matter isocurvature; see \sref{sub:isocurvature}.  

\subsection{Higher-spin particles}
\label{sub:higher_spin}

One may ask how the analysis of CGPP is extended to particle species with spin $s > 2$, here called higher-spin particles, that transform under larger representations of the Lorentz group.  Although no-go theorems forbid higher-spin particles that are massless \cite{Weinberg:1964ew,Coleman:1967ad,Aragone:1979hx,Weinberg:1980kq}, particles that are massive may evade these theorems \cite{Bellazzini:2019bzh}.  Moreover, such particles naturally arise in nuclear physics, in condensed matter physics as composite fermions \cite{Golkar:2016thq}, and play a starring role in string theory (see \eg, \rref{Sagnotti:2011jdy}) where the higher-spin fields originating from massive string excitations, are responsible for curing the UV divergences of gravity.  Further motivation is provided by recent interest in higher-spin particles as candidates for dark matter \cite{Criado:2020jkp,Alexander:2020gmv,Falkowski:2020fsu,Jenks:2022wtj,Capanelli:2023uwv}.  

\para{Higher spin fields in de Sitter spacetime}  
While CGPP in higher-spin theories remains largely unexplored, the results of a few groundbreaking studies have indicated that CGPP may provide an explanation for the origin of higher-spin dark matter.  \rref{Alexander:2020gmv} focused on bosonic fields of spin $s > 2$ that are minimally coupled to gravity and approximated the inflationary spacetime as a de Sitter one with a constant Hubble parameter $H$.  On this background the components $\sigma_{\mu_1\cdots\mu_s}(\eta,\xvec)$ of the higher-spin field solve the equation of motion $[\Box - m^2 + (s^2 - 2s - 2) H^2] \sigma_{\mu_1\cdots\mu_s}$ \cite{Lee:2016vti}.  A ghost instability develops for small masses, and the avoidance of this ghost requires $m^2 \geq s (s-1) H^2$, which is a higher-spin generalization of the Higuchi bound for massive spin-$2$ particles \cite{Higuchi:1986py}.  This wave equation takes the same form as the equation of motion for a minimally coupled scalar field in de Sitter spacetime, and the previously derived analytical results for scalar CGPP during inflation can be carried over with the appropriate replacements.  One may conclude that CGPP provides a viable explanation for the origin of higher-spin dark matter for a wide range of superheavy masses $m > H$ and spins.  

\para{Limitations and possible extensions}  
Treating this as a de Sitter spacetime is a reasonable approach for studying CGPP in modes whose nonadiabaticity occurs during inflation as they leave the horizon, \ie, $k \ll \ae \He$.  For modes on the Hubble scale at the end of inflation $k \approx \ae \He$, which tend to carry most of the particle number, their evolution probes the transition from inflation to reheating and requires a modeling of the inflationary cosmology beyond the de Sitter approximation.  This distinction may be especially important for the helicity-$0$ mode of the higher-spin field, which can mix with the scalar inflaton perturbation, leading to an enhancement of CGPP that is similar to the behavior observed in spin-$2$ fields.  

\section{Recent and ongoing activity}
\label{sec:Recent}

In this section we summarize some of the interesting recent and ongoing activity in the study of CGPP.  

\subsection{Modeling: nonminimal gravitational interaction}
\label{sub:nonminimal}

There is some degree of model building freedom with regard to a particle's gravitational interaction.  A field is said to be minimally coupled to gravity if its covariant action can be obtained from the flat spacetime action by promoting the Minkowski metric to a curved spacetime metric.  If the covariant action contains additional interaction terms, then the field is said to couple nonminimally to gravity.  Some examples of nonminimal interactions include $\Lcal \supset \aphi R$, $\aphi^2 R$, $\bar{\Psi} \Psi R$, and $A_\mu A_\nu R^{\mu\nu}$. 

The presence of a nonminimal interaction can greatly enhance or diminish the efficiency of CGPP.  For example, a conformally coupled scalar field with $\Lcal_\mathrm{int} = \tfrac{1}{2} \xi \aphi^2 R$ and $\xi = \sixth$ experiences less CGPP than a minimally coupled scalar with $\xi = 0$; see \sref{sec:Theory}.  However, a nonminimally coupled scalar with $\xi \gtrsim 1$ can experience significantly more CGPP \cite{Markkanen:2015xuw,Markkanen:2016aes,Fairbairn:2018bsw,Cembranos:2019qlm,Clery:2022wib,Garcia:2023qab,Yu:2023ity}.  This is because the Fourier mode amplitude $\chi_k(\eta)$ evolves in response to the time-dependent oscillator equation \eqref{eq:mode_equation} with frequency $\omega_k^2 = k^2 + a^2 m^2 + (1/6 - \xi) a^2 R$, which is dominated by the $a^2 R$ term for large $\xi$.  During inflation the Ricci scalar is approximately constant ($R \approx - 12 H^2$), but after inflation while the inflaton field oscillates during the epoch of reheating, the Ricci scalar $R$ also oscillates from positive to negative values; see \fref{fig:end_of_inflation}.  Consequently the frequency develops an oscillatory component that increases in amplitude for larger $\xi$ and enhances CGPP.  

\begin{figure}[t]
\begin{center}
\includegraphics[width=0.48\textwidth]{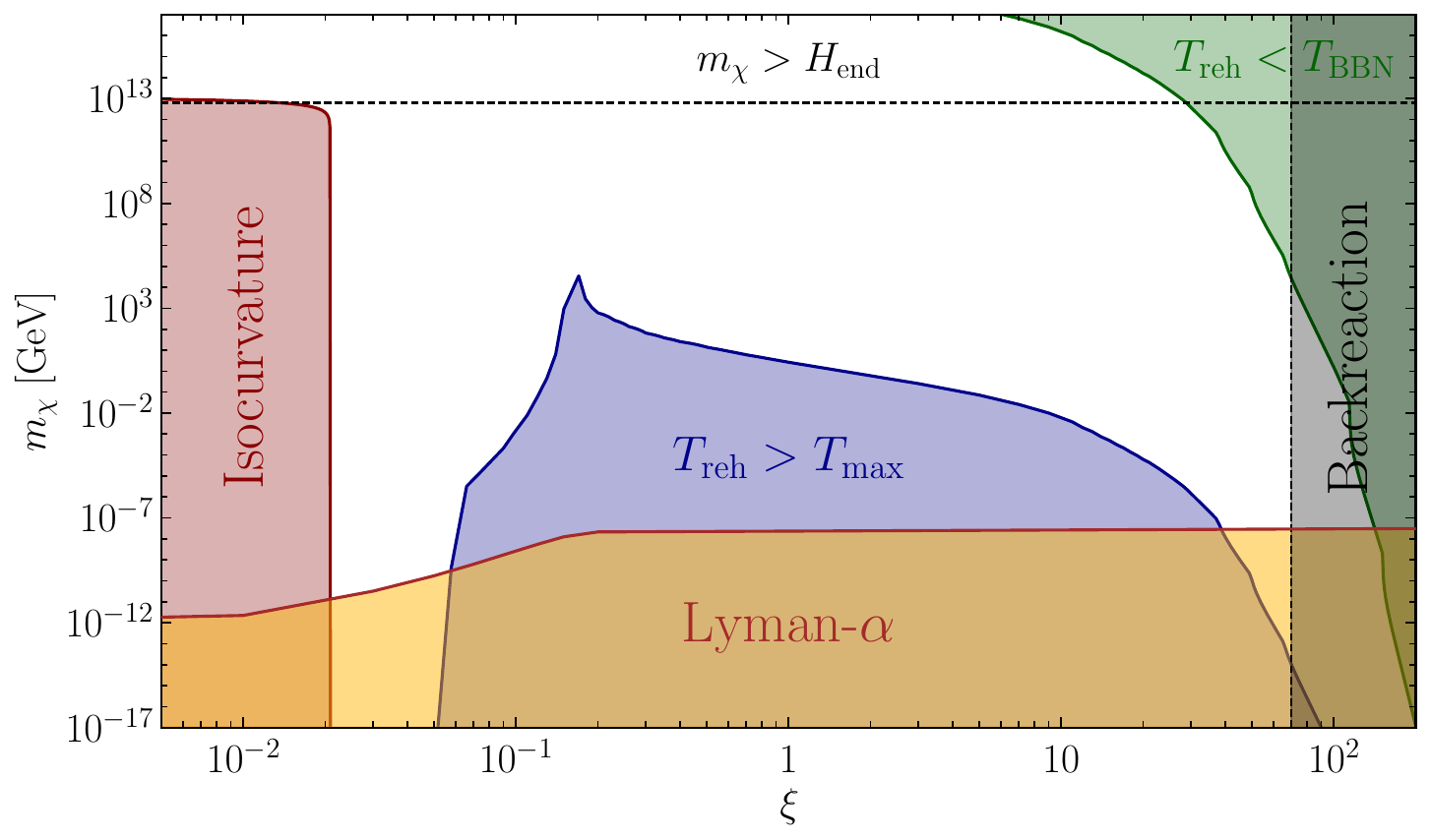} 
\caption{
\label{fig:xi_dependence}
CGPP predictions and constraints across the parameter space of a nonminimally coupled real scalar field.  Cosmological inflation is assumed to arise from the $\mathrm{T}$-Model $\alpha$-Attractor \cite{Kallosh:2013maa} with $\alpha=1$ and $V(\cphi) \approx 3 m_\cphi^2 \Mpl^2 \tanh^2(\cphi / \sqrt{6} \Mpl)$.  From \rref{Garcia:2023qab}.  
}
\end{center}
\end{figure}

A global picture on the parameter space of the nonminimally coupled scalar field is shown in \fref{fig:xi_dependence}.  Within the white region it is possible to find a reheating temperature $T_\mathrm{reh} \equiv \TRH$ such that the predicted relic abundance equals the observed dark-matter relic abundance $\Omega h^2 \propto \TRH$; see \eref{eq:Omegah2}.  In the blue region, which contains the conformally coupled point $\xi = \sixth$ at low mass, CGPP is less efficient and the required reheating temperature $T_\mathrm{reh} > T_\mathrm{max}$ would exceed the maximum possible temperature \eqref{eq:instant_RH}.  In the green region where $\xi \gg 1$ CGPP is more efficient and the needed reheating temperature $T_\mathrm{reh} < T_\BBN$ would fall below the BBN temperature.  

\subsection{Modeling: scale and shape of inflaton potential}
\label{sub:inflation}

One appealing aspect of CGPP as the origin of cosmological relics is that the assumed gravitational interaction implies a limited degree of model dependence.  For the particle being produced, one need only specify its spin, its mass, and the nature of its coupling to gravity.  The discussion in Secs.~\ref{sec:Scalar}~and~\ref{sec:Spin} illustrates how CGPP depends on these parameters.  
Any additional model dependence enters through the physics of inflation and reheating, which controls the time-dependent FLRW background $a(\eta)$.  The energy scale of inflation (parametrized by $\He$) and the duration of reheating (parametrized by $\TRH$) directly impact the efficiency of CGPP, and the dependence of observables on these parameters is easy to assess; see \sref{sub:relic_abundance}.  However, the shape of the inflaton potential typically has a weaker impact on the CGPP observables. 

\begin{figure}[t]
\begin{center}
\includegraphics[width=0.48\textwidth]{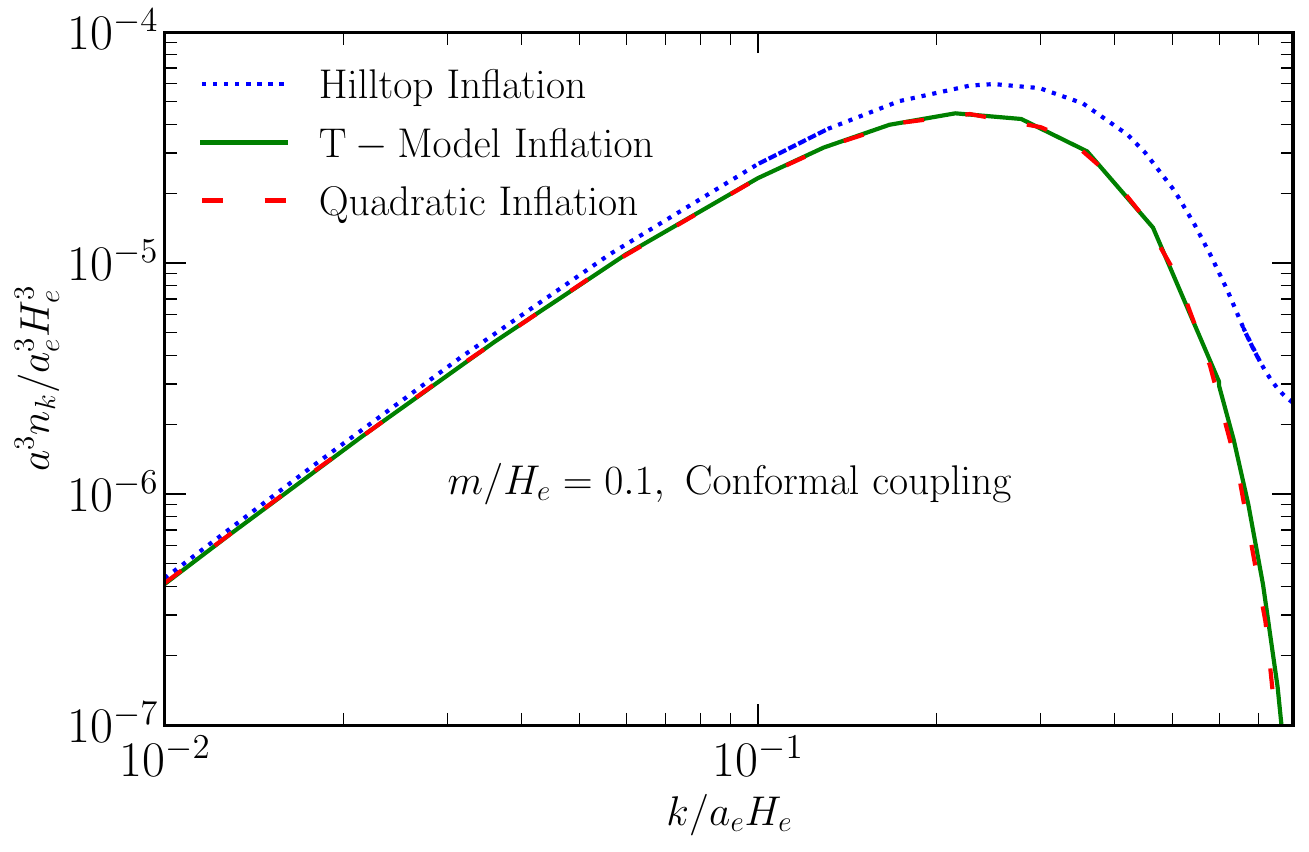} 
\caption{
\label{fig:vary_model_of_inflation}
Illustration of the insensitivity of CGPP to the shape of the inflaton potential.  The three curves correspond to different models: hilltop inflation, $\mathrm{T}$-model $\alpha$-attractor inflation ($\alpha = 1$), and quadratic inflation.  The comoving number density spectrum is shown as a function of $k/a_eH_e$.
}
\end{center}
\end{figure}

To assess the dependence of CGPP on the model of inflation, we consider three models that were discussed in \sref{sub:infcos}: quadratic inflation, hilltop inflation, and $\mathrm{T}$-model $\alpha$-attractor inflation ($\alpha = 1$).  For each model we choose the parameters such that $m/\He$ takes the same value.  In \fref{fig:vary_model_of_inflation} we show the comoving number density spectrum $a^3 n_k$ for a massive scalar ($m / \He = 0.1$) with a conformal coupling to gravity ($\xi = \sixth$).  For quadratic inflation and $\mathrm{T}$-model $\alpha$-attractor inflation with $\alpha = 1$, the spectra are indistinguishable to the eye and hilltop inflation differs by an $O(1)$ factor at the maximum.   Recall that hilltop inflation is not symmetric about the minimum of the potential, whereas sending $\alpha \to \infty$ in $\alpha$-attractor inflation yields the parabolic potential of quadratic inflation.  This numerical example illustrates the insensitivity of CGPP to the shape of the inflaton potential.  For studies of CGPP in models of inflation with a plateau, see \rref{Karam:2020rpa}.

\subsection{Modeling: gravity-mediated thermal freeze-in}
\label{sub:thermal_freeze_in}

During the epoch of reheating, the decay of the inflaton condensate produces the hot primordial plasma.  Consider a scenario in which the theory contains a hidden sector that couples weakly to the inflaton and is not populated via inflaton decays.  Nevertheless, if the mass scale of hidden-sector particles is not much larger than the temperature of the visible-sector radiation, then one can expect the hidden sector to be populated via annihilations of visible-sector particles through gravitational interactions alone.  This is a form of UV-dominated thermal freeze-in \cite{Hall:2009bx} that is similar to gravitino production \cite{Nanopoulos:1983up,Ellis:1983ew,Khlopov:1984pf,Olive:1984bi}.  If the hidden sector contains stable massive particles, they are candidates for dark matter, and the freeze-in provides a viable mechanism for their creation. 

\begin{figure}[t]
\begin{center}
\includegraphics[width=0.48\textwidth]{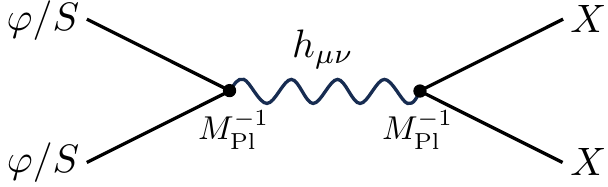} 
\caption{
\label{fig:scattering}
Annihilation channels contributing to gravitational particle production.
}
\end{center}
\end{figure}

To illustrate the parametric dependence of gravity-mediated thermal freeze-in, consider the following simplified example.  Let the visible-sector particles be represented by $S$ and the hidden-sector particles be represented by $X$.  Suppose that $S$ particles are in thermal equilibrium with temperature $T$, and that $X$ particles are initially absent but gradually produced via $SS \to h_{\mu\nu} \to XX$ where the reaction could be mediated by an $s$-channel graviton exchange illustrated in \fref{fig:scattering}.  This reaction is assumed to arise from a gravitational-strength coupling such that $X$ population never becomes too large, and the reverse process can be neglected; \ie, the $X$ particles freeze in.  

The number density $n_X(t)$ of $X$ particles is required to solve the Boltzmann equation \cite{KolbTurner:1990}
\ba{\label{eq:Boltzmann_eqn}
    \frac{\dd n_X}{\dd t} + 3Hn_X = -\langle \sigma v \rangle_{SS\to XX} \bigl[ (n_X)^2 - (n_X^\mathrm{eq})^2 \bigr] 
    \per
}
The reaction $SS \to XX$ has a temperature-dependent thermally averaged cross section $\langle \sigma v \rangle_{SS\to XX}$ \cite{Gondolo:1990dk} that depends on the mass and spin of the particles involved \cite{Garny:2017kha,Garny:2018grs}.  We take $\langle \sigma v \rangle_{SS\to XX} \approx T^2 / \MPl^4$ for these estimates.  The number density $n_X$ of $X$ particles is much smaller than the number density $n_X^\mathrm{eq}$ that the $X$ particles would have if they were in equilibrium at temperature $T$.  If $m_X \ll T$ then $n_X^\mathrm{eq} \approx \zeta(3) T^3 / \pi^2$ up to an $O(1)$ factor that depends on the spin and number of internal degrees of freedom associated with the $X$ particles.  

To solve the equation, it is necessary to specify how the Hubble parameter $H(t)$ and temperature $T(t)$ depend on time.  During the epoch of reheating, these variables depend on the cosmological equation of state and the inflaton's decay rate.  To make a simple estimate, we suppose that the epoch of reheating is brief and can be ignored (\ie, there is instantaneous reheating).  When reheating is completed, the Universe is radiation dominated at a temperature $\TRH$, and subsequently $H(t) \propto a^{-2}(t)$ and $T(t) \propto a^{-1}(t)$.  The solution can then be written as 
\ba{
    n_X(t) 
    & = \frac{1}{a^3(t)} \int_{t_\RH}^t \! \dd t^\prime \ a^3(t^\prime) \, \langle \sigma v \rangle_{SS\to XX}(t^\prime) \, \bigl( n_X^\mathrm{eq}(t^\prime) \bigr)^2 \nonumber \\ 
    & \approx \frac{1}{a(t)^3} \frac{\sqrt{10} \zeta^2(3)}{\pi^5} \frac{\aRH^3 \TRH^6}{\Mpl^3} 
    \per
}
Note that the number density has a power-law sensitivity to the maximum temperature $\TRH$, which is characteristic of UV-dominated freeze-in.  Assuming that the $X$ particles eventually become nonrelativistic, their cosmological energy fraction today is 
\ba{
    \Omega_X h^2 \approx 
    0.6 
    \biggl( \frac{m_X}{10^{12} \GeV} \biggr)^{} 
    \biggl( \frac{\TRH}{10^{13} \GeV} \biggr)^{3} 
}
for $m_X \lesssim \TRH$.  Note that this estimate is compatible with the dark-matter energy fraction $\Omega_\mathrm{dm} h^2 \approx 0.12$ for appropriate choices of $m_X$ and $\TRH$.  

The rough estimate here is provided only for illustrative purposes.  Various studies \cite{Chung:1998rq,Garny:2015sjg,Tang:2016vch,Tang:2017hvq,Garny:2017kha,Bernal:2018qlk,Kaneta:2019zgw,Clery:2021bwz,Garcia:2021iag,Chianese:2020khl,Chianese:2020yjo,Redi:2020ffc,Barman:2021ugy,Mambrini:2022uol,Clery:2022wib,Zhang:2023xcd,Zhang:2023hjk} have more accurately modeled various aspects of the calculation, including replacing $S$ with the full standard-model particle content, which affects $\langle \sigma v \rangle_{SS\to XX}$; modeling the plasma temperature evolution during the epoch of reheating, which affects the scaling $T(a)$ as well as the maximum temperature \cite{Giudice:2000ex}; and extending the analysis to models with $m_X > T$, which introduces a Boltzmann suppression.  The conclusion is that gravity-mediated thermal freeze-in can explain the origin of dark matter and play a significant role in populating weakly coupled sectors.  

\subsection{Modeling:  gravity-mediated inflaton annihilation}
\label{sub:inflaton_annihilations}

During the epoch of reheating, it is often assumed that the inflaton field oscillates around a quadratic minimum of its effective potential $V(\cphi) \approx m_\cphi^2 \cphi^2 / 2$.  This configuration describes a universe that is permeated by a nonrelativistic condensate of inflaton particles.  Each particle carries an energy of approximately $m_\cphi$ (neglecting kinetic energy) and if the maximum field amplitude is $\cphi_\mathrm{max}$ then the number density of particles in the condensate is approximately $m_\cphi \cphi_\mathrm{max}^2 / 2$.  We emphasize that this density can be large.  For instance, if the energy density that drives inflation $\rho_\cphi \approx 3 \Mpl^2 \He^2$ is transferred entirely to the condensate $\rho_\cphi \approx m_\cphi^2 \cphi_\mathrm{max}^2 / 2$ then $\cphi_\mathrm{max} \approx \sqrt{6} \Mpl \He / m_\cphi$ and $n_\cphi \approx 3 \Mpl^2 \He^2 / m_\cphi$ at the end of inflation such that the number of inflaton particles per Hubble volume is $n_\cphi / \He^3 \approx 3 \Mpl^2 / m_\cphi \He \gg 1$.  

The high density of inflaton particles enhances the rate of inflaton annihilations.  Suppose that the theory contains a particle species $X$ in a hidden sector that couples only gravitationally to the inflaton.  Such particles can be produced due to inflaton annihilations via an $s$-channel graviton $\cphi\cphi \to h_{\mu\nu} \to XX$ provided that the $X$ particles are lighter than the inflaton particles $m_X < m_\cphi$ such that the two-to-two channel is not kinematically blocked.  To calculate the relic density, one can solve a Boltzmann equation similar to the one in \eref{eq:Boltzmann_eqn} but with the replacements $H \to \He (a/\ae)^{-3/2}$, $\langle \sigma v \rangle_{SS\to XX} \to (\sigma v)_{\cphi\cphi\to XX} \approx m_\cphi^2 / \Mpl^4$, and $n_X^\mathrm{eq} \to n_\cphi \approx (3 \Mpl^2 \He^2 / m_\cphi) (a/\ae)^{-3}$.  Assuming that the $X$ particles are stable and nonrelativistic, their present-day cosmological energy fraction is 
\ba{
    \Omega_X h^2 & \approx 
    0.7 
    \biggl( \frac{\He}{10^9 \GeV} \biggr) 
    \biggl( \frac{m_X}{10^9 \GeV} \biggr) 
    \biggl( \frac{\TRH}{10^{10} \GeV} \biggr) 
    \nonumber \\[6pt] & \quad \times 
    \biggl( \frac{(\sigma v)_{\cphi\cphi \to XX}}{m_\cphi^2 / \Mpl^4} \biggr) 
    \biggl( \frac{n_\cphi(\ae)}{3 \Mpl^2 \He^2 / m_\cphi} \biggr)^2 
    \com
}
where the dependence on $\TRH$ enters through the redshift factor and not because particles are produced thermally from the plasma.  This corresponds to a number density $n_X(\te) \approx 6 \He^3$ at the end of inflation: a few particles per Hubble volume.  This normalization of the cross section is appropriate if $X$ is a minimally coupled scalar, and for other particle species (such as a conformally coupled scalar) an additional dependence on $m_X$ may enter \cite{Ema:2018ucl}.  The predicted $\Omega_X h^2$ can be comparable to the present-day observed dark-matter energy fraction $\Omega_\mathrm{dm} h^2 \approx 0.12$, indicating that gravity-mediated inflaton annihilations offer a viable explanation for the origin of dark matter.  

Inflaton annihilations lead to a characteristic energy spectrum for the produced $X$ particles.  Since each inflaton particle carries an energy of $E_\cphi \approx m_\cphi$ and assuming $m_X < m_\cphi$, the two-to-two scattering leads to $X$ particles with momentum $p_X \approx m_\cphi$ at the time of production.  Consequently, the first particles to be produced at the end of inflation have a comoving momentum (wave number) of $k = \ae m_\cphi$, and the last particles to be produced at the end of reheating have a larger $k = \aRH m_\cphi$.  Conversely, particles with a comoving wave number $k$ in the range $\ae m_\cphi < k < \aRH m_\cphi$ are produced at a time when the scale factor is $a = k / m_\cphi$.  For a given $k$ integrating the Boltzmann equation over one Hubble time gives 
\ba{
    & a^3 n_{X,k} 
    \approx a^3 \, (\sigma v)_{\cphi\cphi\to XX} \, n_\cphi^2(a) \, H^{-1}(a) \bigr|_{a=k/m_\cphi} \\ 
    & \approx \bigl( 9 \ae^3 \He^3 \bigr) 
    \biggl( \frac{\He}{10^9 \GeV} \biggr)^{-3/2} 
    \biggl( \frac{m_\cphi}{10^9 \GeV} \biggr)^{3/2} 
    \biggl( \frac{k}{\ae \He} \biggr)^{-3/2} 
    \com
    \nonumber
}
which is the comoving number density of $X$ particles with comoving wave numbers around $k$.  Note that $n_{X,k} \propto k^{-3/2}$.  If $X$ particles were heavier such that $m_\cphi < m_X < 3 m_\cphi / 2$, then the two-to-two channel would be blocked and the $\cphi \cphi \cphi \to XX$ channel would be dominant.  Since the production rate would proceed as $n_\cphi^3$, it would decrease more rapidly in time, leading to a steeper spectrum $n_{X,k} \propto k^{-9/2}$.  Moreover, these exponents all assume that $\wRH = 0$, and changing the equation of state during reheating will change the spectral indices.  

Although the preceding discussion characterizes gravitational particle production from inflaton annihilations as a scattering process, the same phenomenon is also captured by the Bogolubov framework.  While the inflaton field oscillates with decreasing amplitude during the epoch of reheating, the FLRW scale factor $a(\eta)$ develops an oscillatory component as well; see \fref{fig:end_of_inflation}.  The mode equations for a spectator field on this FLRW background take the form $\chi_k^\pprime + \omega_k^2(\eta) \chi_k = 0$; for example, a minimally coupled scalar field has $\omega_k^2(\eta) = k^2 + a^2(\eta) m^2 + a^2(\eta) R(\eta) / 6$ as in \eref{eq:mode_equation}.  The Bogolubov coefficient $\beta_k$ is calculated by solving \eref{eq:EOMforalphabeta} for $\tilde{\beta}_k(\eta)$ and then taking $\eta \to \infty$, as in \eref{eq:betak_from_betatildek}.  Provided that particle production is inefficient such that $|\beta_k|^2 \ll 1$ and $|\alpha_k|^2 \approx 1$, the solution is written as the integral 
\bes{\label{eq:Eq116}
    \beta_k & = - \frac{1}{2}  \int_{-\infty}^{\infty} \! \dd\eta \, \frac{\omega_k^\prime(\eta)}{\omega_k(\eta)} \, \ee^{-2 \ii \int_{-\infty}^\eta \! \dd \eta^\prime \omega_k(\eta^\prime)} 
    \per
}
\Eref{eq:Eq116} resembles a Fourier transform, and similar to a Fourier transform it selects out the amplitude of oscillations in $\omega_k^\prime / \omega_k$.  This leads to $\beta_k \approx \beta_k^{(1 \to 2)} + \beta_k^{(2 \to 2)} + \beta_k^{(3 \to 2)} + \cdots$, where $\beta_k^{(n \to 2)}$ corresponds to the contribution from applying the stationary phase approximation at a time $\eta_\ast$ when $n^2 m_\cphi^2 = k^2/a^2(\eta_\ast) + m_X^2$.  Each of these terms corresponds to a scattering channel where $n$ inflaton particles annihilate into two $X$ particles.  

CGPP from homogeneous metric oscillations at the end of inflation was discussed first by \rref{Vilenkin:1985md,Suen:1987gu} and later by \rref{Markkanen:2015xuw}.  The scattering description (gravitational annihilation) was developed by \rref{Ema:2015dka,Ema:2016hlw,Ema:2018ucl}, and its equivalence with the Bogolubov formalism was established by \rref{Chung:2018ayg,Basso:2021whd,Kaneta:2022gug}.  Phenomenological applications for the production of cosmological relics such as dark matter were explored in these studies and many others \cite{Ema:2019yrd,Garcia:2021iag,Clery:2021bwz,Mambrini:2021zpp,Barman:2021ugy,Clery:2022wib,Brandenberger:2023zpx}.  Of particular note is the relative importance of particle production via gravity-mediated thermal freeze-in from the plasma and gravity-mediated inflaton annihilations.  We emphasize that typically the latter dominates over the former since more energy is available in the inflaton at the end of inflation than in the plasma \cite{Clery:2021bwz}. 

\subsection{Modeling: interference effects}
\label{sub:interference}

The discussion in \sref{sub:inflaton_annihilations} explains that, while the inflaton field $\cphi$ oscillates during the epoch of reheating, CGPP can be described as gravity-mediated inflaton annihilations.  Particles of species $X$ are produced via reactions such as $\cphi \cphi \to XX$ and $\cphi \cphi \cphi \to XX$ with an $s$-channel graviton exchange.  In familiar calculations of $S$-matrix elements, interference effects can arise when various intermediate states (usually corresponding to distinct Feynman graphs) contribute to transitions between the same pair of initial and final states; in short, if $\mathcal{M} = \mathcal{M}_1 + \mathcal{M}_2$ then $|\mathcal{M}|^2 = |\mathcal{M}_1|^2 + |\mathcal{M}_2|^2 + 2 \mathrm{Re}[ \mathcal{M}_1^\ast \mathcal{M}_2]$, where the last term is the interference.  It appears that reactions such as $\cphi \cphi \to XX$ and $\cphi \cphi \cphi \to XX$ cannot interfere since they have different initial states (two versus three inflaton particles) as well as different final states ($E_X \approx m_\cphi$ versus $E_X \approx 3 m_\cphi / 2$).  However this conclusion is premature.  

\rref{Basso:2022tpd} argued that interference effects can arise in CGPP and leave their imprint on the spectrum of produced particles.  This conclusion is most easily understood in the Bogolubov formalism since the stationary phase approximation gives $\beta_k \approx \beta_k^{(2 \to 2)} + \beta_k^{(3 \to 2)}$ leading to a comoving number density $a^3 n_k \propto |\beta_k|^2 = |\beta_k^{(2 \to 2)}|^2 + |\beta_k^{(3 \to 2)}|^2 + 2 \mathrm{Re}[ \beta_k^{(2 \to 2) \ast} \beta_k^{(3 \to 2)}]$ where the last term is the interference.  For one to understand the interference in the scattering formalism, it is useful to recognize two important aspects of the calculation.  First, while the inflaton field oscillates after inflation, the quantum mechanical description of this system is a coherent state \cite{Albrecht:1992kf}, \ie, a state of indefinite particle number that is constructed as a coherent superposition of $n$-particle states.  The notation $\cphi \cphi \to XX$ and $\cphi \cphi \cphi \to XX$ is potentially misleading, since it suggests different initial states, whereas both reactions correspond to the same initial state, and they can interfere.  Second, the cosmological expansion reduces the energy of the produced $X$ particles.  As a result, higher-energy $X$ particles produced earlier can interfere with lower-energy $X$ particles produced later since they all have the same energy at late times.  For instance, if production occurs via the three-to-two channel at a time when the redshift is $a_{3\to2}$, then the comoving energy of the $X$ particles is $a_{3\to2} (3 m_\cphi / 2)$, and similarly the two-to-two channel gives $a_{2\to2} m_\cphi$.  These comoving energies are equal provided that $a_{2\to2} = (3/2) a_{3\to2}$, and the scatterings can interfere.   

These interference effects lead to characteristic ``fringes'' in the spectrum $n_k$ of gravitationally produced particles.  A combination of constructive and destructive interference leads to ``wiggly'' spectra, which have been observed in several numerical studies of CGPP; see \rref{Basso:2022tpd} for references and additional details.  Inspection of the curve marked $\tk^{-3/2}$ in \fref{fig:es-conformal} shows such wiggles.  They are not pronounced for quadratic inflation because the symmetry of the potential precludes the $3\to2$ process, so interference first appears in $\beta_k^{(2\to2)} + \beta_k^{(4\to2)}$.  The $3\to2$ process is allowed for other models like hilltop inflation, and the interference fringes are much more pronounced. 

\subsection{Relics: dark matter}
\label{sub:dark_matter}

Gravitational particle production offers an explanation for the origin of the cold dark matter.  
All of the evidence for the presence of dark matter in the Universe today and throughout cosmic history derives from dark matter's gravitational influence on light, ordinary matter, and itself.  Moreover, various observations constrain dark matter's nongravitational interactions with itself and ordinary matter.  Therefore, one can ask whether the gravitational force may have also governed the creation of dark-matter particles.  

Various studies by many groups have explored the idea of gravitationally produced dark matter over the years.  Some of the earliest studies were undertaken by \rref{Chung:1998zb,Kuzmin:1998uv,Chung:1998ua,Chung:1998bt,Kuzmin:1998kk,Chung:2001cb}, which focused on superheavy dark-matter named {\sc WIMPzillas} \cite{Klapdor-Kleingrothaus:1999yie,Kolb:1998ki}.  The phrase {\sc WIMPzilla} was originally used to describe any superheavy and weakly coupled dark-matter candidate regardless of its origin; different production scenarios were considered including thermal freeze-in, nonthermal production during preheating, direct production from inflaton decay, and gravitational production.  The language was intended to contrast with the prevailing weakly interacting massive particle (WIMP) paradigm in which thermal freeze-out of heavier particles requires larger coupling to yield the same relic abundance \cite{Griest:1989wd}.  Here we focus on {\sc WIMPzilla} dark matter created by CGPP.  In the {\sc WIMPzilla} regime, the dark matter's mass is assumed to reside close to the inflationary Hubble scale and the inflaton mass scale $m_\chi \approx \He \approx m_\cphi$.  This is because CGPP is strongly suppressed for $m_\chi \gg \He \approx m_\cphi$, thereby making the production of heavier particles ineffective [however, see \rref{Kannike:2016jfs,Kim:2021ida,Kim:2023wuk}], and because several considerations disfavored $m_\chi \ll \He \approx m_\cphi$.  For spin-$\half$ particles and conformally coupled spin-$0$ particles, CGPP is suppressed for small $m_\chi$, making it difficult to accommodate all the dark matter for masses below the {\sc WIMPzilla} regime; see \sref{sub:spin_one_half}.  However, for minimally coupled spin-$0$ particles CGPP is effective even for $m_\chi \ll \He \approx m_\cphi$, but such models were found to be in conflict with CMB limits on dark-matter isocurvature \cite{Chung:2004nh}, and these constraints were evaded by raising the dark matter's mass up to the {\sc WIMPzilla} regime; see \sref{sub:isocurvature}.  More recent work has explored models that allow for efficient CGPP of light dark matter away from the classic {\sc WIMPzilla} regime.  For scalar dark matter introducing a nonminimal gravitational coupling suppresses isocurvature on scales probed by the CMB \cite{Fairbairn:2018bsw,Kolb:2022eyn,Garcia:2023qab}.  In addition, for spin-$1$ particles the isocurvature constraints are evaded since the evolution of vector and scalar fields differs for nonrelativistic modes outside the horizon \cite{Graham:2015rva}.  

If dark matter were to interact only gravitationally, this could be considered a ``nightmare scenario'' from the perspective of dark-matter phenomenology.  For instance, there would be no sign of dark matter at direct detection experiments searching for nuclear recoil, no detectable signals of dark-matter annihilation in the Galactic halo, and no significant production of dark matter at high-energy collider experiments.  Instead, one would need to search for dark matter's gravitational influence in the laboratory or in cosmological observables; see the forthcoming discussion.  

However, from the perspective of effective field theory one generally expects dark matter to experience nongravitational interactions.  Provided that these interactions are sufficiently weak, they would not play a significant role in dark-matter production, leaving the dark matter's relic abundance to be determined by CGPP.  The relative importance of nongravitational production (for example, thermal freeze-in) and CGPP must be assessed on a model-by-model basis to identify the regions of parameter space where each process dominates \cite{Kolb:2017jvz,Chianese:2020yjo}.  

The presence of nongravitational interactions also threatens to disrupt the dark-matter stability.  This issue is particularly relevant for superheavy dark matter, since a larger mass typically translates into a shorter lifetime for unstable particles.  For illustration consider a real scalar field $\chi$, serving as the dark-matter candidate, that couples to the standard-model Higgs doublet $\aphi$ via $\Lcal_\mathrm{int} = - c_n \Mpl^{3-2n} \chi (\cphi^\dagger \Phi)^n$ where $c_n$ is an $O(1)$ coefficient and $n \geq 1$ is an integer.  Provided that the decay $\chi \to n \Phi + n \bar{\Phi}$ is kinematically accessible, the decay rate is parametrically $\Gamma_\chi \sim m_\chi^{4n-5} \Mpl^{6-4n}$.  Comparing the lifetime $\tau_\chi \sim 1 / \Gamma_\chi$ to the age of the Universe today ($t_U \approx 13.7 \, \mathrm{Gyr}$) gives 
\ba{
    \tau_\chi / t_U 
    \sim \begin{cases}
    10^{-67} \, m_{12} & , \quad n = 1 \\[1ex] 
    10^{-41} \, m_{12}^{-3} & , \quad n = 2 \\[1ex] 
    10^{-16} \, m_{12}^{-7} & , \quad n = 3 \\[1ex] 
    10^{10} \, m_{12}^{-11} & , \quad n = 4 
    \end{cases}
}
where $m_{12} = m_\chi / 10^{12} \GeV$.  For $\chi$ to be cosmologically long-lived ($\tau_\chi / t_U \gg 1$) with a superheavy mass $m_\chi \approx 10^{12} \GeV$, it is necessary to forbid (or to at least strongly suppress) the operators with $n = 1, 2, 3$.  One can argue that when constructing the effective field theory, we should impose a symmetry to forbid operators that would mediate $\chi$ decay, for example, a discrete $\mathbb{Z}_2$ symmetry that takes $\chi \to -\chi$ and leaves other fields unchanged.  However, this approach runs counter to the conjectured absence of global symmetries in theories of quantum gravity \cite{Mambrini:2015sia}.  Alternatively if perturbative decay is forbidden and the decay must proceed through nonperturbative phenomena then the rate for instantonlike transitions can be estimated as $\Gamma_\chi \sim m_\chi \ee^{-4 \pi / \alpha_\chi}$, where $\alpha_\chi$ is the corresponding non-Abelian gauge coupling \cite{Kuzmin:1997jua}; for $m_\chi = 10^{12} \GeV$ and $\alpha_\chi = 0.1$ the lifetime exceeds the current age of the Universe.  

\subsection{Relics: gravitational reheating}
\label{sub:grav_reheating}

It is often assumed that inflation ends when the inflaton field finds a local minimum of its potential and begins oscillating.  During the ensuing epoch of reheating, energy is transferred from the inflaton condensate into a relativistic plasma though a combination of direct perturbative inflaton decay and nonlinear processes (preheating) \cite{Traschen:1990sw,Shtanov:1994ce,Kofman:1994rk,Kofman:1997yn,Amin:2014eta,Lozanov:2019jxc}.  Reheating is said to be complete when the plasma constitutes the Universe's dominant energy component.  In this scenario the plasma inherits most of its energy from the inflaton, but a fraction may have arisen directly from CGPP \cite{Passaglia:2021upk}, although typically this fraction is no larger than order $\Hinf^2 / \Mpl^2$ and is small enough to be neglected.  

However, one can imagine that the inflaton's energy is depleted before it can be transferred into a plasma.  For example, if the inflaton's potential minimum is locally quadratic, then its energy density redshifts down slowly like matter $\rho_\cphi \propto a^{-3}$; however, if the inflaton's potential is flat with $V(\cphi) = 0$, then the inflaton enters a phase of kination while its energy redshifts away quickly as $\rho_\cphi \propto a^{-6}$.  In the latter scenario reheating must be accomplished without relying on energy transfer from the inflaton.  

Gravitational reheating is a scenario in which the plasma's energy arises primarily from CGPP rather than from the inflaton \cite{Ford:1986sy}.  Since the pioneering work by Ford, this idea has been extensively explored \cite{
Copeland:2000hn
,Feng:2002nb
,Liddle:2003zw
,Sami:2003my
,Tashiro:2003qp
,Chun:2009yu
,Nishizawa:2014zra
,Nishi:2016wty
,Dimopoulos:2018wfg
,Hashiba:2018iff
,Figueroa:2018twl
,Haro:2018zdb
,Hashiba:2018tbu
,Opferkuch:2019zbd
,Kamada:2019ewe
,Bettoni:2019dcw
,Bettoni:2021zhq
,Haque:2022kez
,Haque:2023yra
,Laverda:2023uqv
,Barman:2023opy
,Haque:2023zhb
}, frequently in the context of braneworld inflation, quintessential inflation, and curvaton scenarios.  While direct reheating into the standard model (for example, via the Higgs) has been considered \cite{
Figueroa:2016dsc
}, most studies rely on a new, heavy, and unstable particle to experience CGPP and mediate reheating into the standard model.  
A wide range of reheating temperatures is accessible as the particle's mass, and nonminimal coupling to gravity may be varied.  

An important constraint on gravitational reheating involves the amplification of gravitational-wave radiation.  Since a stiff equation of state ($w_\cphi > 1/3$) is required to rapidly deplete the inflaton's energy density, the inflationary gravitational waves develop a blue-tilted spectrum that rises toward higher frequencies \cite{
Giovannini:1998bp
,Giovannini:1999bh
,Riazuelo:2000fc
,Sahni:2001qp
,Boyle:2007zx
}.  
The additional energy in high-frequency gravitational waves is constrained by big-bang nucleosynthesis and the cosmic microwave background anisotropies that probe dark radiation, implying that $\Omega_\mathrm{gw} h^2 \lesssim 10^{-6}$ \cite{
Sendra:2012wh
,Figueroa:2018twl
}.  
The strong implications for gravitational reheating were emphasized by several groups \cite{
Tashiro:2003qp
,Nishizawa:2014zra
,Nishi:2016wty
,Figueroa:2018twl
}, who conclude that the simplest implementations of gravitational reheating are strongly constrained.  Viable gravitational reheating may be possible if there are many particle species experiencing CGPP such that the energy that they collectively carry counteracts the energy carried by the two polarizations of the gravitational-wave radiation.  

\subsection{Relics: baryogenesis}
\label{sub:baryogenesis}

The cosmological excess of matter over antimatter is one of the long-standing mysteries of modern cosmology \cite{Wilczek:1980vy}.  It is usually assumed that this imbalance arose dynamically through a process called baryogenesis as a consequence of the physical laws governing the particle content of the Universe \cite{Kolb:1979qa}.  Successful baryogenesis requires particle interactions that violate certain conservation laws,namely, baryon number $\mathsf{B}$, charge conjugation $\mathsf{C}$, and charge-parity conjugation $\mathsf{CP}$, as well as a departure from thermal equilibrium \cite{Sakharov:1967dj}.  Since CGPP leads to the creation of particles in the early Universe, it is natural to ask whether CGPP may be responsible for the origin of the matter-antimatter asymmetry.  

If the interactions of particles and antiparticles with gravity respect the conservation laws $\mathsf{B}$, $\mathsf{C}$, and $\mathsf{CP}$ then baryogenesis cannot result from CGPP directly since the first two Sakharov criteria are not satisfied.  However, CGPP easily provides the out-of-equilibrium conditions required by the third Sakharov criterion.  Consequently, one can imagine a framework for CGPP baryogenesis in which CGPP creates a population of self-conjugate $X$ particles that decay into particles and antiparticles through interactions violating $\textsf{B}$, $\textsf{C}$, and $\textsf{CP}$, thereby generating a net baryon number \cite{Bambi:2006hp}.  Similar ideas using nonthermal particle production during preheating were first developed by \rref{Kolb:1996jt,Kolb:1998he}.  Alternatively, $X$ may decay by interactions that instead violate the lepton number $\mathsf{L}$, thereby generating a lepton asymmetry that is converted into a baryon asymmetry by electroweak Chern-Simons number diffusion (electroweak sphaleron), as in models of baryogenesis from leptogenesis \cite{Fukugita:1986hr}. 

Recent studies have explored the implications of CGPP for baryogenesis \cite{Hashiba:2019mzm,Bernal:2021kaj,Co:2022bgh,Barman:2022qgt,Fujikura:2022udt}.  Several of these works investigate the role of CGPP for creating a heavy right-handed Majorana neutrino that is long-lived, decays out of equilibrium, and leads to baryogenesis via leptogenesis.  Similar ideas for baryogenesis have been explored using noncosmological GPP via primordial black hole evaporation \cite{Fujita:2014hha,Hooper:2020otu,Perez-Gonzalez:2020vnz,Datta:2020bht,Barman:2022pdo,Aydemir:2020pao,Bernal:2022pue,Barman:2022gjo}.  

\subsection{Relics: dark radiation}
\label{sub:relics_dark_radiation}
 
The present-day Universe may contain an as-yet-unseen population of dark particles that were relativistic at radiation-matter equality and may still be relativistic today or even massless.  Various theories of new physics put forward different particle candidates for this dark radiation, and they offer different explanations for its origin.  CGPP can contribute to the population of dark radiation in two ways: by directly producing the relativistic particles and by producing parent particles that decay into dark radiation.  

For concreteness consider a family of $N$ massless scalar fields that are minimally coupled to gravity and otherwise weakly coupled to one another and other particles.  Suppose that the CMB-scale modes leave the horizon during inflation when $a = a_\CMB$, that inflation ends and reheating begins when $a = \ae$, that the Universe is effectively matter dominated during reheating ($\wRH = 0$), that reheating completes and radiation-domination begins when $a = \aRH$, and that comoving entropy is subsequently conserved.  In this scenario CGPP leads to a nearly scale invariant energy spectrum of radiation for modes with a comoving wave number $\keq \equiv \aeq \Heq < k < \kRH \equiv \ae^{3/2} \He / \aRH^{1/2}$ that leave the horizon during inflation and reenter during the radiation era before radiation-matter equality.  Upon integration of this spectrum, the cosmological energy fraction today is expected to be 
\bes{\label{eq:Omega_dr}
    \Omega_\mathrm{dr} h^2 & \approx 
    \bigl( 3 \times 10^{-14} \bigr) 
    \biggl( \frac{N}{1} \biggr)  
    \biggl( \frac{\He}{10^{14} \GeV} \biggr)^2 
    \biggl( \frac{\log \kRH / \keq}{60} \biggr) 
}
and insensitive to the reheating temperature.  Observational limits on dark radiation are at the level of $\Omega_\mathrm{dr} h^2 \lesssim 0.05 \, \Omega_\gamma h^2 \approx 2 \times 10^{-6}$, and they derive from its would-be gravitational influence on BBN and the CMB.  For instance, it is customary to write $\Omega_\mathrm{dr} h^2 = \Delta N_\mathrm{eff} \tfrac{7}{8} \bigl( \tfrac{4}{11} \bigr)^{4/3} \Omega_\gamma h^2$, and \textit{Planck} gives $\Delta N_\mathrm{eff} \lesssim 0.2$ \cite{Planck:2018vyg}.  Since we have fiducialized $\He$ to its maximal allowed value given $r < 0.1$, achieving a detectably large $\Omega_\mathrm{dr} h^2$ value would require a large number of particle species ($N \gg 1$).  

In other scenarios a larger population of dark radiation could be obtained.  For instance, if the cosmological equation of state during reheating were stiffer, the high-$k$ modes would redshift differently, thereby boosting the relic abundance.  See the discussion in \sref{sub:grav_reheating} regarding a kination-dominated phase of reheating ($\wRH = 1$).  Alternatively dark radiation could arise indirectly via the decay of gravitationally-produced moduli fields into relativistic particles \cite{Ema:2016hlw}.  This scenario has more parametric freedom since one can vary the moduli mass and decay rate.  While the energy is carried by the nonrelativistic moduli field, it redshifts away only as $a^{-3}$, rather than dark radiation's $a^{-4}$, and if the moduli decay is sufficiently delayed, the abundance of dark radiation is correspondingly enhanced. 

\subsection{Relics: gravitational wave radiation}
\label{sub:relics_grav_wave}

Cosmological inflation predicts a broad spectrum of primordial gravitational-wave radiation \cite{Starobinsky:1979ty,Rubakov:1982df,Fabbri:1983us,Abbott:1984fp}.  On an FLRW background the dispersion relation of a massless spin-$2$ field's transverse polarization modes ($\lambda = \pm 2$) is $\omega_k^2(\eta) = k^2 + a^2 R / 6 = k^2 - a^\pprime / a$, which is familiar from studies of inflationary gravitational-wave production; see \sref{sub:spin_two} and \rref{Baumann:2022mni}.  This is the same dispersion relation that one finds for a massless and minimally coupled scalar field \eqref{eq:omegak_scalar}, and the CGPP calculation is thus identical up to a factor of $2$ on account of the gravitational wave's two polarization modes.  A calculation similar to the one that yielded \eref{eq:Omega_dr} now gives a scale invariant spectrum of gravitational waves across the frequencies $f_\mathrm{eq} < f < f_\RH$ and with the present-day amplitude 
\bes{\label{eq:Eq119}
    \frac{1}{\rho_c} \frac{\dd \rho_{\mathrm{gw},0}}{\dd \ln f} h^2 
    & \approx 
    9 \times 10^{-16}  
    \biggl( \frac{\He}{10^{14} \GeV} \biggr)^2 \\ 
    f_\mathrm{eq} & \approx  5 \times 10^{-17} \Hz  \\ 
    f_\RH & \approx 200 \Hz  \biggl( \frac{\TRH}{10^9 \GeV} \biggr) 
    \per
}
\Eref{eq:Eq119} corresponds to a dimensionless strain of about $h_c(f) \approx 4 \times 10^{-26} (f / \mathrm{Hz})^{-1} (\He / 10^{14} \GeV)$.  Although it is customary to say that inflationary gravitational waves arise from quantum fluctuations of the metric, it can also be said that this radiation arises from the phenomenon of CGPP due to the time-dependent dispersion relation.  

Studies of gravitational-wave creation via CGPP often focus on high-frequency gravitational waves \cite{Grishchuk:1974ny,Yajnik:1990un,Schiappacasse:2016nei} rather than the low-frequency component that can be probed on CMB scales.  Notable recent studies by \rref{Ema:2015dka,Ghiglieri:2015nfa,Nakayama:2018ptw,Ghiglieri:2020mhm,Ema:2020ggo,Ringwald:2020ist,Ema:2021fdz,Ghiglieri:2022rfp,Ghoshal:2022kqp,Ebadi:2023xhq,Barman:2023ymn,Bernal:2023wus,Hu:2024awd} explored gravitational-wave production via inflaton decay, inflaton coherent oscillations and annihilations, secondary effects, and thermal freeze-in \cite{Weinberg:1965nx}.  

Over the past decade the experimental opportunities for gravitational-wave detection have expanded with each passing year.  Since the first detection of gravitational waves in 2015, the LIGO-Virgo-KAGRA Collaboration has proven the technology of gravitational-wave interferometry to measure the radiation created by compact object mergers~\cite{LIGOScientific:2016aoc}.  The planning for future interferometers is already under way.  Proposed ground-based observatories, including the Einstein Telescope~\cite{EinsteinTelescope:2020} and Cosmic Explorer~\cite{Evans:2021gyd}, would improve sensitivity at frequencies of $1-100 \Hz$, while space-based interferometers, including LISA~\cite{Colpi:2024xhw} and DECIGO~\cite{Kawamura:2020pcg}, would extend the search down to millihertz frequencies.  In particular, the success of the LISA Pathfinder mission~\cite{Armano:2016bkm} has fueled enthusiasm over the prospect of probing primordial gravitational waves~\cite{Caprini:2018mtu}.  More recently several pulsar timing observatories, including NANOGrav~\cite{NANOGrav:2023gor}, reported the detection of a signal that could arise from the nanohertz stochastic gravitational-wave background created by the mergers of supermassive black hole binaries or may contain evidence of an exotic cosmological source.  Meanwhile, various strategies are being explored~\cite{Aggarwal:2020olq} to detect high-frequency gravitational-wave radiation using tabletop experiments in the laboratory.  Each of these lines of investigation has the potential to be transformative in our understanding of primordial sources of gravitational-wave radiation.  

\subsection{Relics: primordial black holes}
\label{sub:black_holes}

There are several interesting links between CGPP and primordial black holes (PBHs), particularly in regard to their formation and evaporation~\cite{Erfani:2015rqv}.  Whereas astrophysical black holes form when stars collapse, primordial black holes may have formed in the early Universe due to the collapse of Hubble-scale overdensities \cite{Carr:1974nx,Niemeyer:1999ak}.  The mass scale of PBHs formed in this way depends on the amount of energy contained within a Hubble volume at the time when the overdensity enters the horizon and collapses.  Although CGPP typically leads to small energy density inhomogeneities on cosmological scales, it can generate larger inhomogeneities on short scales; see \sref{sub:inhomogeneity}.  Modes on the Hubble scale at the end of inflation have $k \approx \ke \equiv \ae \He$ and, if these modes collapse soon after the end of inflation, then the black hole mass is around 
\ba{
    M_\BH 
    & \approx \rho_c \He^{-3} 
    \approx 
    \bigl( 0.3 \gram \bigr) 
    \biggl( \frac{\He}{10^{14} \GeV} \biggr)^{-1} 
    \com
}
where $\rho_c = 3 \MPl^2 \He^2$ is the cosmological critical density at the end of inflation.  Such minuscule black holes would emit Hawking radiation \cite{Bekenstein:1973ur,Hawking:1974rv} at a temperature of $T_\BH = \MPl^2 / M_\BH = \He / 3$ and evaporate on a timescale of $\tau_\BH \sim M_\BH^3 / \MPl^4$, assuming they have negligible spin and electromagnetic charge.  If the Universe goes directly into radiation domination at the end of inflation (instantaneous reheating), then the black holes would survive for many Hubble times until the plasma temperature drops to $T_d \sim \Mpl^{5/2} / M_\BH^{3/2} \sim (10^{11} \GeV) (\He / 10^{14} \GeV)^{3/2}$.  Unless $\He$ is small, this evaporation would complete long before BBN ($T \approx \mathrm{MeV}$), making detectable signatures challenging to find. 

Even if PBHs are not formed as a direct consequence of CGPP, the evaporation of PBHs that form in other ways would lead to the gravitational production of particles.  Black holes emit thermal Hawking radiation into various species of particles as long as the particle masses are kinematically accessible ($m_\chi < T_\BH$) and the interactions are not conformal.  This radiation can include the known standard-model elementary particles, as well as hypothetical new particles such as dark matter~\cite{Fujita:2014hha,Lennon:2017tqq,Morrison:2018xla,Hooper:2019gtx}.  For instance, \rref{Hooper:2019gtx} studied the gravitational production of dark matter via black hole evaporation and calculated its cosmological energy fraction, finding
\ba{
    \frac{\Omega h^2}{0.1} \sim 
    \biggl( \frac{m_X}{10^9 \GeV} \biggr)^{-1} 
    \biggl( \frac{M_\BH}{10^8 \gram} \biggr)^{-1} 
    \biggl( \frac{f_\BH}{8 \times10 ^{-14}} \biggr) 
    \com
}
where $m_X$ is the dark matter's mass, $M_\BH$ is the black hole's initial mass before evaporation, and $f_\BH = \rho_\BH / \rho_R$ is the cosmological energy fraction in black holes at $10^{10} \GeV$.  These estimates demonstrate that the gravitational production of massive and stable particles via black hole evaporation is a viable mechanism to explain the origin of dark matter. 

\subsection{Observational: laboratory tests}
\label{sub:lab_test}

The phenomenon of CGPP is not expected to have any discernible impact on the scale of laboratories in the Universe today.  Owing to the enormous hierarchy between everyday length scales and the present cosmological horizon, the cosmological expansion plays no role.  Technically speaking, the time-dependent frequency $\omega_k^2(\eta)$ [see \eref{eq:omegak_scalar}] is approximately constant on the timescales of an experiment and mode evolution remains nearly adiabatic without particle production. 

Nevertheless the gravitationally coupled relics produced via CGPP could potentially be probed by experiments on Earth \cite{Carney:2022gse}.  The expected flux $F_\chi \approx \rho_\chi v_\chi / m_\chi$ of $\chi$ particles transiting a detector is 
\ba{\label{eq:Eq122}
    F_\chi \approx 
    \biggl( \frac{200}{(10 \, \mathrm{cm})^2 (1 \, \mathrm{yr})} \biggr) 
    \biggl( \frac{m_\chi}{10^{14} \, \mathrm{GeV}} \biggr)^{-1} 
    f_\rho 
    f_v 
    \per
}
In \eref{eq:Eq122} the local mass density $\rho_\chi$ is assumed to be a fraction $f_\rho$ of the fiducial local dark-matter density $\rho_\mathrm{dm} = 0.3 \GeV / \mathrm{cm}^3$, and the mean velocity $v_\chi$ is assumed to be a fraction $f_v$ of the fiducial local dark-matter velocity $v_\mathrm{dm} = 220 \, \mathrm{km} / \mathrm{sec}$.  At a larger mass $m_\chi = \Mpl = 2.4 \times 10^{18} \GeV$, the flux drops to $0.9 / (100 \, \mathrm{cm})^2 / \mathrm{yr}$.  These fluxes are large enough that several CGPP relics could be incident upon a meter-scale experiment over a year.  

The challenge for such an experiment is achieving the sensitivities needed to detect particles with gravitational-strength interactions while discriminating signal from noise.  An experimental program has been put forward to accomplish this formidable task.  The Windchime Project \cite{Carney:2019pza,Monteiro:2020wcb,Carney:2020xol,Windchime:2022whs} seeks to employ optically or magnetically levitated sensors to precisely measure the location of test masses in a three-dimensional array and thereby detect the passage of dark matter as it exerts a subtle tug on the sensors.  Sensitivity projections indicate that Windchime would be most sensitive to gravitationally interacting ultraheavy dark particles with masses at or above the Planck scale ($\sim 10^{19} \GeV$) \cite{Carney:2019pza}.  Since CGPP is strongly suppressed above the inflaton mass, typically not much larger than $10^{14} \GeV$, it would be difficult to explain the origin of such heavy dark particles via CGPP.  However, if the interaction is nongravitational, then Windchime's sensitivity widens to lower masses, possibly bringing gravitationally produced dark matter within reach.  Optomechanical sensors have long been a powerful tool in precision experiments \cite{Gonzalez-Ballestero:2021gnu}, including gravitational-wave detection, and they may yet offer a unique handle in searches for new physics \cite{Moore:2020awi}.  

\subsection{Observational: dark matter isocurvature}
\label{sub:isocurvature}

Observations of the CMB anisotropies and large-scale structure are compatible with a power spectrum of primordial density perturbations consisting of only an adiabatic mode, and they constrain the amplitude of isocurvature modes.  In other words, the primordial dark-matter inhomogeneities should align in space with the plasma inhomogeneities: overdensities with overdensities, and underdensities with underdensities.  For models of dark matter that is produced out of the plasma, such as thermal freeze-out or freeze-in, there is a negligible isocurvature \cite{Bellomo:2022qbx}.  However, if the dark matter arises from a physical process that is uncorrelated with the origin of the plasma (on cosmological scales), which occurs in some models of CGPP during inflation, then there may be an unacceptably large isocurvature \cite{Chung:2004nh,Chung:2011xd,Chung:2013sla,Tenkanen:2019aij,Herring:2019hbe,Herring:2020cah,Ling:2021zlj,Boyanovsky:2021ija,Kolb:2022eyn,Garcia:2023qab,Garcia:2023awt}.  In this section we define the relevant isocurvature observable, explain how it can be calculated for models of dark matter produced via CGPP, and discuss the constraints imposed by CMB measurements.  See \aref{app:cosmo_perturb} for a review of cosmological perturbations that includes an introduction to adiabatic and isocurvature perturbations.  

The isocurvature perturbation between dark matter (denoted by $X$ with $w_X = 0$) and radiation (denoted by $R$ with $w_R = 1/3$) is given by 
\ba{
     S_{XR} = \frac{\drho_X}{\bar{\rho}_X} - \frac{3}{4} \frac{\drho_R}{\bar{\rho}_R} 
     \per
}
This is a gauge-invariant variable that remains unchanged (to linear order) under a coordinate transformation.  If the initial perturbations were adiabatic, the dark matter's energy density contrast $\delta_X = \drho_X / \bar{\rho}_X$ would be linked to the radiation's $\delta_R = \drho_R / \bar{\rho}_R$ via the relation $\delta_X = (3/4) \delta_R$.  When this relation is violated, a dark-matter-radiation isocurvature is said to be present.  

To calculate the isocurvature it is convenient to work in the uniform density gauge, which is defined by the coordinate system in which the inflaton perturbations vanish [$\drho_\phi(\eta,\xvec) = 0$].  If the plasma arises from the decay of the inflaton, then the radiation perturbations also vanish in this gauge [$\drho_R(\eta,\xvec) = 0$].  It follows that $S_{XR} = \delta_X$ where $\delta_X(\eta,\xvec) = \drho_X / \bar{\rho}_X$ is the dark matter's energy density contrast \cite{Chung:2011xd}.  Thus, the gauge-invariant isocurvature power spectrum equals the power spectrum of the dark matter's energy density contrast in this gauge $\Delta_S^2 = \Delta_\delta^2$; an expression for the latter quantity appears in \eref{eq:Deltadelta}.  

Measurements of the CMB anisotropies are compatible with vanishing isocurvature.  These constraints are typically expressed in terms of the dimensionless isocurvature power spectrum $\Delta_S^2(\eta,k)$.  Assuming that the curvature and isocurvature are uncorrelated (``axion I''), \textit{Planck} reported a constraint of \cite{Planck:2018jri}
\ba{\label{eq:Planck_isocurvature}
    \frac{\Delta_S^2}{\Delta_\Rcal^2 + \Delta_S^2} < 0.038 \quad \text{at $95\%$ CL} 
    \per
}
In \eref{eq:Planck_isocurvature} the power spectra are evaluated at recombination $\eta = \eta_\mathrm{rec}$ for CMB-scale modes $k = 0.05 \, a_0 \Mpc^{-1}$ and $\Delta_\Rcal^2 \approx 2.1 \times 10^{-9}$ is the amplitude of the curvature power spectrum.  In terms of the dark-matter density contrast, \eref{eq:Planck_isocurvature} implies that $\Delta_\delta^2 < 8.0 \times 10^{-11}$ at CMB scales.  

For dark matter arising from CGPP, the scale and shape of the isocurvature power spectrum depends on the spin, mass, and gravitational coupling of particles.  For instance, a conformally coupled scalar field has $\Delta_\delta^2 \propto k^3$, which may give $O(1)$ isocurvature at small scales, but it implies a negligible isocurvature at CMB scales; see \sref{sub:inhomogeneity}.  A similar result is obtained for fields with nonzero spin, which have blue-tilted spectra; see \sref{sec:Spin}.  Therefore, CMB limits on isocurvature are most constraining for models of minimally coupled ($\xi = 0$) scalar dark matter.  If the dark particles are light ($m \ll \He \lesssim \Hinf$), then the power spectrum is nearly scale invariant (see \sref{sub:inhomogeneity}), and its amplitude at CMB scales comes into conflict with the CMB limits, but lifting the dark matter's mass to be $m \gtrsim \He$ tilts the power spectrum toward blue and evades the isocurvature constraints \cite{Chung:2004nh}.  Alternatively, introducing a nonminimal coupling to gravity also leads to a blue tilt and evades these constraints \cite{Kolb:2022eyn,Garcia:2023qab}, even for $\xi = O(0.01)$.  These models have inspired searches for blue-tilted isocurvature in cosmological observables \cite{Chung:2015tha,Chung:2016wvv,Chung:2017uzc}.  

\subsection{Observational: non-Gaussianity}
\label{sub:nonGaussianity}

Observations of the CMB are compatible with a spectrum of primordial curvature perturbations such that the amplitude of each Fourier mode can be modeled as an independent stochastic (random) variable with a Gaussian probability distribution centered at zero.  These Gaussian initial conditions are a hallmark prediction of single-field slow-roll inflation \cite{Mukhanov:1990me}.  Their Gaussian nature follows from the fact that the inflaton is weakly coupled to itself and other fields, and a free scalar field has a Gaussian wave functional in its ground state \cite{Maldacena:2002vr}.  If the theory is extended to include nongravitational interactions, the primordial curvature perturbations develop a larger non-Gaussian component.  This non-Gaussianity manifests in CMB anisotropy observables at the level of three-point (and higher-point) correlation functions, which are known as the bispectrum (trispectrum, etc.).  Measurements of the CMB anisotropies are compatible with Gaussian initial conditions and thereby constrain theories that predict appreciable non-Gaussianity \cite{Planck:2019kim}.  

In theories that extend the inflaton sector to include additional particles at the same mass scale, which is known as quasi-single-field inflation, an appreciable non-Gaussianity can develop in the curvature perturbations \cite{Chen:2009zp,Chen:2009we,Chen:2010xka,Baumann:2011nk,Noumi:2012vr,Arkani-Hamed:2015bza,Baumann:2017jvh,Arkani-Hamed:2018kmz}.  In this way inflationary observables act as a ``cosmological collider'' experiment to probe particles and forces whose high mass scale puts them beyond the reach of terrestrial high-energy experiments.  

In theories where dark matter arises from CGPP, it is natural to consider new particle species having masses around the inflationary Hubble scale $m_\chi \approx \Hinf$.  For instance, the gravitational production of spin-$\half$ particles is suppressed for $m_\chi \ll \Hinf$ since the fermion mass scale is the source of conformal symmetry breaking; see \sref{sub:spin_one_half}.  \rref{Chung:2011xd} [see also \rref{Hikage:2008sk,Chung:2013rda,Li:2019ves}] found that the same theories in which dark matter is produced via CGPP can also lead to an appreciable CMB non-Gaussianity in the limit of the squeezed local triangle form.  For instance, a benchmark parameter point predicts that $f_\text{\sc nl}^\mathrm{local} \sim 30$, which is now excluded by Planck's constraint $f_\text{\sc nl}^\mathrm{local} = -0.9 \pm 5.1$ \cite{Planck:2019kim}.  CGPP can also lead to rare ``hot spots'' in the CMB \cite{Kim:2021ida,Kim:2023wuk}.  These examples illustrate the capability of current and future probes of CMB non-Gaussianity to test CGPP.  

\subsection{Observational: small-scale structure}
\label{sub:small_scale_structure}

If the particles that experience CGPP are going to provide a viable candidate for dark matter, then they must be sufficiently cold at the time of radiation-matter equality.  For superheavy particles with $m_X \approx \He$ this is easily accomplished.  Even if the spectrum after production is peaked at a high comoving wave number, say $k \approx \ae \He$ corresponding to Hubble-scale modes at the end of inflation, the characteristic velocity today is still extremely small: 
\ba{\label{eq:Eq125}
    v_X 
    & = k / a_0 m_X 
    \approx  5 \times 10^{-32} 
    \biggl( \frac{\He}{10^{14} \GeV} \biggr)^{1/3} 
    \\ & \qquad \qquad \qquad \times 
    \biggl( \frac{\TRH}{10^{9} \GeV} \biggr)^{1/3} 
    \biggl( \frac{m_X}{10^{14} \GeV} \biggr)^{-1} 
    \per
    \nonumber 
}
In \eref{eq:Eq125} the redshift factors are calculated by assuming a matter-dominated epoch of reheating ($\wRH = 0$).  However, for models in which gravitational production remains efficient even for $m_X \ll \He$, the velocity can be much larger.  For instance, taking $m_X = 10^{-12} \GeV = \mathrm{meV}$ gives $v_X \approx 0.5 \times 10^{-5} \approx 1.5 \ \mathrm{km} / \mathrm{sec}$.  

Although $v_X \approx 10^{-5}$ still corresponds to extremely nonrelativistic particles, there is nevertheless a problem with such ``fast'' dark matter.  For successful structure formation, dark-matter particles must cluster through their mutual gravitational attraction, thereby allowing overdensities to grow and form halos.  However, if dark matter travels too quickly it will ``escape'' from its initial overdensity before structures can form.  In an FLRW spacetime a free particle travels a distance 
\ba{\label{eq:Eq126}
    l_\mathrm{fs}(t) 
    = a(t) \int_0^t \! \dd t^\prime \, \frac{v(t^\prime)}{a(t^\prime)} 
    = a \int_0^a \! \dd a^\prime \, \frac{v(a^\prime)}{(a^\prime)^2 H(a^\prime)} 
}
in time $t$ called the free-streaming length.  For the values of $\He$, $\TRH$, and $m_X = \mathrm{meV}$ used in \eref{eq:Eq126}, the free-streaming length today is $l_\mathrm{fs}(t_0) \approx 20 \Mpc$, and it accumulates mostly during the radiation era.  The free-streaming length indicates the size of structures whose formation are inhibited by the dark matter's speediness.  

There are various probes of cosmologically small-scale structure in the spatial distribution of matter \cite{Tulin:2017ara,Drlica-Wagner:2022lbd}.  For instance, observations of Lyman-$\alpha$ absorption due to intergalactic neutral hydrogen in the spectra of distant quasars, \ie, the Lyman-$\alpha$ forest, are compatible with an $\Lambda$CDM-like matter power spectrum at wave numbers below $k \approx 10 a_0 h \Mpc^{-1}$ \cite{Viel:2013fqw}.  These measurements place upper bounds on the free-streaming length of dark matter, which are frequently used to constrain warm dark matter \cite{Irsic:2017ixq}, but they also lead to constraints on $m_X$ in the CGPP production scenario.  A recent study by \rref{Garcia:2023qab} assessed the impact of Lyman-$\alpha$ observations on a theory of nonminimally coupled scalar dark matter that is formed by CGPP during inflation and reheating; see also \rref{Haque:2021mab,Garcia:2022vwm}.  Their parameter space constraints are reproduced here as \fref{fig:xi_dependence}.  Observe that gravitationally produced dark matter with a mass below $m_\chi \sim 10^{-12} \GeV$ (rising to $10^{-7} \GeV$ for larger nonminimal coupling $\xi$) would be in conflict with Lyman-$\alpha$ observations and is thereby ruled out.  

\subsection{Observational: signatures of late time decay}
\label{sub:late_decay}

If the particles arising from CGPP are unstable but long-lived, then their late-time decay into visible-sector radiation could lead to signatures in cosmological and astrophysical observables.  Here we remark on three possible probes of such an energy injection: nucleosynthesis, CMB spectrum and anisotropies, and cosmic rays.  

The observed abundances of the light elements are compatible with the predictions of BBN \cite{RevModPhys.88.015004}.  The standard BBN calculation assumes that the universe is filled with a plasma of standard-model particles at temperatures around a few mega-electron-volts.  This good agreement between observation and theory leads to constraints on the presence of an additional particle species that decay and inject energy into the plasma \cite{Cyburt:2002uv,Kawasaki:2017bqm,Fong:2022cmq}.  

The energy spectrum of the CMB is observed to agree well with a Planck thermal blackbody distribution \cite{Chluba:2019kpb}.  In the standard $\Lambda$CDM cosmology, this thermal distribution arises before recombination because photons couple to the electron-ion plasma through single-Compton and double-Compton scattering, which leads to a common temperature for all species and a vanishing chemical potential for the photon.  However, double-Compton scattering goes out of equilibrium before single-Compton scattering, and during this time interval the photon number is approximately conserved.  An energy injection would then lead to a nonzero photon chemical potential (spectral distortion) \cite{Hu:1993gc}, which is constrained by measurements of the CMB energy spectrum by COBE/FIRAS \cite{Fixsen:1996nj}.  Similarly, energy injection from large-scale inhomogeneities can leave an imprint on CMB anisotropies \cite{Ali-Haimoud:2016mbv,Acharya:2019uba}.  

If superheavy particles are decaying in the present-day Universe, their visible-sector decay products would correspond to highly energetic photons, protons, electrons, positrons, and neutrinos.  Of particular interest is the production of ultra-high-energy cosmic ray (UHECR) protons above the Greisen-Zatsepin-Kuzmin (GZK) cutoff $E_\mathrm{cr} \sim 10^{10} \GeV$ \cite{Greisen:1966jv,Zatsepin:1966jv}.  Since protons scatter on CMB photons via a $\Delta$ resonance, the Universe is not transparent to UHECR protons on cosmological distances.  Conversely, a detection of UHECR protons could be interpreted as evidence for a nearby source, such as superheavy dark-matter decays within the local halo \cite{Kuzmin:1997jua,Kuzmin:1998uv,Kuzmin:1998kk}.  More generally, searches with cosmic ray, gamma ray, and neutrino telescopes are probing long-lived superheavy dark matter \cite{Aloisio:2015lva,Kachelriess:2018rty,Ishiwata:2019aet,Chianese:2021htv,Chianese:2021jke,Das:2023wtk}. 

\section{Outlook}
\label{sec:Outlook}

How does a strong gravitational field lead to particle creation?  What are the cosmological implications of these particles and their observational signatures?  For nearly the past 100 years, these questions have guided the study of cosmological gravitational particle production.  Today, a vibrant community of researchers is working to enrich our knowledge of CGPP as a mechanism and to explore novel applications and implications of this phenomenon.  With this review we hope to provide a tool kit that another generation of researchers can employ to pursue their own studies of CGPP, derive predictions for cosmological relics, explore the implications for new directions beyond the standard model, and propose observational tests.  

\begin{acknowledgments}

We thank our colleagues and collaborators for feedback over the many years while this review was taking shape.  We appreciate Mustafa Amin, Daniel Chung, Leah Jenks, Siyang Ling, Evan McDonough, and Sarunas Verner for taking the time to provide comments on a draft of the article.  E.W.K. acknowledges the hospitality of The Institute for Theoretical Physics in Madrid, where part of this work was done.  The work of E.W.K. was supported in part by U.S.~Department of Energy Contract No.~DE-FG02-13ER41958.  A.J.L. is supported in part by the National Science Foundation under Award No.~PHY-2114024. 

\end{acknowledgments}

\appendix

\section{General relativity}
\label{app:general_relativity}

This appendix provides a refresher on general relativity (GR), the relativistic field theory of gravity that describes its interactions with radiation and matter.  There are many notable textbooks on GR, including:  
\textit{Gravitation and Cosmology} \cite{Weinberg:1972kfs}, 
\textit{Gravitation} \cite{Misner:1973prb}, 
\textit{The Classical Theory of Fields} \cite{Landau:1975pou}, 
\textit{General Relativity} \cite{Wald:1984rg}, 
\textit{Quantum field theory in curved spacetime and black hole thermodynamics} \cite{Wald:1995yp}
\textit{Geometry, Topology, and Physics} \cite{Nakahara:2003nw}, \textit{Supergravity} \cite{Freedman:2012zz}, and 
\textit{Spacetime and Geometry} \cite{Carroll:2004st}.  
 
\para{Sign and unit conventions} 
Different researchers adopt different conventions when writing the equations of GR.  A table summarizing frequently encountered sign choices was provided by \rref{Misner:1973prb}, and an appendix of \rref{Freedman:2012zz} included additional signs and phases that arise in the frame-field formalism.   In these appendixes we follow \rref{Freedman:2012zz} and use the symbols $\alpha$, $\beta$, $\s{\ast}$, $\s{1}$, $\s{2}$, $\cdots$, $\s{8}$ to simultaneously track the various sign choices; one can set $\alpha = \pm 1, \pm \ii$, $\beta = \pm 1$, and $\s{i} = \pm 1$.  In the body of the review we take $-\ii \alpha = \beta = -\s{\ast} = - \s{1} = \s{2} = \s{3} = \s{4} = \s{6} = \s{8} = 1$; we do not use $\s{5}$ or $\s{7}$.  This corresponds to $(-,+,+)$ in the notation of \rref{Misner:1973prb}.  In addition, we work in natural units, setting the speed of light $c$ and the reduced Planck constant $\hbar$ equal to $1$.  

\para{Spacetime and metric} 
The coordinates of four-dimensional spacetime are denoted by $x^\mu$ for $\mu = 0,1,2,3$.  Gravity is described by a type $(0,2)$ tensor field $\metric_{\mu\nu}(x)$, called the metric, that may vary throughout spacetime.  The inverse metric field is denoted by $\metric^{\mu\nu}(x)$ and defined by $\metric^{\mu\nu}(x) \metric_{\nu\rho}(x) = \delta_\rho^\mu$, where $\delta_\rho^\mu$ is the Kronecker delta symbol.  The determinant of the metric is denoted by $\metdet$ and appears in the factor $\sqrtg$.  The metric gives the differential invariant line element $(\dd s)^2 = \metric_{\mu\nu}(x) \dd x^\mu \dd x^\nu$.  In Minkowski spacetime the metric is static, homogeneous, and isotropic; it can be written as 
\ba{
    \metric_{\mu\nu}^\mathrm{Mink}(x) = \metricMink_{\mu\nu} = \s{1} \, \mathrm{diag}(-1,1,1,1)
    \com
}
and one can take $\s{1} = \pm 1$.  Its inverse is $\metric_\mathrm{Mink}^{\mu\nu} = \eta^{\mu\nu}$. 

\para{Curvature definitions}
Objects that arise in the definition of spacetime curvature include the Levi-Civita connection (torsion-free affine connection) 
\ba{
	\Gamma_{\mu\nu}^\rho(x)
	& = \frac{1}{2} \metric^{\rho\lambda} \bigl( \partial_\mu \metric_{\nu\lambda} + \partial_\nu \metric_{\mu\lambda} - \partial_\lambda \metric_{\mu\nu} \bigr) 
    \com
}
the Riemann curvature tensor  
\bes{
	\tensor{R}{^\rho_{\sigma\mu\nu}}(x) 
    & = \s{2} \, \Bigl( \partial_\mu \Gamma^\rho_{\nu\sigma} - \partial_\nu \Gamma^\rho_{\mu\sigma} 
    \\ & \qquad \quad 
    + \Gamma^\rho_{\mu\lambda} \Gamma^\lambda_{\nu\sigma} - \Gamma^\rho_{\nu\lambda} \Gamma^{\lambda}_{\mu\sigma} \Bigr) 
    \com
}
the Ricci tensor $R_{\mu\nu}(x) = (\s{2} \s{3})^{-1} \tensor{R}{^\alpha_{\mu\alpha\nu}}$, the Ricci scalar $R(x) = \metric^{\mu\nu} R_{\mu\nu}$, and the Einstein tensor $G_{\mu\nu}(x) = R_{\mu\nu} - \tfrac{1}{2} \metric_{\mu\nu} R$.  The totally-antisymmetric Levi-Civita connection $\varepsilon_{\mu\nu\rho\sigma}$ is normalized as $\varepsilon_{0123} = \s{5} = \pm 1$.  

\para{Covariant derivatives} 
Derivatives that transform covariantly under coordinate transformations (diffeomorphisms) are denoted by $\del{\mu}$.  Covariant derivatives of scalar, vector, and rank-2 tensor fields are defined by 
\bsa{}{
    & \del{\sigma} S = \partial_\sigma S \\ 
    & \del{\sigma} V^\mu = \partial_\sigma V^\mu + \Gamma^\mu_{\sigma\rho} V^\rho \\ 
    & \del{\sigma} V_\mu = \partial_\sigma V_\mu - \Gamma^\rho_{\sigma\mu} V_\rho \\ 
    & \del{\sigma} \tensor{T}{^\mu^\nu} = \partial_\sigma \tensor{T}{^\mu^\nu} + \Gamma^\mu_{\sigma\rho} \tensor{T}{^\rho^\nu} + \Gamma^\nu_{\sigma\rho} \tensor{T}{^\mu^\rho} \\ 
    & \del{\sigma} \tensor{T}{^\mu_\nu} = \partial_\sigma \tensor{T}{^\mu_\nu} + \Gamma^\mu_{\sigma\rho} \tensor{T}{^\rho_\nu} - \Gamma^\rho_{\sigma\nu} \tensor{T}{^\mu_\rho} \\ 
    & \del{\sigma} \tensor{T}{_\mu^\nu} = \partial_\sigma \tensor{T}{_\mu^\nu} - \Gamma^\rho_{\sigma\mu} \tensor{T}{_\rho^\nu} + \Gamma^\nu_{\sigma\rho} \tensor{T}{_\mu^\rho} \\ 
    & \del{\sigma} \tensor{T}{_\mu_\nu} = \partial_\sigma \tensor{T}{_\mu_\nu} - \Gamma^\rho_{\sigma\mu} \tensor{T}{_\rho_\nu} - \Gamma^\rho_{\sigma\nu} \tensor{T}{_\mu_\rho} 
    \com
}
and the generalization to higher-rank tensors is obvious.  When applied to a type $(p,q)$ tensor with $p$ contravariant (up) indices and $q$ covariant (down) indices, the covariant derivative produces a type $(p,q+1)$ tensor.  The Levi-Civita connection parallel transports the metric $\del{\sigma} \metric_{\mu\nu} = 0$.  The d'Alembertian operator of a scalar field $\aphi$ is 
\bes{
    \Box \aphi 
     = \metric^{\mu\nu} \del{\mu} \del{\nu} \aphi  
     = (\sqrtg)^{-1} \partial_\mu (\sqrtg \, \metric^{\mu\nu} \partial_\nu \aphi) 
    \per 
}

\para{Einstein-Hilbert action and matter} 
The dynamics of gravity and its interactions with radiation and matter are encoded in an action $S$ using the formalism of Lagrangian mechanics.  We treat the radiation and matter as classical fields, and denote them collectively by $F_i(x)$.  We study a class of theories whose actions take the form 
\ba{
	S[F_i(x), \metric_{\mu\nu}(x)] = S_\EH[\metric_{\mu\nu}(x)] + S_\M[F_i(x), \metric_{\mu\nu}(x)] 
    \com
}
where $S_\EH$ is the Einstein-Hilbert (EH) action and $S_\M$ is the radiation-matter action.  The EH action is 
\ba{\label{eq:SEH_def}
	S_\EH[\metric_{\mu\nu}(x)] = \s{1} \s{3} \, \frac{1}{2} \Mpl^2 \int \! \dd^4 x \, \sqrtg \, R 
    \com
}
where $\Mpl$ is the reduced Planck mass, $G_N = 1/8 \pi \Mpl^2$ is Newton's gravitational constant, and $\kappa^2 = 8 \pi G_N$ is the gravitational coupling.  The radiation-matter action takes the form 
\ba{\label{eq:SM_def}
    S_\M[F_i(x), \metric_{\mu\nu}(x)] = \int \! \dd^4x \, \sqrtg \, \Lcal 
    \com
}
where $\Lcal$ is the radiation-matter Lagrangian density. 

\para{Einstein's equation}
The equation of motion for the metric, known as Einstein's equation, is derived by applying the variational principle.  Imposing $\delta S / \delta \metric^{\mu\nu} = 0$ leads to Einstein's equation \cite{Einstein:1916vd}
\ba{\label{eq:Einstein_eqn}
	\s{3} \, G_{\mu\nu} = 8\pi G_N \, T_{\mu\nu} = \Mpl^{-2} \, T_{\mu\nu} = \kappa^2 \, T_{\mu\nu} 
    \per
}
When evaluating the variational derivative of the Einstein-Hilbert term, the following identities prove useful: 
\bsa{}{
    \delta \metric^{\mu\nu} 
    & = - \metric^{\mu\rho} \metric^{\nu\sigma} \, \delta \metric_{\rho\sigma} \\ 
    \delta \metric_{\mu\nu} & = - \metric_{\mu\rho} \metric_{\nu\sigma} \delta \metric^{\rho\sigma} \\ 
    \delta \metdet 
    & = \metdet \, \metric^{\mu\nu} \delta \metric_{\mu\nu} \\ 
	\delta \sqrtg 
    & = \tfrac{1}{2} \sqrtg \, \metric^{\mu\nu} \, \delta \metric_{\mu\nu} \\ 
	\delta R 
    & = - R^{\mu\nu} \delta \metric_{\mu\nu} + \s{3} \bigl( \tfrac{1}{2} \metric^{\rho\mu} \metric^{\sigma\nu} 
    \\ & \qquad 
    + \tfrac{1}{2} \metric^{\rho\nu} \metric^{\sigma\mu} - \metric^{\rho\sigma} \metric^{\mu\nu} \bigr) \del{\nu} \del{\mu} \delta \metric_{\rho\sigma}
    \nonumber 
    \com
}
and it follows that $\delta S_\EH / \delta \metric^{\mu\nu} = \s{1} \s{3} (\sqrtg/2) \Mpl^2 G_{\mu\nu}$.  

\para{Stress-energy tensor}
The stress-energy tensor field $T_{\mu\nu}(x)$ is defined as the variational derivative of the radiation-matter action 
\ba{\label{eq:Tmunu}
    T_{\mu\nu} 
    & = - \s{1} \frac{2}{\sqrtg} \, \frac{\delta S_\M}{\delta \metric^{\mu\nu}}  
    \per
}
Note that the stress-energy tensor is manifestly symmetric since the inverse metric is symmetric.  The scalar $\tensor{T}{^\mu_\mu} = \metric^{\mu\nu} T_{\nu\mu}$ is called the trace of the stress-energy tensor.  For theories with fields of half-integer spin, the matter action is expressed in terms of the frame fields $e_a^\mu(x)$ (see the forthcoming discussion) rather than the metric, and the stress-energy tensor is calculated using 
\ba{
    T{_a^\mu} = \frac{1}{\dete} \frac{\delta S_\M}{\delta e{^a_\mu}} 
    \quad \text{and} \quad 
    T_{\mu\nu} = \s{1} \metric_{\mu\rho} \metric_{\nu\sigma} \eta^{ab} e_a^\rho T_b^\sigma 
    \per
}

\para{Perfect fluid stress-energy tensor} 
In many applications the radiation-matter sector is modeled as a perfect fluid.  In such systems the stress-energy tensor takes the form 
\ba{\label{eqTmunu}
    T_{\mu\nu} = \bigl( \rho + \pressure \bigr) u_\mu u_\nu + \s{1} \pressure \, \metric_{\mu\nu}
    \com
}
where $\rho(x)$ is the scalar energy density field, $\pressure(x)$ is the scalar pressure field, and $u_\mu(x)$ is the vector velocity field.  The vector field is normalized such that $\metric^{\mu\nu} u_\mu u_\nu = - \s{1}$.  It follows that $\rho = T^{\mu\nu} u_\mu u_\nu$ and $3\pressure = \s{1} \tensor{T}{^\mu_\mu} + \rho$

\para{Canonical stress-energy tensor} 
For this discussion we restrict our attention to Minkowski spacetime.  If spacetime translations are a symmetry of the theory, the four Noether currents form a rank-2 tensor called the canonical stress-energy tensor, which we denote by $\Theta^{\mu\nu}$.  These currents may be derived by applying Noether's theorem; see \rref{Weinberg:1995mt} for a review.  For a theory of fields $F_i(x)$ one finds that 
\ba{
    \Theta^{\mu\nu} = \frac{\partial \Lcal}{\partial(\partial_\mu F_i)} \eta^{\nu\rho} \partial_\rho F_i - \metricMink^{\mu\nu} \Lcal 
    \per
}
Although the canonical stress-energy tensor is conserved ($\partial_\mu \Theta^{\mu\nu} = 0$), it need not be symmetric or even gauge invariant.  However if Lorentz invariance is also a symmetry of the theory, then one can construct a conserved and symmetric tensor, the Belinfante-Rosenfeld stress-energy tensor, and denote it by $t^{\mu\nu}$.  The relation takes the form $t^{\mu\nu} = \Theta^{\mu\nu} + \partial_\rho B^{\rho\mu\nu}$, where $B^{\rho\mu\nu} = - B^{\mu\rho\nu}$ is antisymmetric in its \textit{first} two indices \cite{Belinfante1,Belinfante2,Rosenfeld2}.  In Minkowski spacetime the Belinfante-Rosenfeld stress-energy tensor $t^{\mu\nu}$ coincides with the Hilbert stress-energy tensor $T^{\mu\nu}$ defined in \eref{eq:Tmunu} when the action is varied with respect to the metric.  While this discussion is intended to illuminate the relation between $T^{\mu\nu}$ and conserved quantities in the sense of Noether's theorem, ultimately the quantity of interest for cosmology is $T^{\mu\nu}$, which appears in Einstein's equation. 

\para{Frame-field formalism} 
For theories whose matter action contains particles of half-integer spin, a covariant coupling to gravity cannot be accomplished within the metric formulation of general relativity.  Instead, for these purposes it is customary to employ the frame-field formalism,\footnote{The frame field was introduced by \cite{Cartan:1926}, who called them \textit{rep\'{e}res mobiles} (moving frames); in German they are referred to as \textit{vierbeins} (four leg), and in the English literature they are usually called \textit{tetrads} (Greek for ``set of four'').  Here we refer to them as frame fields.} which is an elegant and general formulation of general relativity that allows field theory with spinors to be extended to curved spacetime.  Hereafter we follow the notation of \rref{Freedman:2012zz}.  

At any point on the spacetime manifold, one can find a basis that locally diagonalizes the metric, and with an appropriate rescaling this yields the metric for Minkowski spacetime $\eta_{ab} = \s{1} \, \mathrm{diag}(-1,1,1,1)$.  Requiring this diagonalization to be smooth leads to the relation 
\bes{
    & \metric_{\mu\nu}(x) 
    = e_\mu^a(x) \, e_\nu^b(x) \, \eta_{ab} 
    \com
}
where $e_\mu^a(x)$ for $a=1,2,3,4$ is called the frame field.\footnote{Frame indices, also called local Lorentz indices, are denoted by the Latin letters $a$, $b$, etc., while coordinate indices are denoted by the Greek letters $\mu$, $\nu$, etc.  Frame indices are raised and lowered by the Minkoski metric and its inverse $\eta_{ab}$ and $\eta^{ab}$, while coordinate indices are raised and lowered by the local metric and its inverse $\metric_{\mu\nu}$ and $\metric^{\mu\nu}$.}  The decomposition of the metric $\metric_{\mu\nu}(x)$ into a pair of frame fields is not unique, and equivalent expressions are related by the local Lorentz transformations $e_\mu^{\prime a}(x) = \tensor{(\Lambda^{-1})}{^a_b}(x) e_\mu^b(x)$ under which the frame index transforms and the coordinate index is inert.  However, under coordinate transformations (diffeomorphisms) the frame field transforms as a set of four covariant vectors $e_\mu^{\prime a}(x^\prime) = (\partial x^\rho / \partial x^{\prime \mu}) e_\rho^a(x)$ while the frame index is inert.  Note that $\mathrm{det}(e) \equiv \dete = \sqrtg$.  There exists an inverse of the frame field, denoted by $e_a^\mu(x)$, such that $e_\mu^a e_a^\nu = \delta_\mu^\nu$ and $e_a^\mu e_\mu^b = \delta_a^b$.  

\para{Spin connection} 
Tensor fields of arbitrary rank can be expressed in the frame-field formalism.  For example a contravariant vector field $V^\mu(x)$ can be written as $V^\mu(x) = V^a(x) e_a^\mu(x)$ where $V^a(x) = e_\mu^a(x) V^\mu(x)$ are the frame components of the vector field, which transform as scalars under coordinate transformations and as a vector under local Lorentz transformations.  The same holds for covariant vectors $V_\mu(x) = e_\mu^a(x) V_a(x)$ where $V_a(x) = e_a^\mu(x) V_\mu(x)$.  Derivatives of frame vector fields that transform covariantly under the local Lorentz transformations are given by 
\bes{
    D_\mu V^a & = \partial_\mu V^a + (\s{2} \s{4})^{-1} \, \tensor{\omega}{_\mu^a_b} V^b \\ 
    D_\mu V_a & = \partial_\mu V_a - (\s{2} \s{4})^{-1} \, \tensor{\omega}{_\mu^b_a} V_b 
    \com
}
where the appropriate affine connection, also known as the spin connection, is given by 
\ba{
    \tensor{\omega}{_\mu^a^b} = \s{2} \s{4} \, \Bigl[ 2 \ee^{\nu[a} \partial_{[\mu} \tensor{e}{_{\nu]}^{b]}} - \ee^{\nu[a} \ee^{b]\sigma} e_{\mu c} \partial_\nu e_\sigma^c \Bigr] 
    \per
}
Here $X^{[a}Y^{b]} = (X^a Y^b - X^b Y^a)/2$.  The spin connection is required to be antisymmetric in its frame indices: $\tensor{\omega}{_\mu^{ab}} = - \tensor{\omega}{_\mu^{ba}}$.  The derivatives $\del{\mu} V^\nu = e_a^\nu D_\mu V^a$ and $\del{\mu} V_\nu = e_\nu^a D_\mu V_a$ are covariant under coordinate transformations, transforming as type $(1,1)$ and $(0,2)$ tensors, respectively.  The Levi-Civita connection is related to the spin connection by $\Gamma_{\mu\nu}^\rho = e_a^\rho \partial_\mu e_\nu^a + (\s{2} \s{4})^{-1} \, e_a^\rho \tensor{\omega}{_\mu^a_b} e_\nu^b$.  

\para{Spinor matrices} 
Let $\gamma^a$ for $a=0,1,2,3$ denote the set of four complex $4 \times 4$ spinor matrices.  They obey the Clifford algebra of anticommutation relations 
\bes{
    \{ \gamma^a, \gamma^b \} 
    & = \s{6} \, 2 \eta^{ab} \bbone 
     = (\s{6}/\s{1}) \, 2 \, \mathrm{diag}(-1,1,1,1) \, \bbone 
     \com
}
where $\bbone$ is the identity matrix on the spinor indices.  For instance, in the Dirac representation 
\ba{\label{eq:Dirac_representation}
    \gamma^0 = \sqrt{\frac{-\s{6}}{\s{1}}} \begin{pmatrix} \bbone & 0 \\ 0 & -\bbone \end{pmatrix} 
    , \quad 
    \gamma^i = \sqrt{\frac{-\s{6}}{\s{1}}} \begin{pmatrix} 0 & \sigma^i \\ -\sigma^i & 0 \end{pmatrix} 
    \com
}
where $\sigma^i$ for $i=1,2,3$ are the Pauli matrices.  One can also define 
\bsa{}{
    \gamma_5 & = \s{7} \, \ii \gamma^0 \gamma^1 \gamma^2 \gamma^3 \\ 
    \gamma^{ab} & = \gamma^{[a} \gamma^{b]} = \tfrac{1}{2} (\gamma^a \gamma^b - \gamma^b \gamma^a) \\ 
    \gamma^{abc} & = \gamma^{[a} \gamma^{b} \gamma^{c]} = \tfrac{1}{2} (\gamma^a \gamma^b \gamma^c - \gamma^c \gamma^b \gamma^a) 
    \com
}
etc.; in addition, $\Sigma^{ab} = \gamma^{ab}/2$.  Using the frame field, one defines a set of local gamma matrices 
\ba{
    \gamma^\mu(x) = e_a^\mu(x) \, \gamma^a 
    \com
}
that satisfy the Clifford algebra locally 
\ba{
    \{ \gamma^\mu(x), \gamma^\nu(x) \} = \s{6} \, 2 \metric^{\mu\nu}(x) \bbone 
    \com
}
and from which one may construct $\gamma^{\mu\nu}(x)$ and $\gamma^{\mu\nu\rho}(x)$ in analogy with the previous expressions.  

\para{Spinor fields} 
For spinor fields $\Psi(x)$ or vector-spinor fields $\Psi_\mu(x)$ one can define the covariant derivatives 
\bsa{}{
    D_\mu \Psi & = \partial_\mu \Psi + \s{2} \s{4} \, \tfrac{1}{4} \omega_{\mu ab} \gamma^{ab} \Psi \\ 
    \del{\mu} \Psi & = D_\mu \Psi \\ 
    D_\mu \Psi_\nu & = \partial_\mu \Psi_\nu + \s{2} \s{4} \, \tfrac{1}{4} \omega_{\mu ab} \gamma^{ab} \Psi_\nu \\ 
    \del{\mu} \Psi_\nu & = D_\mu \Psi_\nu - \Gamma_{\mu\nu}^\rho \Psi_\rho 
    \per
}
All four quantities transform as spinors under local Lorentz transformations, whereas only $\del{\mu} \Psi_\nu$ transforms as a type $(0,2)$ tensor under coordinate transformations.  

\para{Conjugation}
The Hermitian conjugate of the spinor matrices, either $\gamma^a$ or $\gamma^\mu(x)$, can be written as 
\ba{
    (\gamma^a)^\dagger = (\s{1} \s{\ast} / \s{6}) \, \gamma^0 \gamma^a \gamma^0 
    \per
}
The Clifford algebra implies that $(\gamma^0)^\dagger = - \s{\ast} \gamma^0$ and $(\gamma^i)^\dagger = \s{\ast} \gamma^i$ for $i=1,2,3$, such that $\gamma^0$ is Hermitian for $\s{\ast} = -1$ and anti-Hermitian for $\s{\ast} = +1$.  Regardless of the sign choices, $(\gamma_5)^\dagger = \gamma_5$.  One can define the Dirac adjoint spinors 
\ba{
    \bar{\Psi} = (\ii / \alpha) \, \Psi^\dagger \gamma^0 
    \quad \text{and} \quad 
    \bar{\Psi}_\mu = (\ii / \alpha) \, \Psi_\mu^\dagger \gamma^0 
    \com
}
where $\gamma^0 = \gamma^a|_{a=0}$ is the constant spinor matrix and the phase $\alpha$ takes values $\pm 1$ or $\pm \ii$.  Covariant derivative of the Dirac adjoint spinors are defined by
\bsa{}{
    \bar{\Psi} \overleftarrow{D_\mu} & = \partial_\mu \bar{\Psi} - \s{2} \s{4} \, \tfrac{1}{4} \omega_{\mu ab} \bar{\Psi} \gamma^{ab} \\ 
    \bar{\Psi} \overleftarrow{\del{\mu}} & = \bar{\Psi} \overleftarrow{D_\mu} \\ 
    \bar{\Psi}_\nu \overleftarrow{D_\mu} & = \partial_\mu \bar{\Psi}_\nu - \s{2} \s{4} \, \tfrac{1}{4} \omega_{\mu ab} \bar{\Psi}_\nu \gamma^{ab} \\ 
    \bar{\Psi}_\nu \overleftarrow{\del{\mu}} & = \bar{\Psi}_\nu \overleftarrow{D_\mu} - \Gamma_{\mu\nu}^\rho \Psi_\rho 
    \per
}

\para{Bilinear products} 
A final sign ambiguity arises in the complex conjugation of spinor products.  If $\chi_1$ and $\chi_2$ are two anticommuting variables, such as components of a spinor field $\Psi$, then $\chi_1 \chi_2 = - \chi_2 \chi_1$.  The complex conjugate of their product may be evaluated as $(\chi_1 \chi_2)^\ast = - \beta \chi_1^\ast \chi_2^\ast = \beta \chi_2^\ast \chi_1^\ast$ where the sign $\beta = \pm 1$ controls whether complex conjugation also exchanges order.  For example the bilinear product of two spinors complex conjugates to $(\bar{\psi} \chi)^\ast = \s{8} \, \bar{\chi} \psi$, where $\s{8} \equiv \beta \s{\ast} \alpha / \alpha^\ast = \pm 1$.  If $\s{8} = +1$, then $\bar{\psi} \psi$ is real, \ie, complex conjugates to itself, but if $\s{8} = -1$ then $\bar{\psi} \psi$ is imaginary and an additional factor of $\ii = \sqrt{\s{8}}$ must be added when one constructs the Lagrangian mass term.  A similar argument applies for kinetic terms. 

\para{Radiation-matter action} 
The frame-field formalism may be used to study radiation-matter sectors containing fields of either integer or half-integer spin, but we employ it only in the study of spinor fields.  The radiation-matter action for free massive fields of spin $s=0$, $\half$, $1$, and $\threehalf$ can be written as 
\ba{\label{eq:wheretheactionis}
    S_\M & = \int \! \dd^4x \, \sqrtg \, \sum_{s} \Lcal^{(s)} 
    \\ 
    \Lcal^{(0)} & = - \s{1} \tfrac{1}{2} \metric^{\mu\nu} \del{\mu} \aphi \del{\nu} \aphi - \tfrac{1}{2} m^2 \Phi^2 - \s{1} \s{3} \tfrac{1}{2}\xi R \aphi^2 
    \nn  
    \Lcal^{(\half)} & = 
    \sqrt{\s{8} \s{\ast}} \bigl( 
    \tfrac{1}{4} \bar{\Psi} \gamma^\mu \del{\mu} \Psi 
    - \tfrac{1}{4} \bar{\Psi} \overleftarrow{\del{\mu}} \gamma^\mu \Psi \bigr) 
    - \sqrt{\s{8}} \, \tfrac{1}{2} m \bar{\Psi} \Psi 
    \nn 
    \Lcal^{(1)} & = - \tfrac{1}{2} \metric^{\mu\rho} \metric^{\nu\sigma} F_{\mu\nu} F_{\rho\sigma} - \s{1} \tfrac{1}{2} m^2 \metric^{\mu\nu} A_\mu A_\nu 
    \nn 
    \Lcal^{(\threehalf)} & = 
    \sqrt{\s{8} \s{\ast}^3} \bigl( 
    \tfrac{1}{4} \bar{\Psi}_\mu \, \gamma^{\mu\nu\rho} \del{\nu} \Psi_\rho 
    - \tfrac{1}{4} \bar{\Psi}_\mu \overleftarrow{\del{\nu}} \gamma^{\mu\nu\rho} \Psi_\rho \bigr) 
    \nn & \quad 
    - \sqrt{\s{8} \s{\ast}^2} \, \tfrac{1}{2} m \bar{\Psi}_\mu \gamma^{\mu\nu} \Psi_\nu 
    \com
    \nonumber
}
where $m^2$, $m$, and $\xi$ are real parameters, and where 
\bsa{}{
    \del{\mu} \Phi & = \partial_\mu \Phi \\ 
    \del{\mu} \Psi & = \partial_\mu \Psi + \s{2} \s{4} \, \tfrac{1}{4} \omega_{\mu ab} \gamma^{ab} \Psi \\ 
    \bar{\Psi} & = (\ii / \alpha) \, \Psi^\dagger \gamma^0 \\ 
    \bar{\Psi} \overleftarrow{\del{\mu}} & = \partial_\mu \bar{\Psi} - \s{2} \s{4} \, \tfrac{1}{4} \omega_{\mu ab} \bar{\Psi} \gamma^{ab} \\ 
    F_{\mu\nu} & = \del{\mu} A_\nu - \del{\nu} A_\mu \\ 
    \del{\mu} A_\nu & = \partial_\mu A_\nu - \Gamma_{\mu\nu}^\rho A_\rho \\ 
    \del{\mu} \Psi_\nu & = \partial_\mu \Psi_\nu + \s{2} \s{4} \, \tfrac{1}{4} \omega_{\mu ab} \gamma^{ab} \Psi_\nu - \Gamma_{\mu\nu}^\rho \Psi_\rho \\ 
    \bar{\Psi}_\mu & = (\ii / \alpha) \, \Psi_\mu^\dagger \gamma^0 \\ 
    \bar{\Psi}_\mu \overleftarrow{\del{\nu}} & = \partial_\mu \bar{\Psi}_\nu - \s{2} \s{4} \, \tfrac{1}{4} \omega_{\mu ab} \bar{\Psi}_\nu \gamma^{ab} - \Gamma_{\mu\nu}^\rho \bar{\Psi}_\rho 
    \per
}
Note that the $\Gamma_{\mu\nu}^\rho$ terms in $\del{\mu} A_\nu$ and $\del{\mu} \Psi_\nu$ do not contribute to the Lagrangian, since the symmetric Levi-Civita connection is contracted with an antisymmetric tensor.  In the body of the review, we take $-\ii \alpha = \beta = -\s{\ast} = - \s{1} = \s{2} = \s{3} = \s{4} = \s{6} = \s{8} = 1$.  

\section{FLRW spacetime}
\label{app:FLRW_spacetime}

This appendix summarizes key relations that arise frequently in the FLRW spacetime.  We provide expressions using both coordinate time $t$ and conformal time $\eta$ so that one can move between these two conventions.  We consider only the spatially flat FLRW spacetime ($\mathrm{k}=0$). 

\para{Metric}  
The differential spacetime interval for the homogeneous and isotropic FLRW spacetime \cite{Friedmann:1922,Friedmann:1924,Lemaitre:1931,Robertson:1935,Robertson:1936a,Robertson:1936b,Walker:1937} can be written as  
\bes{\label{eq:FLRW_metric}
    (\dd s)^2 
    & = \s{1} \bigl[ - (\dd t)^2 + a^2(t) \, |\dd \xvec|^2 \bigr] 
    \quad \text{or} \\ 
    (\dd s)^2 & = \s{1} \bigl[ - a^2(\eta) (\dd \eta)^2 + a^2(\eta) |\dd \xvec|^2 \bigr] 
    \per
}
We call $t$ the time coordinate, $\eta$ the conformal time coordinate, $\xvec$ the comoving spatial coordinate, and $a$ the scale factor.  Note that $\dd t = a(\eta) \dd \eta$ and $\dd \eta = \dd t/a(t)$, as well as $\partial_t = a^{-1}(\eta) \partial_\eta$ and $\partial_\eta = a(t) \partial_t$.  In other words, the nonzero components of the metric are 
\bes{
    \metric_{00}(t) = - \s{1} 
    \quad & \text{or} \quad 
    \metric_{00}(\eta) = - \s{1} a^2(\eta) \\ 
    \metric_{ii}(t) = \s{1} a^2(t) 
    \quad & \text{or} \quad 
    \metric_{ii}(\eta) = \s{1} a^2(\eta) 
}
for $i=1,2,3$ (no summation).  The metric determinant gives 
\ba{
    \sqrtg = a^3(t) 
    \quad \text{or} \quad 
    \sqrtg = a^4(\eta) 
    \per
}
When conformal time is used, the FLRW metric can be written compactly as $\metric_{\mu\nu}(\eta) = a^2(\eta) \, \metricMink_{\mu\nu}$, where $\metricMink_{\mu\nu}$ is the Minkowski metric.  Hence, the (spatially flat) FLRW spacetime is conformally related to Minkowski spacetime.  The components of the corresponding frame fields are 
\bes{
    e_\mu^0(t) = \delta_\mu^0 
    \quad & \text{or} \quad 
    e_\mu^0(\eta) = a(\eta) \, \delta_\mu^0 \\ 
    e_\mu^i(t) = a(t) \, \delta_\mu^i 
    \quad & \text{or} \quad 
    e_\mu^i(\eta) = a(\eta) \, \delta_\mu^i 
}
for $\mu=0,1,2,3$ and $i=1,2,3$.  

\para{Hubble parameter}  
Evolution of the scale factor is measured with the Hubble parameter $H$, which is defined equivalently by both of the following relations: 
\ba{\label{eq:B5}
	H(t) = \frac{\dot{a}(t)}{a(t)} 
    \quad \text{or} \quad 
    H(\eta) = \frac{a^\prime(\eta)}{a^2(\eta)} 
    \per
}
In \eref{eq:B5}, the dot denotes the derivative with respect to $t$ and the prime denotes the derivative with respect to $\eta$.  One can also define the comoving Hubble parameter $\Hcal(\eta) = a^\prime(\eta) / a(\eta)$.  The present-day value of the Hubble parameter, which is also known as the Hubble constant \cite{Hubble:1929ig}, is denoted by $H_0$.  It is customary to write  $H_0 = 100 h \ \mathrm{km}\,\sec^{-1}\, \Mpc^{-1}$, where the dimensionless factor $h \approx 0.7\simeq 1/\sqrt{2}$.

\para{Distances}  
The distance between any two points in the expanding Universe changes in time in proportion to the scale factor $a$.  If at some time the scale factor is $a_1$ and the physical distance between two points is $l_{\mathrm{phys},1}$, then at some other time when the scale factor is $a$, the distance will be $l_\mathrm{phys}(a) = l_{\mathrm{phys},1} \, (a/a_1)$.  It is often useful to consider a comoving distance $l = l_\mathrm{phys} / a$; the comoving distance is constant in expansion.  Similarly, the momentum of a particle redshifts in expansion as $\pvec_\mathrm{phys}(a) = \pvec_{\mathrm{phys},1} \, (a_1/a)$, and it is useful to define a comoving momentum $\pvec = a \, \pvec_{\rm phys}$.  Analogous relations arise when spatially varying fields are represented in Fourier space.  A Fourier mode with a comoving wavevector $\kvec$ and a comoving wave number $k=|\kvec|$ corresponds to a comoving wavelength of $\lambda = 2\pi/k$, and it corresponds to a physical wavelength of $\lambda_\mathrm{phys} = 2\pi a / k$.  We define the comoving Hubble length scale as 
\ba{
    d_H(t) = \frac{2\pi}{a(t) H(t)} 
    \quad \text{or} \quad 
    d_H(\eta) = \frac{2\pi}{a(\eta) H(\eta)} 
    \per
}
Modes with a comoving wavelength $\lambda$ and a comoving wave number $k = 2 \pi / \lambda$ are said to ``cross'' the Hubble scale (\ie, ``enter'' or ``leave'' the horizon) when $k = aH$, corresponding to $\lambda = d_H$. 

\para{GR tensors}  
In the FLRW spacetime, the only nonzero components of the Levi-Civita connection $\Gamma_{\mu\nu}^\rho$ are 
\bes{
    \Gamma^0_{00}(t) = 0 
    \quad & \text{or} \quad 
    \Gamma^0_{00}(\eta) = a H \\ 
    \Gamma^0_{ii}(t) = a^2 H 
    \quad & \text{or} \quad 
    \Gamma^0_{ii}(\eta) = a H \\ 
    \Gamma^i_{0i}(t) = H 
    \quad & \text{or} \quad 
    \Gamma^i_{0i}(\eta) = a H \\ 
    \Gamma^i_{i0}(t) = H 
    \quad & \text{or} \quad 
    \Gamma^i_{i0}(\eta) = a H 
    \com
}
for $i=1,2,3$ (no summation).  Similarly the nonzero components of the spin connection $\tensor{\omega}{_\mu^a^b}$ are 
\bes{
    \frac{\s{1}}{\s{2}\s{4}} \, \tensor{\omega}{_i^0^i}(t) = aH 
    \quad & \text{or} \quad 
    \frac{\s{1}}{\s{2}\s{4}} \, \tensor{\omega}{_i^0^i}(\eta) = aH \\ 
    \frac{\s{1}}{\s{2}\s{4}} \, \tensor{\omega}{_i^i^0}(t) = -aH 
    \quad & \text{or} \quad 
    \frac{\s{1}}{\s{2}\s{4}} \, \tensor{\omega}{_i^i^0}(\eta) = -aH 
    \per
}
The nonzero components of the Ricci tensor are 
\ba{
    \s{3} \, R_{00}(t) = -3 \frac{\ddot{a}}{a} 
    \quad & \text{or} \quad 
    \s{3} \, R_{00}(\eta) = -3 \frac{a^{\prime\prime}}{a} + 3 \frac{a^{\prime 2}}{a^2} \nn 
    \s{3} \, R_{ii}(t) = a \ddot{a} + 2 \dot{a}^2 
    \quad & \text{or} \quad 
    \s{3} \, R_{ii}(\eta) = \frac{a^\pprime}{a} + \frac{a^{\prime 2}}{a^2} 
}
for $i=1,2,3$ (no summation).  The Ricci scalar is 
\ba{
    \s{1} \s{3} \, R(t) = 6 \frac{\ddot{a}}{a} + 6 \frac{\dot{a}^2}{a^2}
    \quad \text{or} \quad 
    \s{1} \s{3} \, R(\eta) = 6 \frac{a^\pprime}{a^3} 
    \per
}
The nonzero components of the Einstein tensor are
\ba{
    \s{3} \, G_{00}(t) = 3 \frac{\dot{a}^2}{a^2} 
    \quad & \text{or} \quad 
    \s{3} \, G_{00}(\eta) = 3 \frac{a^{\prime 2}}{a^2} \\ 
    \s{3} \, G_{ii}(t) = - 2 a \ddot{a} - \dot{a}^2 
    \quad & \text{or} \quad 
    \s{3} \, G_{ii}(\eta) = - 2 \frac{a^\pprime}{a} + \frac{a^{\prime 2}}{a^2} 
    \nonumber 
}
for $i=1,2,3$ (no summation).  For a perfect fluid with energy density $\rho$ and pressure $\pressure$, the stress-energy tensor in the fluid's rest frame can be written as
\bes{
    T_{00}(t) = \rho
    & \quad \text{or} \quad 
    T_{00}(\eta) = a^2 \rho \\ 
    T_{ii}(t) = a^2 \pressure 
    & \quad \text{or} \quad 
    T_{ii}(\eta) = a^2 \pressure 
}
for $i=1,2,3$ (no summation).  The diagonal entries of Einstein's equation give 
\ba{\label{eq:B13}
    \frac{\dot{a}^2}{a^2} = \frac{1}{3 \Mpl^2} \rho 
    \quad & \text{or} \quad 
    \frac{a^{\prime 2}}{a^4} = \frac{1}{3 \Mpl^2} \rho \\ 
    \frac{\ddot{a}}{a} = - \frac{1}{6 \Mpl^2} (\rho + 3 \pressure) 
    \quad & \text{or} \quad 
    \frac{a^\pprime}{a^3} = \frac{1}{6 \Mpl^2} (\rho - 3 \pressure) 
    \com
    \nonumber
}
which are Friedmann's first and second equations.  Differentiating the first equation in \eref{eq:B13} gives the continuity equation 
\ba{
    \dot{\rho} + 3 H (\rho + \pressure) = 0 
    \quad \text{or} \quad 
    \rho^\prime + 3 a H (\rho + \pressure) = 0 
    \com
}
which expresses energy nonconservation due to the cosmological expansion $H \neq 0$.  Note that Friedmann's equations allow the Ricci scalar to be written as 
\ba{
    \s{1} \s{3} R = \Mpl^{-2} \bigl( \rho - 3 \pressure \bigr) 
    \com
}
which holds using either coordinate or conformal time.  

\para{Constant equation of state}  
Although the energy density $\rho$ and pressure $\pressure$ may vary in time, it is often the case that their ratio, the equation of state $w = \pressure / \rho$, is a constant.  In that case the continuity equation is solved to find 
\bes{
    \rho(a) & = \rho_1 \bigl( a/a_1 \bigr)^{-3(1+w)} \\ 
    \pressure(a) & = \pressure_1 \bigl( a/a_1 \bigr)^{-3(1+w)} \\ 
    H(a) & = H_1 \bigl( a/a_1 \bigr)^{-3(1+w)/2} 
    \com
}
where $\pressure_1 = w \rho_1$.  Frequently encountered equations of state include: 
\begin{itemize}
    \item \textit{Kination:}  $w=1$, $\rho \propto a^{-6}$, and $H \propto a^{-3}$.  
    \item \textit{Radiation:}  $w=\ssfrac{1}{3}$, $\rho \propto a^{-4}$, and $H \propto a^{-2}$.  
    \item \textit{Matter:}  $w=0$, $\rho \propto a^{-3}$, and $H \propto a^{-3/2}$.  
    \item \textit{Vacuum:} $w=-1$, $\rho \propto a^0$, and $H \propto a^0$.
\end{itemize}
The cases labeled as \textit{radiation} and \textit{matter} refer more generally to relativistic and nonrelativistic particles.  

\section{Cosmological perturbations}
\label{app:cosmo_perturb}

This appendix contains an introduction to cosmological perturbations.  The FLRW spacetime models an idealized homogeneous and isotropic universe, and describing our messy Universe requires a departure from FLRW.  Provided that this departure is small, it can be treated perturbatively.  Introductions to cosmological perturbation theory are given in various textbooks and review articles; we follow \rref{Baumann:2022mni}.  

\para{Metric perturbations}
Allowing for perturbations on the FLRW metric, the spacetime interval may be written as 
\bes{\label{eq:perturb_FLRW}
    & (\dd s)^2 
    = \s{1} \, a^2 \bigl[ 
    - \bigl( 1 + 2 A \bigr) (\dd \eta)^2 
    + 2 B_i \, \dd x^i \dd \eta 
    \\ & \qquad \qquad \qquad 
    + \bigl( \delta_{ij} + 2 E_{ij} \bigr) \, \dd x^i \dd x^j 
    \bigr]
    \\ & \quad \text{where} \quad 
    B_i = \partial_i B + \hat{B}_i 
    \\ & \quad \text{and} \quad 
    E_{ij} = C \delta_{ij} + \partial_{\langle i} \partial_{j \rangle} E + \partial_{(i} \hat{E}_{j)} + \hat{E}_{ij} 
    \per
}
Note that $\partial_{\langle i} \partial_{j \rangle} E = ( \partial_i \partial_j - \tfrac{1}{3} \delta_{ij} \nabla^2 ) E$ and $\partial_{(i} \hat{E}_{j)} = \tfrac{1}{2} (\partial_i \hat{E}_j + \partial_j \hat{E}_i)$.  The functions $A(\eta,\xvec)$, $B(\eta,\xvec)$, $C(\eta,\xvec)$, and $E(\eta,\xvec)$ are called the scalar metric perturbations; $\hat{B}_i(\eta,\xvec)$ and $\hat{E}_i(\eta,\xvec)$ are called the vector metric perturbations; and $\hat{E}_{ij}(\eta,\xvec)$ is called the tensor metric perturbation.  The vector and tensor perturbations obey the transverse and traceless conditions: $\partial^i \hat{B}_i = \partial^i \hat{E}_i = \partial^i \hat{E}_{ij} = \tensor{\hat{E}}{^i_i} = 0$.

\para{Matter perturbations}
During and after inflation the Universe contained various particle species, including the inflaton, the standard-model particles, and the dark matter.  We label the particle species with an index $s$.  When interactions are neglected, the total stress-energy tensor can be expressed as a sum over the component stress-energy tensors associated with each species: $T_{\mu\nu} = \sum_s (T_s)_{\mu\nu}$.  The component stress-energy tensors may be decomposed as
\bes{
    \tensor{(T_s)}{^0_0} & = - \s{1} \, \bigl( \bar{\rho}_s + \drho_s \bigr) \\ 
    \tensor{(T_s)}{^0_i} & = \s{1} \, \bigl( \bar{\rho}_s + \barpressure_s \bigr) v_{s,i} \\ 
    \tensor{(T_s)}{^i_0} & = - \s{1} \, \bigl( \bar{\rho}_s + \barpressure_s \bigr) v_{s,i} \\ 
    \tensor{(T_s)}{^i_j} & = \s{1} \, \bigl[ \bigl( \barpressure_s + \dpressure_s \bigr) \delta_i^j + \tensor{(\Pi_s)}{^i_j} \bigr] 
    \\ & \hspace{-0.5cm} \text{where} \quad 
    v_{s,i} = \partial_i v_s + \hat{v}_{s,i} 
    \\ & \hspace{-0.5cm} \text{and} \quad 
    q_{s,i} = \partial_i q_s + \hat{q}_{s,i} 
    \\ & \hspace{-0.5cm} \text{and} \quad 
    (\Pi_s)_{ij} = \partial_{\langle i} \partial_{j\rangle} \Pi_s + \partial_{(i} \hat{\Pi}_{s,j)} + \hat{\Pi}_{s,ij} 
    \per
}
The functions $\bar{\rho}_s(\eta)$ and $\barpressure_s(\eta)$ are the background energy density and pressure, $\drho_s(\eta,\xvec)$ and $\dpressure_s(\eta,\xvec)$ are the energy density and pressure perturbations, $v_{s,i}(\eta,\xvec)$ is the velocity perturbation, $q_{s,i}(\eta,\xvec) = ( \bar{\rho}_s + \barpressure_s) v_{s,i}$ is the momentum density perturbation, and $\tensor{[\Pi_s(\eta,\xvec)]}{^i_j}$ is the anisotropic stress perturbation.  Note that the vector and tensor perturbations are transverse and traceless: $\partial^i \hat{v}_{s,i} = \partial^i \hat{q}_{s,i} = \partial^i (\Pi_s)_{ij} = \tensor{(\Pi_s)}{^i_i} = 0$.  One often frequently encounters the velocity divergence $\theta_s = \partial^i v_{s,i}$, the scaled anisotropic stress $(\sigma_s)_{ij} = 2 (\Pi_s)_{ij} / [3 (\bar{\rho}_s + \barpressure_s)]$, and the energy density contrast $\delta_s = \drho_s / \bar{\rho}_s$.  

\para{Continuity equation}
The stress-energy tensor, which is expressed as $\del{\mu} \tensor{T}{^\mu_\nu} = 0$, is covariantly conserved.  At the background level this leads to the energy continuity equation 
\bes{\label{eq:energy_continuity_equation}
    \partial_\eta \bar{\rho}_s = - 3 \Hcal (1 + w_s) \bar{\rho}_s
    \com
}
where $w_s = \barpressure_s / \bar{\rho}_s$ is the equation of state. 

\para{Coordinate transformations}
Under a general coordinate transformation
\ba{
    x^\mu \mapsto x^\mu + \xi^\mu(x) 
}
where we can write $\xi^0 = T$ and $\xi^i = \partial^i L + \hat{L}^i$ where $\partial^i \hat{L}_i = 0$.  The functions $T(\eta,\xvec)$ and $L(\eta,\xvec)$ parametrize the two scalar gauge freedoms, and the function $\hat{L}^i(\eta,\xvec)$ parametrizes the single vector gauge freedom.  Invariance of the spacetime interval to linear order in $\xi^\mu$ implies the following transformation rules for the metric perturbations: 
\bsa{}{
    A & \mapsto A - \partial_\eta T - \Hcal T \\ 
    B & \mapsto B + T - \partial_\eta L \\ 
    C & \mapsto C - \Hcal T - \tfrac{1}{3} \nabla^2 L \\ 
    E & \mapsto E - L \\ 
    \hat{B}_i & \mapsto \hat{B}_i - \partial_\eta \hat{L}_i \\ 
    \hat{E}_i & \mapsto \hat{E}_i - \hat{L}_i \\ 
    \hat{E}_{ij} & \mapsto \hat{E}_{ij} 
    \per
}
Recall that $\Hcal = a H$ is the comoving Hubble parameter.  Note that the tensor perturbations remain invariant.  In addition, the matter perturbations transform as 
\bsa{}{
    \drho_s & \mapsto \drho_s - \partial_\eta \bar{\rho}_s \, T \\ 
    \dpressure_s & \mapsto \dpressure_s - \partial_\eta \barpressure_s \, T \\ 
    q_{s,i} & \mapsto q_{s,i} + \bigl( \bar{\rho}_s + \barpressure_s \bigr) \, \partial_\eta L_i \\ 
    v_{s,i} & \mapsto v_{s,i} + \partial_\eta L_i \\ 
    (\Pi_s)_{ij} & \mapsto (\Pi_s)_{ij} 
    \per
}
Note that the anisotropic stress remains invariant. 

\para{Gauge fixing}
A judicious choice of the scalar transformation parameters $T(\eta,\xvec)$ and $L(\eta,\xvec)$ may be used to remove some of the scalar metric perturbations; this procedure is known as gauge fixing.  Popular gauge choices have special names; for example, Newtonian gauge sets $B = E = 0$, synchronous gauge sets $A = B = 0$, spatially flat gauge sets $C = E = 0$, uniform density gauge sets $\drho_s = 0$, and comoving gauge sets $q_s = 0$ for some species $s$ (typically the inflaton during inflation).  

\para{Gauge-invariant variables}
Whenever possible, it is best to work with quantities that are left unchanged under a coordinate transformation, at least to linear order in $\xi^\mu$; these are known as first-order gauge-invariant variables~\cite{Bardeen:1980kt,Bardeen:1983qw}.  Useful gauge-invariant variables include the comoving density contrast $\Delta(\eta,\xvec) = \sum_s \Delta_s$, the curvature perturbation $\zeta(\eta,\xvec) = \sum_s \zeta_s$, and the curvature perturbation $\Rcal(\eta,\xvec) = \sum_s \Rcal_s$.  Each of these quantities can be expressed as a sum over gauge-invariant components for each particle species, 
\begin{subequations}\label{eq:gauge_invariant_variables}
\ba{
    \Delta_s & = \frac{\drho_s}{\bar{\rho}_s} + \frac{\partial_\eta \bar{\rho}_s}{\bar{\rho}_s} \, \bigl( v_s + B \bigr) \\ 
    \zeta_s & = - C + \tfrac{1}{3} \nabla^2 E + \Hcal \frac{\drho_s}{\partial_\eta \bar{\rho}_s} \\ 
    \Rcal_s & = - C + \tfrac{1}{3} \nabla^2 E - \Hcal \, \bigl( v_s + B \bigr) 
    \per
}
Similarly, the relative curvature perturbation carried by two different particle species, denoted by $S_{sr}(\eta,\xvec)$ for species $s$ and $r$, is a gauge-invariant variable known as the isocurvature perturbation, 
\ba{
    S_{sr} 
    = -3 \bigl( \zeta_s - \zeta_r \bigr) 
    = \frac{1}{1 + w_s} \frac{\drho_s}{\bar{\rho}_s} - \frac{1}{1 + w_r} \frac{\drho_r}{\bar{\rho}_r} 
    \com
}
\end{subequations}
where we have used the energy continuity equation \eqref{eq:energy_continuity_equation} in the second equality. 

\para{Adiabatic and isocurvature modes} 
To solve for the evolution of the cosmological perturbations, it is necessary to set initial conditions for the various gauge-invariant variables.  The initial condition is said to be in the adiabatic mode if $S_{sr}(\eta_\mathrm{init},\xvec) = 0$ for all species $s$ and $r$.  This is a well-motivated assumption since it arises naturally in single-field inflation, where all of the various particle species arise from a common source, namely, the inflaton field.  However, if any of the $S_{sr}$ are allowed to be nonzero initially, then an isocurvature mode is said to be present.  

\para{Power spectra}
The two-point correlations of the curvature and isocurvature are encoded in the corresponding power spectra.  We can write the power spectra as 
\bes{
    \Delta_{XY}^2(\eta,k) = \frac{k^3}{2\pi^2} \int \! \dd^3 \rvec \, \expval{0^\IN}{X(\eta,\xvec) Y(\eta,\yvec)}{0^\IN} \, \ee^{- \ii \kvec \cdot \rvec}  
}
where $\yvec = \xvec + \rvec$ and where $X, Y$ stand for either $\Rcal$ or $S$.  For CMB studies it is customary to parametrize the initial curvature power spectrum as $\Delta_\Rcal^2 \equiv \Delta_{\Rcal\Rcal}^2 = A_s \, (k / k_\CMB)^{n_s - 1}$ where the fiducial wave number $k_\CMB = 0.05 \, a_0 \Mpc^{-1}$ corresponds to the typical scales probed by CMB observations. 

\section{de Sitter spacetime}
\label{app:dS_spacetime}

This appendix contains expressions for solutions of the mode equations in de Sitter spacetime, a discussion of why CGPP does not take place in a de Sitter universe, and a discussion of applications to CGPP in an inflationary cosmology.  This appendix is partly adapted from \rref{Birrell:1982ix}.  

\para{de Sitter spacetime}
The de Sitter spacetime \cite{deSitter:1916zza,deSitter:1916zz,deSitter:1917zz} is defined as the set of points $(z_0, z_1, z_2, z_3, z_4)$ that lie on the four-dimensional hyperboloid $z_0^2 - z_1^2 - z_2^2 - z_3^2 - z_4^2 = -1/H^2$ that is embedded in five-dimensional Minkowski spacetime with a length element 
\ba{
    (\dd s)^2 = \s{1} \biggl[ -(\dd z_0)^2 + \sum_{i=1}^4 (\dd z_i)^2 \biggr] 
    \per
}
Points in de Sitter space can be written as 
\bes{
    z_0 & = (2\eta)^{-1}(\eta^2 - x_1^2 - x_2^2 - x_3^2 - 1/H^2) \\ 
    z_1 & = - (H\eta)^{-1}x_1 \\ 
    z_2 & = - (H\eta)^{-1} x_2 \\ 
    z_3 & = - (H\eta)^{-1} x_3 \\ 
    z_4 & = (2\eta)^{-1}(x_1^2 + x_2^2 + x_3^2 - \eta^2 - 1/H^2)
}
using the four-dimensional coordinates $\eta,x_1,x_2,x_3 \in (-\infty,\infty)$, and the metric is 
\ba{
    (\dd s)^2 = \frac{\s{1}}{H^2\eta^2} \Bigl[ -(\dd \eta)^2 + \sum_{i=1}^3 (\dd x_i)^2 \Bigr] 
    \per
}
For $\eta \in (-\infty,0)$ the distance between points of fixed comoving separation is growing like $-1/H\eta$ as $\eta$ increases, at $\eta = 0$ there is a coordinate singularity, and for $\eta \in (0,\infty)$ the distance is shrinking.  The de Sitter spacetime metric takes the same form as the spatially flat FLRW metric in \eref{eq:FLRW_metric} with $a(\eta) = -1/H\eta$.  One can verify that the Hubble parameter and the Ricci scalar are constant: $H = a^\prime/a^2$ and $\s{1} \s{3} R = 12 H^2$. 

\para{Mode functions}
In de Sitter spacetime with Hubble parameter $H$, the scalar field's mode equation \eqref{eq:mode_equation} becomes 
\bes{
    & \chi_k^\pprime(\eta) + \omega_k^2(\eta) \, \chi_k(\eta) = 0 
    \qquad \text{where} 
    \\ & \quad 
    \omega_k^2(\eta) = k^2 + \bigl( m^2/H^2 - 2 + 12 \xi \bigr) / \eta^2
    \per
}
There is an infinite one-parameter family of models with $m^2/H^2 - 2 + 12 \xi = 0$ for which the dispersion relation is independent of time $\omega_k^2 = k^2$, the mode functions are the usual $\chi_k \propto \ee^{\mp \ii k \eta}$ of Minkowski spacetime, and no particle production occurs.  To study solutions for general parameters, one can first put this equation into a familiar form by defining $x = k\eta \in (-\infty,\infty)$, $y = \chi_k(\eta) / \sqrt{k\eta}$, and 
\ba{\label{eq:nu_def}
    \nu = \sqrt{\frac{9}{4} - \frac{m^2}{H^2} - 12 \xi}
    \per
}
With these substitutions the mode equation is recast in the form of Bessel's equation, 
\ba{
    x^2 y^\pprime(x) + x y^\prime(x) + \bigl( x^2 - \nu^2 \bigr) \, y(x) = 0 
    \per
}
Its solutions are linear combinations of Bessel functions of the first and second kinds $J_\nu(x)$ and $Y_\nu(x)$.  In particular, we focus on Hankel functions of the first and second kinds $H_\nu^{(1,2)}(x) = J_\nu(x) \pm \ii Y_\nu(x)$ and write the solutions as 
\bes{
    \chi_k(\eta) & = C_1 \, \sqrt{-k\eta} \, H_\nu^{(1)}(-k\eta) 
    \\ & \quad 
    + C_2 \ \sqrt{-k\eta} \ H_\nu^{(2)}(-k\eta) 
    \per
}
Note that both the square-root function and the Hankel functions are double valued in the complex $k\eta$ plane with a single branch point at $k\eta = 0$, and we can orient their branch cuts to lie along the positive real axis, $k\eta \in \mathbb{R}^{>0}$.  Similarly, as $m/H$ and $\xi$ are analytically continued into the complex plane, $\nu$ is interpreted as the double-valued square-root function, and the Hankel functions may be evaluated with the complex index.  Note that Hankel functions satisfy the relation $H_{-\nu}^{(1,2)}(z) = \ee^{\pm \ii \pi \nu} H_\nu^{(1,2)}(z)$ for $z,\nu \in \mathbb{C}$.  

\para{Bunch-Davies vacuum condition} 
The $\IN$ and $\OUT$ mode functions $\chi_k^\IN(\eta)$ and $\chi_k^\OUT(\eta)$ correspond to the solutions that have the asymptotic forms in \eref{eq:chik_IN_OUT_def}.  The complex integration constants $C_1$ and $C_2$ can be chosen to yield the $\IN$ mode function $\chi_k^\IN(\eta)$.  Hankel functions are asymptotic to \cite{abramowitz+stegun}
\bes{
	H_{\nu}^{(1)}(z) & \sim \sqrt{ \tfrac{2}{\pi z} } \, \ee^{\ii (z - \nu \pi/2 - \pi/4)} 
    \quad \text{as $|z| \to \infty$} \\ 
    H_{\nu}^{(2)}(z) & \sim \sqrt{ \tfrac{2}{\pi z} } \, \ee^{-\ii (z - \nu \pi/2 - \pi/4)} 
    \quad \text{as $|z| \to \infty$} 
}
for $\mathrm{arg}(z)$ throughout the first sheet, and for fixed $\nu$.  Choosing $C_1 = \sqrt{\pi/4k} \, \ee^{\ii(\nu \pi/2 + \pi/4)}$ and $C_2 = 0$ gives 
\bes{\label{eq:chikIN_dS}
    \chi_k^\IN(\eta) & = \sqrt{\frac{\pi}{4k}} \sqrt{-k\eta} \, H_\nu^{(1)}(-k\eta) \, \ee^{\ii(\nu \pi/2 + \pi/4)} 
    \com
}
which is asymptotic to 
\ba{\label{eq:dSearly}
    \chi_k^\IN(\eta) \sim \frac{1}{\sqrt{2k}} \, \ee^{-\ii k \eta} 
    & \quad \text{as} \quad k\eta \to - \infty 
}
at early times for any value of $\nu$.  For $k\eta \in (0,\infty)$ we evaluate these functions above the cut on the first Riemann sheet.  These mode functions solve the scalar field's equation of motion in de Sitter spacetime along with the Bunch-Davies initial condition. 

\para{Special cases}
For models involving a massless scalar, or more generally $m \ll H$, the mode functions simplify.  Here we regain expressions that are familiar from studies of scalar inflaton perturbations during inflation \cite{Baumann:2022mni}.  For the massless and conformally coupled scalar, setting $m=0$ and $\xi=\ssfrac{1}{6}$ yields 
\bes{
    \chi_k^\IN(\eta) 
    & = \ii \sqrt{\frac{\pi}{4k}} \sqrt{z} \, H_{1 \mkern-2.5mu / \mkern-1.5mu 2}^{(1)}(z)  = \frac{1}{\sqrt{2k}} \, \ee^{\ii z} 
    \com
}
whereas for the massless and minimally coupled scalar, setting $m=0$ and $\xi=0$ yields
\bes{ \label{eq:eoiminimal}
    \chi_k^\IN(\eta) 
    = - \sqrt{\frac{\pi}{4k}} \sqrt{z} \, H_{3 \mkern-2.5mu / \mkern-1.5mu 2}^{(1)}(z) 
    = \frac{1}{\sqrt{2k}} \left( 1 + \frac{\ii}{z} \right) \ee^{\ii z} 
}
where $z \equiv - k \eta$.  
As expected, the massless and conformally coupled scalar's mode functions are equivalent to the positive-frequency mode functions in Minkowski spacetime; see the discussion in \sref{sub:conformal_symmetry}.  However, the minimally coupled scalar's mode functions grow relatively by a factor of $1/k\eta$ as modes leave the horizon, and $-1 \ll k \eta < 0$.  

\para{Connection to inflation}
During the epoch of inflation, the Hubble parameter varies slowly ($\epsilon = -H^\prime / a H^2 \ll 1$), and the expanding branch of de Sitter spacetime is a good approximation for the spacetime.  Consequently, the de Sitter mode functions \eqref{eq:chikIN_dS} are a good approximation of the evolution of the spectator field.  At early times they are compatible with the Bunch-Davies initial condition \eqref{eq:dSearly}.  At late times they progress to 
\bsa{}{
    \label{eq:dSeoi}
    \chi_k^\IN(\etae) 
    & = \sqrt{\frac{\pi}{4 \ae \He}} \, H_\nu^{(1)}(k / \ae \He) \, \ee^{\ii (\nu \pi/2 + \pi/4)} \\ 
    \label{eq:dSeoione} 
    |\chi_k^\IN(\etae)| & = \dfrac{1}{\sqrt{2k}} 
    \quad \text{for $m=0$, $\xi=\ssfrac{1}{6}$} \\ 
    \label{eq:dSeoitwo} 
    |\chi_k^\IN(\etae)| & = \dfrac{1}{\sqrt{2k}} \sqrt{1 + \dfrac{a_e^2 H_e^2}{k^2}} 
    \quad \text{for $m=0$, $\xi=0$} 
    \com
}
where we have defined the ``end of inflation'' as $\etae = - 1 / \ae \He$.  Although the Hubble parameter begins to vary significantly at the end of inflation, thus invalidating the de Sitter solution, these expressions typically provide reliable approximations that can be used as initial conditions for the mode equations during reheating after inflation.  

\section{JWKB Method}
\label{app:jwkb_method}

This appendix contains a review of the Jeffreys-Wentzel-Kramers-Brillouin method \cite{Jeffreys:1925,Wentzel:1926aor,Kramers:1926njj,Brillouin:1926blg}.  The JWKB method can be applied to derive approximate solutions to equations of the form 
\ba{\label{eq:WKB_diff_eqn}
    y^\pprime(x) + Q(x) \, y(x) = 0 
}
where $Q(x)$ is a differentiable function.  See \rref{BenderOrszag:1999} for an introduction to asymptotic methods and perturbation theory, and see \rref{Parker:2009uva} for applications of JWKB to QFT in curved spacetime (such as adiabatic subtraction and trace anomaly).  

\para{Motivation}
Equations of this form arise in the description of both classical and quantum physical systems.  In classical mechanics the displacement of a one-dimensional harmonic oscillator with a time-dependent natural frequency is governed by $d^2 x/dt^2 + \omega^2(t) x(t) = 0$.  In quantum mechanics the wave function of a nonrelativistic mass moving in a one-dimensional static potential is governed by $d^2 \psi/dx^2 + (2m/\hbar^2) [E-V(x)] \, \psi(x) = 0$.  In quantum field theory the Fourier mode amplitudes of a free scalar field in an FLRW spacetime are governed by $\chi_k^\pprime(\eta) + \omega_k^2(\eta) \, \chi_k(\eta) = 0$, which we encountered in \eref{eq:mode_equation}.  Similar second-order linear equations appear in the study of free fields with spin; see \sref{sec:Spin}.  

\para{Strategy}
The JWKB method draws upon perturbation theory and asymptotic series to tackle equations of the form of \eref{eq:WKB_diff_eqn} (as well as more general equations).  To appreciate the strategy note that if $Q(x) = q^2$ is a constant, one can immediately identify the four-dimensional space of solutions $y(x) = C_1 \, \ee^{-\ii q x} + C_2 \, \ee^{\ii q x}$ for arbitrary complex coefficients $C_1, C_2 \in \mathbb{C}$.  This observation motivates one to seek a solution for general $Q(x)$ using perturbation theory, with the constant case taken as the expansion point.  The JWKB method entails the construction of a power series.  With the exception of special cases, such as the constant $Q(x) = q^2$, the resultant power series is divergent.  However the first few terms in this series typically provide a good approximation to the desired solution.  In this appendix we explain the conceptual underpinnings of the JWKB method and remark upon its application to QFT in curved spacetime. 

\para{Adiabatic parameter}
Since \eref{eq:WKB_diff_eqn} is easily solved for the constant $Q(x)$, this suggests that approximate solutions should be available for slowly varying $Q(x)$ with small time derivatives.  To seek out such solutions, introduce an expansion parameter $\varepsilon \in \mathbb{R}$ at each appearance of a derivative: $\partial_x \to \varepsilon \, \partial_x$.  In principle $Q(x)$ may be constructed from derivatives as well, so we should write $Q(x) = \sum_{n=0}^\infty \varepsilon^n Q_n(x)$.  With these substitutions one obtains an infinite one-parameter family of equations,\footnote{It may appear that introducing $\varepsilon$ has made the problem more difficult since, rather than needing to solve the single equation \eqref{eq:WKB_diff_eqn}, it is now necessary to solve an infinite family of equations.  However, if we assume that the solutions of \eref{eq:WKB_diff_eqn_with_epsilon} can be expressed as a Taylor series in powers of an arbitrary $\varepsilon$, then we can solve order by order in powers of $\varepsilon$.  In this way perturbation theory substitutes one hard problem for an infinite number of easy problems \cite{BenderOrszag:1999}.}
\ba{\label{eq:WKB_diff_eqn_with_epsilon}
    \varepsilon^2 \, y^\pprime(x;\varepsilon) + \sum_{n=0}^\infty \varepsilon^n Q_n(x) \, y(x;\varepsilon) = 0 
    \per
}
Once these equations are solved, the desired solution is $y(x;1)$.  Since $\varepsilon$ tracks time derivatives, it is called the adiabatic parameter (or $T = \varepsilon^{-1}$, the slowness parameter), and $O(\varepsilon^n)$ terms are of $n$th adiabatic order.  

\para{JWKB ansatz}
Since the equation with the constant $Q(x)$ is solved using an exponential function, this observation motivates one to search for solutions that can be written as 
\ba{
    \tilde{y}(x;\varepsilon) = \mathrm{exp}\biggl( \frac{1}{\varepsilon} \sum_{n=0}^\infty \varepsilon^n S_n(x) \biggr) 
    \per
}
It may not be the case that solutions of \eref{eq:WKB_diff_eqn_with_epsilon} can be written in this form.  Nevertheless, it is valuable to search for such solutions since, even if they do not solve the equation, they may provide a good approximation to the desired solution.  Putting this ansatz into the equation for $y(x;\varepsilon)$ and solving order by order in powers of $\varepsilon$ yields 
\bes{\label{eq:S_solution}
    [S_0^\prime(x)]^2 & = - Q_0(x) \\ 
    2 S_0^\prime(x) S_1^\prime(x) & = - Q_1(x) - S_0^\pprime(x) \\ 
    2 S_0^\prime(x) S_2^\prime(x) & = - Q_2(x) - S_1^\pprime(x) - [S_1^\prime(x)]^2 
\com
}
etc.  Note that there are two solution branches corresponding to $S_0(x) = \mp \int^x \! \dd x^\prime \sqrt{-Q_0(x^\prime)}$.

\para{Discussion}
If the series $\tilde{y}(x;\varepsilon)$ is convergent at $\varepsilon = 1$, then $y(x) = \tilde{y}(x;1)$ is the desired solution to the original equation.  However, for the situations of interest, the series typically diverges for any nonzero $\varepsilon$.  In that case the formal series should be understood as an asymptotic expansion of the solution at small adiabatic parameter, 
\ba{\label{eq:y_asymptotic}
    y(x;\varepsilon) \sim \tilde{y}(x;\varepsilon) 
    \quad \text{as} \quad 
    \varepsilon \to 0 
    \per 
}
This outcome may seem like a failure since the goal was to solve the equation with $\varepsilon = 1$, but \eref{eq:y_asymptotic} gives the asymptotic form of the solution as $\varepsilon \to 0$.  Nevertheless, it is often the case that an asymptotic series, when truncated to a low order, provides a good approximation.  For instance, truncating the series at $O(\varepsilon^N)$ and taking $\varepsilon = 1$ results in the function 
\ba{
    y^{(N)}(x) = \mathrm{exp}\biggl( \, \sum_{n=0}^N S_n(x) \biggr) 
    \per
}
To be concrete suppose that $Q_0(x) = [q(x)]^2$ and $Q_{n}(x) = 0$ for $n \geq 1$.  For $N=1$ we then have 
\ba{
    y^{(1)}(x) = \frac{1}{\sqrt{2 q(x)}} \, \mathrm{exp}\biggl( -\ii \int^x \! \dd x^\prime \, q(x) \biggr) 
    \com
}
which is the leading-order JWKB approximation.  Although $y^{(1)}(x)$ is not a solution of \eref{eq:WKB_diff_eqn} (except for special cases like the constant $q(x)$), it typically provides a good approximation.  Adding a few more terms in the asymptotic series will typically improve the approximation, but eventually adding more terms will expose the divergence of the series and cause the approximation to degrade.  The onset of the divergence is controlled by the magnitude of the derivatives of $q(x)$: since $y^{(1)}(x)$ is an exact solution for the constant $q(x)$, one expects $y^{(1)}(x)$ to be a good approximation when $q(x)$ is slowly varying, \ie, adiabatic.  It is customary to define a dimensionless measure of the adiabaticity of the system as 
\ba{
    A(x) = |q^\prime(x) / q(x)^2| 
    \com
}
which corresponds to the fraction by which $q(x)$ changes during each oscillation cycle.  Smaller $A(x)$ corresponds to a more adiabatic system, and $y^{(1)}(x)$ provides a better approximation of the desired solution. 

\para{Applications to QFT in curved spacetime}
In the remainder of this appendix, we focus on applications of the JWKB method to studies of QFT in curved spacetime.  For concreteness we use the scalar field in FLRW spacetime as an example.  The scalar field's Fourier mode amplitudes $\chi_k(\eta)$ obey the second-order, linear differential equation \eqref{eq:mode_equation}.  When the adiabatic parameter $\varepsilon$ is introduced, the equation of motion becomes 
\ba{
    & 
    \varepsilon^2 \, \chi^\pprime(\eta;\varepsilon) + \omega^2(\eta;\varepsilon) \, \chi(\eta;\varepsilon) = 0 
    \quad \text{with} \quad 
    \\ & 
    \omega^2(\eta;\varepsilon) = \bigl[ k^2 + a^2(\eta) m^2 \bigr] 
    + \bigl[ (\tfrac{1}{6} - \xi) a^2(\eta) R(\eta) \bigr] \, \varepsilon^2 
    \com
    \nonumber
}
where we have dropped the subscript $k$ to simplify the notation.  Since the Ricci scalar $R(\eta)$ involves two time derivatives, this term develops a factor of $\varepsilon^2$.  For studies of QFT in curved spacetime, it is customary to write the JWKB ansatz as \cite{Birrell:1982ix}
\bes{
    & \tilde{\chi}(\eta;\varepsilon) = \frac{1}{\sqrt{2 W(\eta; \varepsilon)}} \, \exp\left(- \frac{\ii}{\varepsilon} \int^\eta \! \dd \eta^\prime \, W(\eta^\prime; \varepsilon) \right) 
}
where the power series $W(\eta; \varepsilon) = \sum_{n=0}^\infty \varepsilon^n W_n(\eta)$ satisfies 
\ba{
    \bigl[ W(\eta,\varepsilon) \bigr]^2 = \omega^2(\eta; \varepsilon) - \frac{\varepsilon^2}{2} \biggl[ \frac{W^\pprime(\eta; \varepsilon)}{W(\eta; \varepsilon)} - \frac{3}{2} \biggl( \frac{W^\prime(\eta; \varepsilon)}{W(\eta; \varepsilon)} \biggr)^2 \biggr] 
    \per
}
Equating order by order in powers of $\varepsilon$ yields 
\ba{
    \bigl[ W_0(\eta) \bigr]^2 & 
    = k^2 + a^2(\eta) m^2 \equiv [\omega_0(\eta)]^2 \nonumber \\  
    2 W_0(\eta) W_2(\eta) & 
    = (\tfrac{1}{6} - \xi) a^2(\eta) R(\eta) 
    \nonumber\\ & \quad 
    + \tfrac{3}{4} \left( \frac{W_0^\prime(\eta)}{W_0(\eta)} \right)^2 
    - \tfrac{1}{2} \frac{W_0^\pprime(\eta)}{W_0(\eta)} 
    \com
}
and $W_n(\eta) = 0$ for odd $n$.  Once again there are two solution branches, with $W_0(\eta) = \pm \omega_0(\eta)$, which correspond to positive- and negative-frequency modes, respectively.  The $N$th-order JWKB approximation is obtained by truncating the series at $O(\varepsilon^N)$ and setting $\varepsilon = 1$, 
\ba{
    \chi^{(N)}(\eta) = \frac{\exp\left(- \ii \int^\eta \! \dd \eta^\prime \, \sum_{n=0}^N W_n(\eta^\prime) \right)}{\sqrt{2 \sum_{n=0}^N W_n(\eta)}}  
    \per
}
For instance, the leading-order JWKB approximation is 
\ba{
    \chi^{(0)}(\eta) = \frac{\exp\Bigl(- \ii \int^\eta \! \dd \eta^\prime \, \sqrt{k^2 + a^2(\eta^\prime) m^2} \Bigr)}{\bigl[ 2 \sqrt{k^2 + a(\eta)^2 m^2} \bigr]^{1/2}}  
    \per
}

\para{Adiabatic vacuum}
One application of the JWKB method for QFT in curved spacetime is the definition of the adiabatic vacuum \cite{Birrell:1982ix,Mukhanov:2007zz}.  Although the JWKB approximation is not a solution of the original equation, it can be used as an initial condition at conformal time $\eta_0$ to define an exact solution, 
\ba{
    \chi(\eta_0) = \chi^{(N)}(\eta_0) 
    \quad \text{and} \quad 
    \chi^\prime(\eta_0) = \frac{\dd \chi^{(N)}}{\dd \eta} \Bigr|_{\eta=\eta_0} 
    \per
}
The solution consistent with these initial conditions is known as the $N$th-order adiabatic mode function at time $\eta_0$.  The vacuum state annihilated by the corresponding ladder operators is known as the $N$th-order adiabatic vacuum, $\ket{0_{\mathrm{ad},\eta_0}}$.  If the system is strongly adiabatic at early times, \ie, $\omega^\prime(\eta) / \omega^2(\eta) \to 0$ as $\eta \to - \infty$, then all the vacuum prescriptions agree asymptotically, thereby providing a naturally unique vacuum state.  

\para{Adiabatic regularization}
The construction of adiabatic orders through the JWKB method also proves useful for handling UV divergences through the technique of adiabatic regularization \cite{Parker:2009uva}.  Calculating observables that are nonlinear in the fields, such as expectation values of the stress-energy tensor, generally leads to UV divergences.  Adiabatic regularization entails expressing the observable as a power series in the adiabatic parameter $\varepsilon$ and removing all of the terms at a given order if any of those terms contain a UV divergence.  For a schematic example suppose that we calculate an observable and find that it can be written as $O = (a_0) \varepsilon^0 + (a_2 + b_2 + c_2) \varepsilon^2 + (a_4 + b_4) \varepsilon^4 + \cdots$, where the coefficients represent momentum integrals.  Suppose that $a_0$ and $b_2$ are integrals with UV divergences while the other terms are finite.  Adiabatic regularization then requires the removal of all terms at $\varepsilon^0$ and $\varepsilon^2$, leaving only $O_\mathrm{reg} = (a_4 + b_4) \varepsilon^4 + \cdots$.  Adiabatic regularization proves useful in several derivations, including the conformal anomaly in the trace of the stress-energy tensor.  

\para{Extensions}
Whereas this discussion has focused on the JWKB method for equations of the form of \eref{eq:WKB_diff_eqn}, which arise in bosonic field theories~\cite{Ferreiro:2022ibf,Ferreiro:2023uvr,Maranon-Gonzalez:2023efu}, a similar method is available for the linear, coupled equations that arise in theories of fermions.  See \rref{Durrer:2009ii,Dumlu:2010ua,Landete:2013axa,Landete:2013lpa,delRio:2014cha,Dabrowski:2014ica,Dabrowski:2016tsx,Enomoto:2020xlf,Sou:2021juh,Enomoto:2021hfv,Hashiba:2022bzi,Corba:2022ugu}.  See \rref{SolaPeracaula:2022hpd} and references therein for a discussion of the implications of the cosmological constant problem.  

\bibliographystyle{apsrmp4-1}
\bibliography{RMP}

\end{document}